\documentclass[11pt,a4paper]{article}

\usepackage{jheppub}
\usepackage{bm,longtable,enumerate}
\usepackage{indentfirst}
\usepackage{setspace}
\usepackage{multicol}
\usepackage{multirow}

\def\beq{\begin{equation}}
\def\eeq{\end{equation}}
\def\beqa{\begin{eqnarray}}
\def\eeqa{\end{eqnarray}}
\usepackage{array}

\title{Melting of heavy vector mesons and quasinormal modes in a finite density plasma from holography}
\author[a]{Luis A. H. Mamani,}
\author[b]{Defu Hou,}
\author[c]{and Nelson R. F. Braga}
\affiliation[a]{Centro de Ci\^encias Exatas Naturais e Tecnológicas,\\ Universidade Estadual da Regi\~ao Tocantina do Maranh\~ao,\\ Rua Godofredo Viana 1300, 65901- 480, Imperatriz, MA, Brazil.}
\affiliation[b]{Institute of Particle Physics and Key Laboratory of Quark and Lepton Physics (MOS), \\
Central China Normal University, Wuhan 430079, P.R. China.}
\affiliation[c]{Instituto de F\'{i}sica, Universidade
Federal do Rio de Janeiro,\\
Caixa Postal 68528, RJ 21941-972, Brazil.}
\emailAdd{luis.mamani@uemasul.edu.br}
\emailAdd{houdf@mail.ccnu.edu.cn}
\emailAdd{braga@if.ufrj.br}

\abstract{
In this work, we investigate the melting of charmonium states within a holographic QCD model in the context of Einstein-Maxwell-Dilaton (EMD) theory. In the dual field theory, the model describes the heavy mesons inside a finite temperature and density medium. First, we calculate the spectrum at zero temperature. Then, at finite temperature, we obtain the spectral functions, where the heavy vector meson are represented by peaks. We show that the charmonium  melts down at temperatures above the confinement/deconfinement temperature of the quark-gluon plasma. We also observe that the chemical potential speeds up the melting process. This finding is in agreement with results previously reported in the literature. 
In the gravitational side of the theory we solve the perturbation equations in the hydrodynamics limit. From this result we read off the diffusion coefficient by comparing the dispersion relation against the corresponding result obtained in the dual field theory. We also investigate the behavior of the diffusion coefficient as a function of the temperature. The perturbation equations are solved numerically, in order to get the quasinormal frequencies. We report the emergence of a new mode whose real part increases rapidly at a certain value of the chemical potential while its imaginary part decreases with the increasing of the chemical potential. Finally, by comparing against results obtained in the conformal plasma, we observe that the real part of the frequency increases, while the imaginary part decreases when we consider the non-conformal plasma.
}
\arxivnumber{}
\begin{document}
\maketitle
\section{Introduction}
\noindent

Heavy-ion collisions allow us to investigate quantum chromodynamics (QCD) in the laboratory. The medium created after ($A+A$) collisions, known as the quark gluon plasma (QGP),  is very hot and dense with extremely short lifetime ($\sim 5–10 $ fm/c). In this plasma, light quarks and gluons  interact strongly but are not confined inside hadrons. It is believed that one can use heavy mesons as probes in order to extract relevant information of the medium in such extreme conditions \cite{Matsui:1986dk, Ma:2018tmg, Karsch:2005ex}. The idea is that, in contrast to hadrons made of the light quarks: $u$ (up), $d$ (down), and $s$ (strange), that  dissociate at the critical temperature~\cite{Shuryak:1988ck} when the plasma is formed, heavy mesons, made of $c$  (charm) or $b$ (bottom) quarks  survive at higher temperatures. The fraction of heavy mesons produced in a heavy ion collision may serve as an important source of information about the pre-existing QCD. That is the motivation for understanding how the properties of the QGP, like temperature and density, affect the dissociation of charmonium.

An important framework  to investigate the dissociation of heavy vector mesons is the use of  holographic models inspired in the anti-de Sitter/Conformal Field Theory (AdS/CFT) correspondence. In its original form, the AdS/CFT correspondence states a duality between Super Yang-Mills (SYM) theory living on a flat 4-dimensional spacetime, with a supergravity theory living on an AdS$_5 \times S^5$ spacetime \cite{Maldacena:1997re} (see also \cite{Witten:1998qj, Gubser:1998bc}). A  phenomenological approach to gauge/gravity duality,  now called AdS/QCD,   was proposed in Refs.~\cite{Polchinski:2001tt,Boschi-Filho:2002wdj,Boschi-Filho:2002xih}. Since then, a considerable amount of papers were published with similar phenomenological models, see for instance Refs.~\cite{Cherman:2008eh, Abidin:2009aj, Gherghetta:2009ac, Chelabi:2015gpc, Ballon-Bayona:2020qpq, Ghoroku:2005vt,Colangelo:2008us,Grigoryan:2007wn,Vega:2008af,Kwee:2007dd} and references therein. It is worth stressing that, in this, so called bottom up AdS/QCD approach,  the geometry is kept as AdS space-time,  neglecting back-reactions of the fields introduced in the models on the geometry. 
The investigation of hadron dissociation in a thermal medium in the framework of holography was carried out, for instance,  in Refs.~\cite{Colangelo:2009ra, Miranda:2009uw, Mamani:2013ssa, Dudal:2014jfa, Mamani:2018uxf, Bartz:2013asa}, see also references therein. Finite temperature effects in the dual field theory are related to black hole thermodynamics in the gravitational field theory, while finite density effects are related to the charge of the black hole solution. Following the holographic dictionary one may extract relevant information about the dissociation process in the dual field theory. 
Heavy vector mesons have been studied following a bottom up holographic approach in  \cite{Braga:2015jca, Braga:2015lck, Braga:2016wkm, Braga:2017bml, Braga:2017oqw, Braga:2019xwl, Braga:2019yeh}.

 On the other hand, a different approach is followed in the construction of the so called top down holographic models. In this case the gravitational backgrounds are obtained solving Einstein's equations. In other words, back-reaction of the dilaton field on the metric is not neglected. Examples of such Einstein-Dilaton models can be found for instance in Refs.~\cite{Csaki:2006ji, Gursoy:2007er, Gursoy:2008za, Gubser:2008yx, Li:2013oda, Ballon-Bayona:2017sxa, Ballon-Bayona:2021tzw, dePaula:2008fp, Li:2014dsa} and references therein. Investigations of finite density and magnetic field effects in the context of the Einstein-Maxwell-Dilaton models  appear, for example, in Refs.~\cite{He:2013qq, Yang:2014bqa, Dudal:2018rki, Dudal:2017max, Mamani:2020pks, Ballon-Bayona:2020xls, Chen:2018vty, He:2020fdi}.

 In this work we follow the Einstein-Maxwell-Dilaton holographic approach in order to analyse the dissociation of heavy charmonium in a plasma with finite temperature and density. We analyse the thermal spectrum and the quasinormal modes  and compare our findings with results available in the literature.
The paper is organized as follows. In section \ref{Sec:HolographicModel} we present a brief review of the holographic model we are going to work with. Section \ref{Sec:HeavyVector} is devoted to investigate the charmonium states within the holographic model. We calculate the spectrum at zero temperature, then, we introduce finite temperature effects through a black hole embedded in the dual gravitational background. We get the equations of motion describing two sectors: longitudinal and transverse which we write in the Schr\"odinger-like form. In turn, in section \ref{Sec:EffectivePotential} we investigate finite temperature and density effects on the effective potential arising in the Schr\"odinger-like equation. The analysis of the spectral functions for selected values of the temperature and chemical potential are presented and discussed in section \ref{Sec:SpectralFunctions}. It is also interesting to solve the equations of motion using perturbative techniques. This is possible in the so called hydrodynamic limit where the energy and wave-number are smaller than the temperature. We present this analysis in section \ref{Sec:HydrodynamicLimit}. From the solutions in the hydrodynamic limit we calculate the correlation functions in the dual field theory. These results allow us to calculate the quark-number susceptibility that we present in section \ref{Sec:QuarkNumber}. Moreover, it is worth to solve the equations of motion numerically to get the quasinormal frequencies in the gravitational side of the duality. We implement this procedure  in section \ref{Sec:QuasinormalModes}. Finally, our conclusions are presented in section \ref{Sec:Conclusion}. We present complementary material in Appendix \ref{Sec:SpectrumT0}.

\section{Holographic model}
\label{Sec:HolographicModel}

In the following we define the holographic QCD model we are going to work with proposed in Ref.~\cite{He:2013qq}.
The five-dimensional action describing the finite density medium in the dual field theory is given by
\noindent
\begin{equation}\label{EqAction}
S_b=\frac{1}{16\pi G_5}\int d^{5}x\,\sqrt{-g}\left(
R-\frac{f(\phi)}{4}F^{2}-\frac12\,(\partial^{m}\phi)(\partial_{m}\phi)-V(\phi)\right),
\end{equation}
\noindent
where $G_5$ is the gravitational constant in five dimension, $\phi$ is the scalar field and $V(\phi)$ its potential, $f(\phi)$ represents the kinetic function (non-minimal coupling) and $F^2=F_{mn}F^{mn}$, with $ F_{mn} = \partial_m A_n -\partial_n A_m $ and $ A_m$ is the gauge field. The corresponding equations of motion are given by
\noindent
\begin{subequations}
\begin{align}
G_{mn}-\frac{1}{2}(\partial_m\phi)(\partial_n\phi)+\frac{g_{mn}}{4}(\partial^p\phi)(\partial_p\phi)+\frac{g_{mn}}{2}V+\frac{f}{2}\left(\frac{g_{mn}}{4}F^2-F_{mp}F_n^{p}\right)=&\,0,\label{Eq:Einstein} \\
\partial_{m}\left(\sqrt{-g}fF^{nm}\right)=&\,0,\label{Eq:Maxwell} \\
\frac{1}{\sqrt{-g}}\partial_{m}\left(\sqrt{-g}\partial^{m}\phi\right)-\partial_{\phi}V-\frac{f}{4}F^2=&\,0,\label{Eq:KG}
\end{align}
\end{subequations}
where $G_{mn}$ is the Einstein tensor. Eqs.~\eqref{Eq:Einstein},  \eqref{Eq:Maxwell} and  \eqref{Eq:KG}  are the Einstein equations, the Maxwell equations and the Klein-Gordon equation, respectively. As can be seen, these equations are coupled and must be solved simultaneously. 

As we are interested in the finite temperature and density plasma, we need to consider the black hole solution of these set of equations. We consider the ansatz
\noindent
\begin{equation}\label{EqBHmetric}
\begin{split}
ds^2=&\,\frac{1}{\zeta(z)^2}\left(-g(z)dt^2
+\frac{1}{g(z)}dz^2+dx_idx^i\right),\\
A_t=&\,A_t(z),\quad A_{x^1}=A_{x^2}=A_{x^3}=A_{z}=0,\\
\phi=&\,\phi(z).
\end{split}
\end{equation}
\noindent
where $g(z)$ is the horizon (blackening) function, $\zeta(z)$ is a function related to the warp factor, while $A_t(z)$ is the nonzero component of the gauge field which gives rise to finite density in the dual field theory. The black hole solutions are characterized by the presence of an event horizon, $z_h$, where the horizon function vanishes, $g(z_h)=0$. Thus, the holographic coordinate belongs to the interval $0\leq z\leq z_h$. Considering the ansatz \eqref{EqBHmetric} the Einstein equations \eqref{Eq:Einstein} reduce to 
\noindent
\begin{equation}\label{EqsBH}
\begin{split}
\frac{\zeta''}{\zeta}
-\frac{1}{6}\phi'^{\,2}=&\,0,\\
g''-\frac{3\zeta'}{\zeta}g'-f(A_t'\,\zeta)^2=&\,0,\\
V-\frac{f}{2}(\zeta^2\,A_t')^2+3\zeta^5\left(\frac{g\,\zeta'}{\zeta^4}\right)'=&\,0.
\end{split}
\end{equation}
\noindent
Meanwhile, the nontrivial Maxwell equation is given by
\noindent
\begin{equation}\label{Eq:Maxwell2}
\left(\frac{f}{\zeta}A_t'\right)'=0.
\end{equation}
As is usual in this kind of holographic models, the Klein-Gordon equation becomes redundant and can be obtained from the Einstein equations. We point out that these equations are the same as presented in Ref.~\cite{He:2013qq} written in a compact form.

In turn, regularity conditions imposed on the horizon function and gauge field at the horizon requires that
\noindent
\begin{equation}
g(z_h)=0,\qquad\text{and}\qquad A_t(z_h)=0.
\end{equation}
\noindent
Meanwhile, at the boundary the horizon function must reduce to the unity, $g(0)=1$, while the asymptotic expansion of the gauge field takes the form
\noindent
\begin{equation}\label{Eq:GaugeBoundary}
A_t=\mu-\rho z^2+\mathcal{O}(z^4),\qquad\qquad z\to 0 
\end{equation}
\noindent
where $\mu$ is the chemical potential, and $\rho$ the baryon density. Thus, once we solve Eq. \eqref{Eq:Maxwell2} we expand the solution close to the boundary to read off the chemical potential and baryon density by comparing the solution against the asymptotic expansion \eqref{Eq:GaugeBoundary}.

\begin{figure}[ht!]
\centering
\includegraphics[width=7cm]{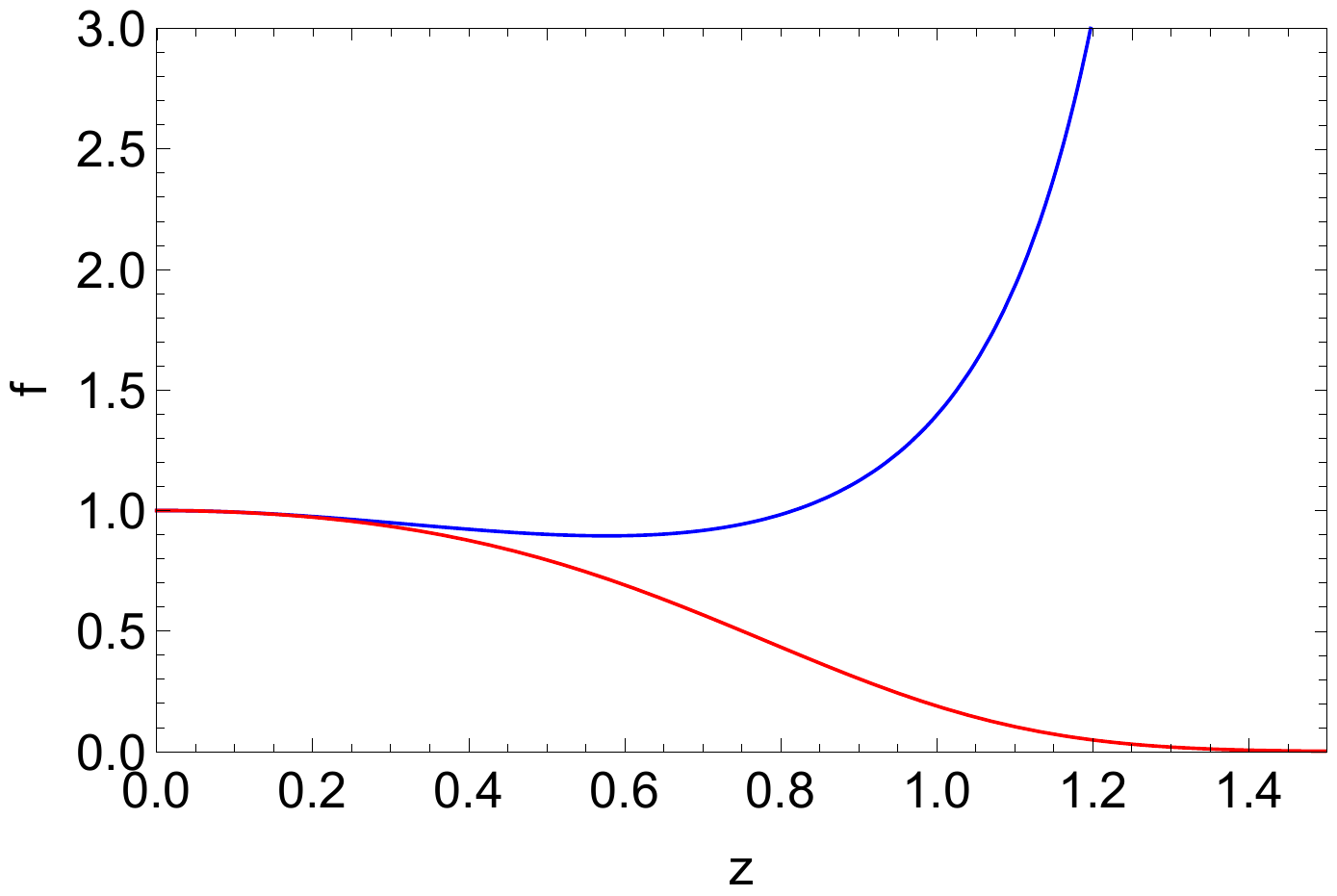}\hfill
\includegraphics[width=7cm]{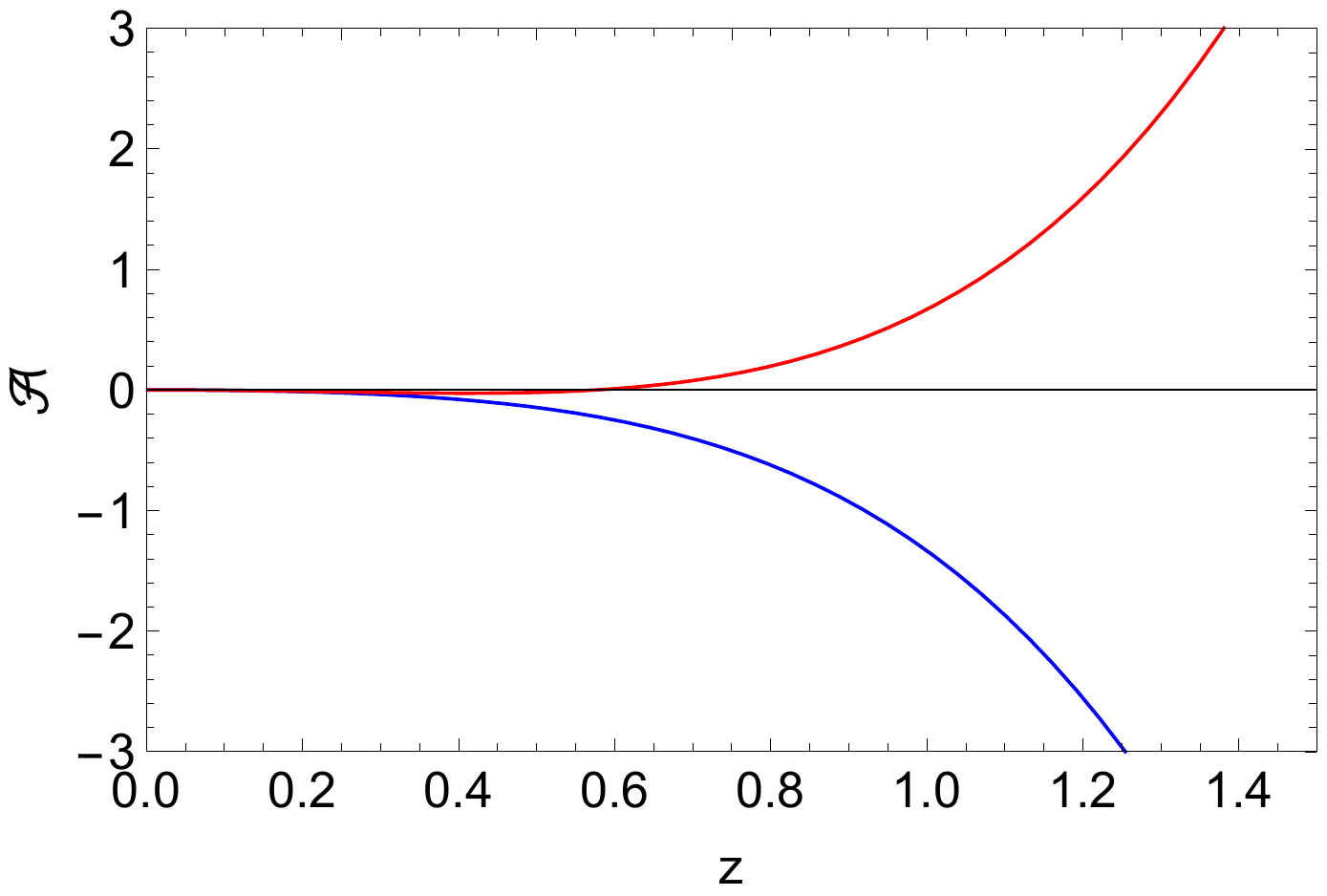}
\caption{
Left: The kinetic function as a function of $z$ for $b=1$ (blue) and $b=-1$ (red) setting $c=1$. Right: The function $\mathcal{A}$ as a function of $z$ for $b=1$ (blue) and $b=-1$ (red) setting $c=1$.
}
\label{Fig1:fA}
\end{figure}

The coupled Eqs. \eqref{EqsBH} may be solved following different approaches, see the discussion in Refs \cite{Ballon-Bayona:2017sxa, Li:2013oda, Ballon-Bayona:2021tzw} and references therein. Thus, the warp factor and the kinetic function are given by \cite{He:2013qq}\footnote{Note that we are using the negative sign of the kinetic function exponent. This is motivated by the original holographic soft wall model \cite{Karch:2006pv}, for a recent discussion on the sign of the soft wall model see Ref.~\cite{Ballon-Bayona:2021ibm}}
\noindent
\begin{equation}\label{Eq:HologrphicModel}
\zeta=\frac{z}{\ell}e^{-\mathcal{A}(z)},\qquad\qquad f=e^{-c\,z^2-\mathcal{A}(z)},
\end{equation}
\noindent
where $\mathcal{A}(z)$ is a function defined by
\noindent
\begin{equation}
\mathcal{A}(z)=-\frac{c}{3}z^2-bz^4.
\end{equation}
\noindent
It is interesting to calculate the asymptotic expansion of the functions $\zeta$ and $f$ close to the boundary, which are given by
\noindent
\begin{equation}
\begin{split}
\zeta=\,&\frac{z}{\ell}\left(1+\frac{c}{3}z^2+\left(b+\frac{c^2}{18}\right)\,z^4\cdots\right),\qquad z\to 0\\
f=\,&1-\frac{2}{3}c\,z^2+\left(b+\frac{2c^2}{9}\right)\,z^4\cdots.\qquad z\to 0
\end{split}
\end{equation}
\noindent
In Fig. \ref{Fig1:fA} we display a plot for functions $\mathcal{A}$ (right panel) and $f$ (left panel) setting $c=1$ and considering two values for the parameter $b$, positive ($b=1$) and negative ($b=-1$). This plot was motivated by the discussion of Ref. \cite{Yang:2014bqa} where a negative signal for $b$ was considered. As can be seen, the kinetic function increases with $z$ for $b>0$ (blue line), while it decreases with $z$ for $b<0$ (red line). In turn, the function $\mathcal{A}$ decreases with $z$ for $b>0$ (blue line), while it increases with $z$ for $b<0$ (red line).
\noindent

\begin{figure}[ht!]
\centering
\includegraphics[width=7cm]{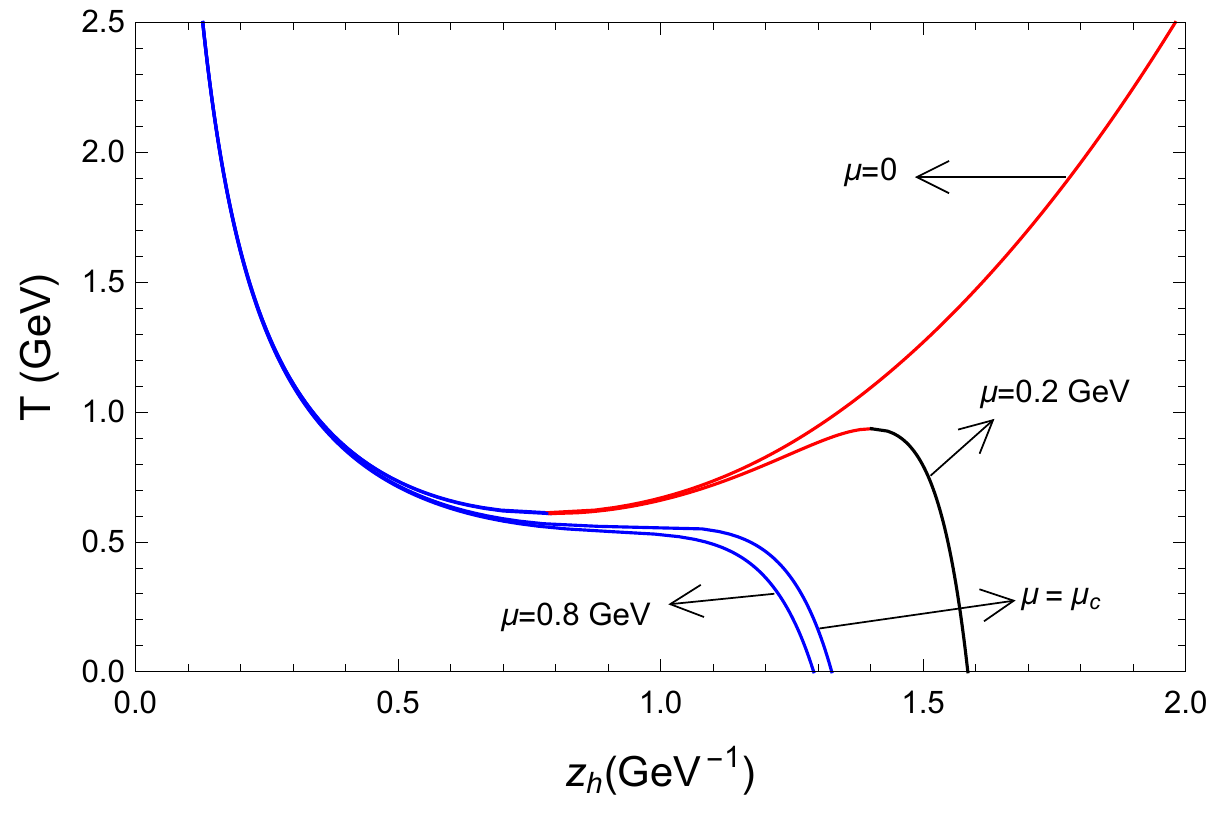}\hfill
\includegraphics[width=7cm]{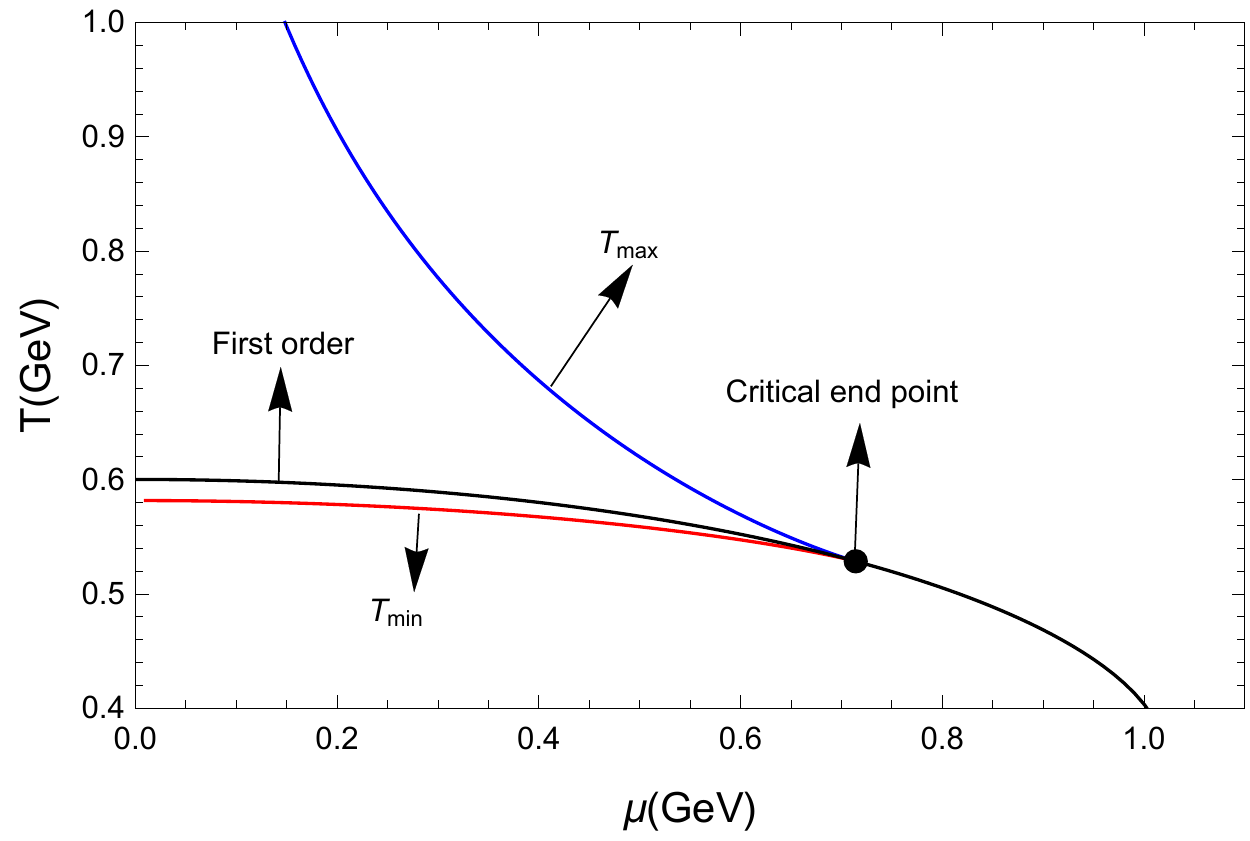}
\caption{
Left: The temperature as a function of $z_h$ for different values of the chemical potential. Right: The phase diagram of the holographic model, where the critical end point is highlighted.
}
\label{Fig1:TF}
\end{figure}

Let us turn our attention to the free parameters $c$ and $b$. They were fixed by phenomenology in Ref. \cite{He:2013qq}, their corresponding values are:
\noindent
\begin{equation}\label{Eq:Parameters}
c=1.16\,\text{GeV}^2,\qquad\qquad b=0.273\,\text{GeV}^4.
\end{equation}
\noindent
We can now solve the background equations by plugging \eqref{Eq:HologrphicModel} into \eqref{Eq:Maxwell2}. Thus, we get a solution for the gauge field
\noindent
\begin{equation}
A_t=c_2+\frac{c_1}{2c\,\ell}e^{c\,z^2}.
\end{equation}
\noindent
We fix the constants using the boundary conditions. Hence, the gauge field and its expansion close to the boundary are given by
\noindent
\begin{equation}\label{Eq:AtCloseBoundary}
A_t=\frac{e^{c\,z^2}-e^{c\,z_h^2}}{1-e^{c\,z_h^2}}\mu,\qquad\qquad A_t=\mu-\frac{c\,\mu}{e^{c\,z_h^2}-1}z^2+\mathcal{O}(z^4).
\end{equation}
\noindent
From the last expression we read off the baryon (charge) density $\rho$ by comparing against \eqref{Eq:GaugeBoundary}. It is worth pointing out that the gauge field does not depend on the parameter $b$. Analogously, we can get a solution for the horizon function,  $g(z)$. Hence,  the thermodynamic variables like the temperature and entropy density are defined by
\noindent
\begin{equation}\label{Eq:Temperature}
T=-\frac{g'(z_h)}{4\pi},\,\qquad\qquad s=\frac{1}{4G_5\,\zeta^3(z_h)}.
\end{equation}
\noindent
A plot of the temperature as a function of $z_h$ is displayed in the left panel of Fig.~\ref{Fig1:TF}. As can be seen, the behavior of the temperature depends on the value of the chemical potential. For $\mu=0$ there is a global minimum, this point splits up the large black hole phase (stable phase) and the small black hole phase (unstable phase). Moreover, for $\mu>0$ there are a  local minimum and a local  maximum,  which merge in the same point for a critical value of the chemical potential, $\mu_{\text{CEP}}$, with the corresponding critical temperature, $T_{\text{CEP}}$. The point  $(\mu_{\text{CEP}},T_{\text{CEP}})$ defines the critical end point in the $\mu-T$ plane.
\noindent

To calculate the phase diagram we need the free energy density, which is calculated using the first law of thermodynamics
\noindent
\begin{equation}
\frac{d\mathcal{F}}{dT}=-s
\end{equation}
\noindent
Then, the integral representation for the free energy density is given by
\noindent
\begin{equation}
\mathcal{F}=\int_{z_h}^{\infty}s(\tilde{z})\left(\frac{dT(\tilde{z})}{d\tilde{z}}\right)d\tilde{z}.
\end{equation}
\noindent
The last result considers the free energy of the thermal gas which is considered to be zero. The numerical results of the phase diagram are displayed in the right panel of Fig.~\ref{Fig1:TF}. In this way we finish the short review of the holographic model we are going to work with, for additional discussions and details see Ref. \cite{He:2013qq}. In the following, we are going to calculate the spectrum of heavy-vector mesons on this background, then, we investigated their melting.

\section{Heavy vector mesons}
\label{Sec:HeavyVector}

The heavy-vector mesons in the dual field theory are described by  five-dimensional gauge field whose action is given by
\noindent
\begin{equation}\label{Eq:5Daction}
S_m=-\frac{1}{16\pi G_5}\int d^{5}x\,\sqrt{-g}\frac{f(\phi)}{4}F^{2}_{V},
\end{equation}
\noindent

where the gauge field is defined by ${F_V}_{mn}=\partial_m A_n-\partial_n A_m$ and $f(\phi)$ the kinetic (non-minimal) function defined in Eq.~\eqref{Eq:HologrphicModel}. The equations of motion obtained from this action are given by
\noindent
\begin{equation}\label{Eq:VectorMesons}
\partial_{m}\left(\sqrt{-g}fF_V^{mn}\right)=0.
\end{equation}
Let us focus in the zero temperature case where the background metric \eqref{EqBHmetric} reduces to
\noindent
\begin{equation}\label{Eq:metric}
\begin{split}
ds^2=&\,\frac{1}{\zeta(z)^2}\left(dz^2+dx_\mu dx^\mu\right).
\end{split}
\end{equation}
\noindent
To simplify the analysis we are going to work in the radial gauge $A_z=0$. Setting $n=z$ in \eqref{Eq:VectorMesons} we get the constraint $\partial_{\alpha}A^{\alpha}=0$. In turn, setting $n=\nu$ we get the equation describing the heavy-vector mesons which may be written as
\noindent
\begin{equation}
\begin{split}
\frac{\zeta}{f}\partial_{z}\left(\frac{f}{\zeta}\partial_z\,A^{\nu}\right)+\square\,A^{\nu}=0.
\end{split}
\end{equation}
\noindent
Introducing the Fourier transform on the gauge field
\noindent
\begin{equation}\label{Eq:FourierTransf}
A_{\nu}(z,x^\mu)=\int\frac{d^4k}{(2\pi)^4}e^{ik_{\alpha}\,x^{\alpha}}\,A_{\nu}(z,k),
\end{equation}
\noindent
it transforms as $A^{\nu}(x^{\mu},z)\to A^{\nu}(k^{\mu},z)$. The equation may be rewritten in the Schr\"odinger-like form using the transformation $A_{\nu}=\xi_{\nu}e^{-B}\psi$, where $\xi_{\nu}$ is a polarization vector and $2B=\ln{\left(f/\zeta\right)}$, thus, the equation becomes
\noindent
\begin{equation}\label{Eq:SchrodingerT0}
-\partial_z^2\psi+V\,\psi=m_n^2\,\psi,
\end{equation}
\noindent
where we have replaced $\square\to m_n^2$, $V$ is the potential given by
\noindent
\begin{equation}\label{Eq:PotentialT0}
V=\left(\partial_z\,B\right)^2+\partial_z^2B.
\end{equation}
\noindent

\begin{figure}[ht!]
\centering
\includegraphics[width=7cm]{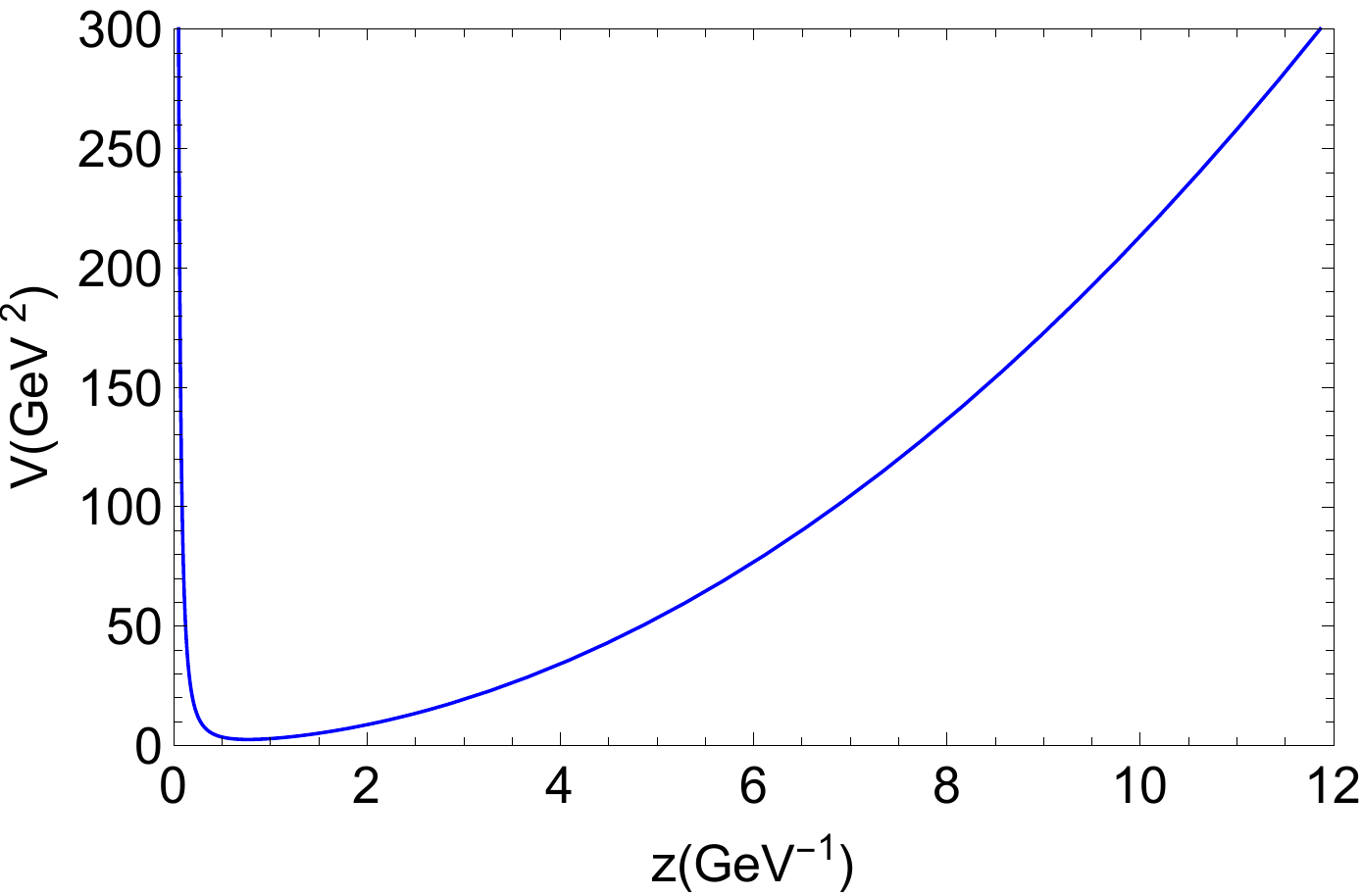}
\caption{
The potential of the Schr\"odinger-like equation.
}
\label{Fig1:PotT0}
\end{figure}

As the background was already fixed, we may solve the eigenvalue problem using a shooting method, for example. It is worth pointing out that the ratio $f/\zeta$ does not depend on the parameter $b$, for that reason the spectrum is insensitive to this parameter. The way this holographic model was built allows us to get an analytic solution for the mass spectrum, which is given by
\noindent
\begin{equation}
m_n^2=4\,c\,(n+1),\qquad\qquad n=0,1,2,\cdots
\end{equation}
\noindent
In the sequence, we fix the free parameter by fitting this formula with the first two resonances of charmonium available from experimental data \cite{Tanabashi:2018oca}, by doing so we get $c=1.46\,\text{GeV}^2$.\footnote{Note that the value of $c=1.46\,\text{GeV}^2$ is slightly different from the value used in \cite{He:2013qq}, see Appendix \ref{Sec:SpectrumT0} for discussion about the value we are using.} A plot of the potential is displayed in Fig.~\ref{Fig1:PotT0}.
\noindent

While the numerical results for the spectrum compared against charmonium experimental data are displayed in Table~\ref{Tab:VectorT0}.

\begin{table}[ht]
\centering
\begin{tabular}{l |c|c}
\hline 
\hline
 $n$ & Holographic model& Charmonium  experimental \cite{Tanabashi:2018oca} \\
\hline 
 $0$ & 2420   & $3096.916\pm 0.011$  \\
 $1$ & 3422  & $3686.109\pm 0.012$  \\
 $2$ & 4191  & $4039\pm 1$ \\
 $3$ & 4839  & $4421\pm 4$  \\
\hline\hline
\end{tabular}
\caption{
The mass of the vector mesons (in MeV) obtained in the holographic model compared against the experimental results from PDG \cite{Tanabashi:2018oca}.
}
\label{Tab:VectorT0}
\end{table}

Having fixed the parameter $c$, the critical end point in the phase diagram lies in a different position in relation to the one obtained in Ref.~\cite{He:2013qq}. Considering $c=1.46\,\text{GeV}$ it lies at $(\mu_{\text{CEP}},T_{\text{CEP}})=(0.708\,\text{GeV}, 0.559\,\text{GeV})$. Finally, the eigenvalue problem solved at zero temperature has real solutions. However, we shall see below that the black hole embedded in the geometry will change the eigenvalues into complex. These states shall be interpreted as quasiparticles characterized by the complex frequencies  whose real part is interpreted as the thermal mass, while their imaginary part related to the decay rate of these states. The corresponding field solutions are called the quasinormal modes. They are the finite temperature version of the normal modes that describe the states at zero temperature.

On the other hand, in the black hole background the problem changes completely due to Poincar\'e symmetry breaking.
To simplify the analysis we are going to work in the radial gauge, $A_z=0$, and considering plane wave solutions in the form $A_{\mu}(x^{\nu},z)=e^{-i\omega\,t+iq\,x^3}A_{\mu}(\omega,q,z)$, where we are considering the direction of propagation $q^{\mu}=(\omega,0,0,q)$. Thus, the equations of motion \eqref{Eq:VectorMesons} can be written in the form:
\noindent
\begin{subequations}
\begin{align}
\partial_{z}\left(\frac{f}{\zeta}\partial_{z}A_{t}\right)-\frac{q\,f}{g\,\zeta}\left(qA_{t}+\omega\,A_{x^{3}}\right)=&\,0,\\
\omega\,\partial_{z}A_{t}+q\,g\,\partial_{z}A_{x^3}=&\,0,\\
\partial_{z}\left(\frac{g\,f}{\zeta}\partial_{z}A_{x^{3}}\right)+\frac{\omega\,f}{g\,\zeta}\left(qA_{t}+\omega\,A_{x^{3}}\right)=&\,0,\\
\frac{g\,\zeta}{f}\partial_{z}\left(\frac{f\,g}{\zeta}\partial_{z}A_{\alpha}\right)+\left(\omega^2-q^2g\right)A_{\alpha}=&\,0.\quad (\alpha=x^1,\,x^2)
\end{align}
\end{subequations}
\noindent
Next, we write the last equations in terms of gauge invariant fields defined by $E_{x^{1}}=\omega\,A_{x^{1}}$, $E_{x^{2}}=\omega\,A_{x^{2}}$, and $E_{x^{3}}=q\,A_{t}+\omega\,A_{x^{3}}$, as
\noindent
\begin{subequations}
\begin{align}
\frac{g\,\zeta}{f}\partial_{z}\left(\frac{f\,g}{\zeta}\partial_{z}E_{\alpha}\right)+\left(\omega^2-q^2g\right)E_{\alpha}=&\,0,\quad (\alpha=x^1,\,x^2)\label{Eq:Transverse},\\
\frac{g\,\zeta}{f}\partial_{z}\left(\frac{f\,g}{\zeta\left(\omega^2-q^2\,g\right)}\partial_{z}E_{x^3}\right)+E_{x^3}=&\,0\label{Eq:Longitudinal}.
\end{align}
\end{subequations}
Eqs.~\eqref{Eq:Transverse} represent the propagation in the transverse direction, while Eq.~\eqref{Eq:Longitudinal} the propagation along the longitudinal direction. It is also possible to rewrite each of these equations into a Schr\"odinger-like form. The Schr\"odinger-like form allows us to investigate the potential and how it will be deformed by the temperature and chemical potential, which is interpreted as the melting of the quasiparticle states. To get the Schr\"odinger-like equation we need to define the tortoise coordinate, $\partial_{r_*}=-g(z)\partial_z$, and the transformation $E_{\alpha}=e^{-B_{T}}\,\psi_{\alpha}$. Thus, Eq.~\eqref{Eq:Transverse} becomes
\noindent
\begin{equation}\label{Eq:SchrodingerTrans}
-\partial^2_{r_*}\psi_{\alpha}+V_T\,\psi_{\alpha}=\omega^2\,\psi_{\alpha},
\end{equation}
\noindent
where $V_T$ is the transverse potential defined by
\noindent
\begin{equation}
V_T=q^2g+\left(\partial_{r_*}B_T\right)^2+\partial^2_{r_*}B_T,
\end{equation}
\noindent
with $2B_T=\ln{\left(f/\zeta\right)}$. By restoring the holographic coordinate, the transverse potential becomes
\noindent
\begin{equation}\label{Eq:SchrodingerPotTrans}
V_T=g\left(q^2+g\left(\partial_{z}B_T\right)^2+\partial_{z}\left(g\,\partial_z B_T\right)\right).
\end{equation}
As can be seen, the transverse potential is zero at the horizon, where $g(z_h)=0$. In the same way, we may write Eq.~\eqref{Eq:Longitudinal} in the Schr\"odinger-like form by using the tortoise coordinate and the transformation $E_{x^3}=e^{-B_L}\psi_{x^3}$ getting
\noindent
\begin{equation}\label{Eq:SchrodingerLong}
-\partial^2_{r_*}\psi_{x^3}+V_L\,\psi_{x^3}=\omega^2\,\psi_{x^3},
\end{equation}
\noindent
where $V_L$ is the longitudinal potential defined by
\noindent
\begin{equation}
V_L=q^2g+\left(\partial_{r_*}B_L\right)^2+\partial^2_{r_*}B_L,
\end{equation}
\noindent
with $2B_L=\ln{\left(f/\left[\zeta(\omega^2-q^2\,g)\right]\right)}$. Restoring the holographic coordinate, the longitudinal potential becomes
\noindent
\begin{equation}\label{Eq:SchrodingerPotLong}
V_L=g\left(q^2+g\left(\partial_{z}B_L\right)^2+\partial_{z}\left(g\,\partial_z B_L\right)\right).
\end{equation}
\noindent
Note that $V_T$ and $V_L$ are the same when $q=0$. Note also that the longitudinal potential vanishes at the horizon. The Schr\"odinger-like form of the differential equations \eqref{Eq:SchrodingerTrans}, \eqref{Eq:SchrodingerLong} 
may be solved close to the horizon where $g(z_h)=0$. Thus, we have the following solutions for both sectors
\noindent
\begin{equation}
\psi_{j}\sim \mathcal{C}_{j}\,e^{-i\,\omega\,r_*}+\mathcal{D}_{j}\,e^{+i\,\omega\,r_*},\qquad (j=\alpha, x^3)
\end{equation}
\noindent
where the first solution is interpreted as an incoming wave falling into the black hole, while the second one as an outgoing wave coming from the black hole interior. To be more precise, we may calculate the incoming $\psi^{(-)}_{j}$ and outgoing $\psi^{(+)}_{j}$ solutions including a few subleading terms in the form
\noindent
\begin{subequations}
\begin{align}
\psi^{(+)}_{j}=\,&e^{+ i\,\omega\,r_*}\left(a^{(+)}_{0j}+a^{(+)}_{1j}(z_h-z)+a^{(+)}_{2j}(z_h-z)^2+\cdots\right)\\
\psi^{(-)}_{j}=\,&e^{- i\,\omega\,r_*}\left(a^{(-)}_{0j}+a^{(-)}_{1j}(z_h-z)+a^{(-)}_{2j}(z_h-z)^2+\cdots\right).
\end{align}
\end{subequations}
\noindent
The coefficients $a_{0j}^{(\pm)}, a_{1j}^{\pm}, \cdots$, are given by $a_{0j}^{(\pm)}=1$, 
\noindent
\begin{equation}
\begin{split}
a_{1j}^{(\pm)}=&\frac{a_{0j}^{(\pm)}}{2z_h(f'(z_h)\mp 2 i \omega)}\left(2q^2z_h-(1+2\,c\,z_h^2)f'(zh)+\delta_{jx^3}\frac{q^2z_h(f'(z_h))^2}{\omega^2}\right),\\
a_{2\alpha}^{(\pm)}=&\frac{a_{0\alpha}^{(\pm)}}{8z_h^2(2\omega\pm if'(z_h))(5\omega\pm4if'(z_h))}\bigg(6\,\omega^2-8\,q^4z_h^2+8c^2z_h^4\omega^2+\\
&+\left[8\,q^2(z_h+2c\,z_h^3)\pm i(23-16\,c\,z_h^2+20\,c^2z_h^4)\omega\right]f'(z_h)-\\
&-4(3+4\,c^2z_h^4)f'(z_h)^2+8z_h\left[q^2z_h\mp i(\omega+2\,c\,z_h^2\omega)\right]f''(z_h)\bigg),\\
a_{2\,x^3}^{(\pm)}=&\frac{a_{0\,x^3}^{(\pm)}}{8z_h^2\omega^4(2\omega\pm if'(z_h))(5\omega\pm4if'(z_h))}\bigg((6+8c^2z_h^4)\,\omega^6-8\,q^4z_h^2\omega^4+\\
&+q^2z_h\omega(\pm25 iq^2z_h+8(1+2cz_h^2)\omega)f'(z_h)^3-12q^4z_h^2f'(z_h)^4+\\
&+4z_h\omega^4(3q^2z_h\mp 2i\omega(1+2cz_h^2))f''(z_h)+[-8q^2z_h^2\omega^2f''(z_h)+\\
&+2\omega^2(q^4z_h^2\mp 5iq^2z_h(1+2cz_h^2)\omega-2(3+4c^2z_h^4)\omega^2)]f'(z_h)^2+\\
&+[\omega(4q^2z_h(1+2cz_h^2)\pm i(23-16cz_h^2+20c^2z_h^4)\omega)\\
&\pm26iq^2z_h^2f''(z_h)]\omega^3f'(z_h)\bigg).
\end{split}
\end{equation}
\noindent

On the other hand, we may solve the Schr\"odinger-like equations close to the boundary, where the normalizable $\psi^{(1)}_{k}$ and non-normalizable $\psi^{(2)}_{k}$ solutions are given by
\noindent\begin{subequations}
\begin{align} 
\psi^{(1)}_{j}=\,& z^{3/2}\left(b_{0j}+b_{2j}z^2+b_{4j}z^4+\cdots\right),\label{Eq:NormSol}\\  
\psi^{(2)}_{j}=\,& z^{-1/2}\left(c_{0j}+c_{2j}z^2+c_{4j}z^4\cdots\right)+d_j\,\psi^{(1)}_{j}\ln{\left(z\right)},\label{Eq:NonNormSol}
\end{align}
\end{subequations}
\noindent
where the coefficients are given by
\noindent
\begin{equation}
\begin{split}
b_{2j}=&\frac{(q^2-\omega^2)}{8}b_{0j},\qquad
c_{4\alpha}=\frac{1}{64}\left(8c^2-3b_{0\alpha}(q^2-\omega^2)^2\right)c_{0\alpha},\qquad\\ d_j=&\frac{(q^2-\omega^2)}{2}c_{0j},
\end{split}
\end{equation}
\noindent
the coefficients $b_{4j}$ and $c_{4x^3}$ are complicated exressions for that reason we do not write them here. Meanwhile, we have the freedom to set $c_{2j}=0$. The next step forward is to write the incoming and outgoing solutions as a linear combination of the normalizable and non-normalizable solutions:
\noindent
\begin{subequations}
\begin{align}
\psi^{(+)}_j=\,& \mathcal{A}^{(+)}_j\,\psi^{(2)}_j+\mathcal{B}^{(+)}_j\,\psi^{(1)}_{j},\\
\psi^{(-)}_j=\,& \mathcal{A}^{(-)}_j\,\psi^{(2)}_j+\mathcal{B}^{(-)}_j\,\psi^{(1)}_{j}
\end{align}
\end{subequations}
\noindent
Analogously, the normalizable and non-normalizable solutions may be written as a linear combination of the incoming and outgoing solutions in the form:
\noindent
\begin{subequations}
\begin{align}
\psi^{(2)}_j=\,& \mathcal{C}^{(2)}_j\,\psi^{(-)}_j+\mathcal{D}^{(2)}_j\,\psi^{(+)}_{j},\\
\psi^{(1)}_j=\,& \mathcal{C}^{(1)}_j\,\psi^{(-)}_j+\mathcal{D}^{(1)}_j\,\psi^{(+)}_{j}.
\end{align}
\end{subequations}
\noindent
The coefficients of the last equations are related through

\begin{equation}\label{Eq:MatrixCoeff}
\begin{pmatrix}
\mathcal{A}^{(+)}_j & \mathcal{B}^{(+)}_j\\
\mathcal{A}^{(-)}_j & \mathcal{B}^{(-)}_j
\end{pmatrix}
=
\begin{pmatrix}
\mathcal{C}^{(2)}_j & \mathcal{D}^{(2)}_j\\
\mathcal{C}^{(1)}_j & \mathcal{D}^{(1)}_j
\end{pmatrix}^{-1}
\end{equation}
\noindent
These relations shall be useful below when we calculate the spectral functions, see Refs~\cite{Miranda:2009uw, Mamani:2013ssa, Mamani:2018uxf} for additional details.

\section{The effective potential}
\label{Sec:EffectivePotential}

\begin{figure}[ht!]
\centering
\includegraphics[width=7cm]{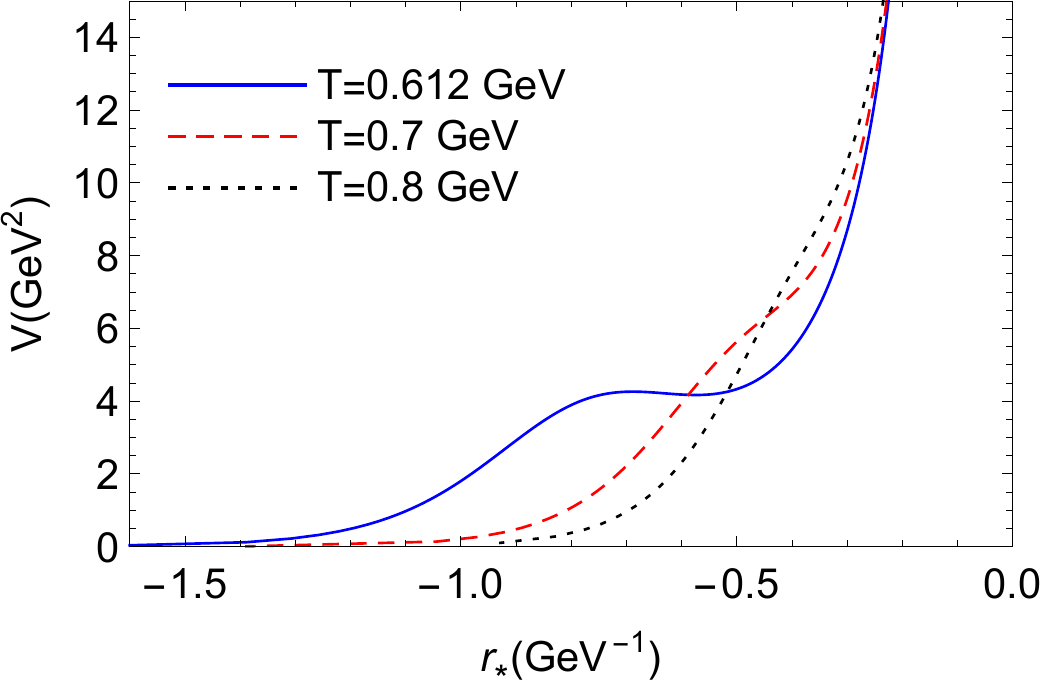}\hfill
\includegraphics[width=7cm]{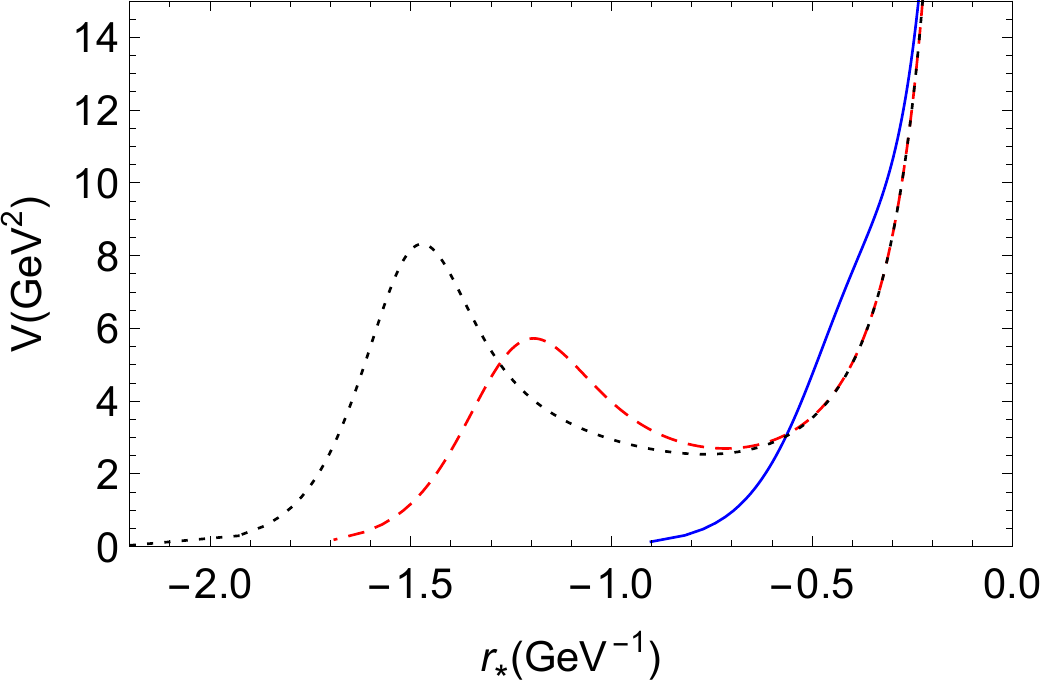}
\caption{
Left: The potential as a function of the tortoise coordinate $r_*$ for $\mu=0$, and $q=0$. Blue line represents the results for $T=T_{\text{min}}=0.612\,\text{GeV}$, red dashed line for $T=0.7\,\text{GeV}$, and black dashed line for $T=0.8\,\text{GeV}$. Right: The figure shows the potential as a function of $r_*$ for $\mu=0.2\,\text{GeV}$ and the isotherm $T=0.8\,\text{GeV}$ setting $q=0$. Blue line represents the large black hole branch, red dashed line the second branch, and black dashed line the third one. These branches can be seen on the left panel of Fig.\ref{Fig1:TF}.
}
\label{Fig1:Potential}
\end{figure}

Let us start by investigating the case when $\mu=0$. In this case the plot of the temperature as a function of $z_h$, see left panel of Fig.~\ref{Fig1:TF}, has two branches: large and small black holes. Considering the stable regime, i.e., the large  black hole branch, the temperature belongs to the interval $T_{\text{min}}\leq T <\infty$. A plot of the potential as a function of the tortoise coordinate for selected values of the temperature setting $q=0$ is displayed in the left panel of Fig.~\ref{Fig1:Potential}. As can be seen, for $T=T_{\text{min}}=0.612\,\text{GeV}$ (blue line), the potential has a small potential well, this means that probably we will not find quasiparticle states for this temperature. In turn, for $T=0.7\,\text{GeV}$ (red dashed line) and $T=0.8\,\text{GeV}$ (black dashed line) there is no potential well meaning that the probability of finding quasiparticle states should be practically zero. It is worth comparing the potential at finite temperature against the potential at zero temperature displayed in Fig.~\ref{Fig1:PotT0}. As can be seen, the temperature deforms the potential well. This deformation is interpreted as the dissociation of bound states, that are thermally at zero temperature.
\noindent

\begin{figure}[ht!]
\centering
\includegraphics[width=7cm]{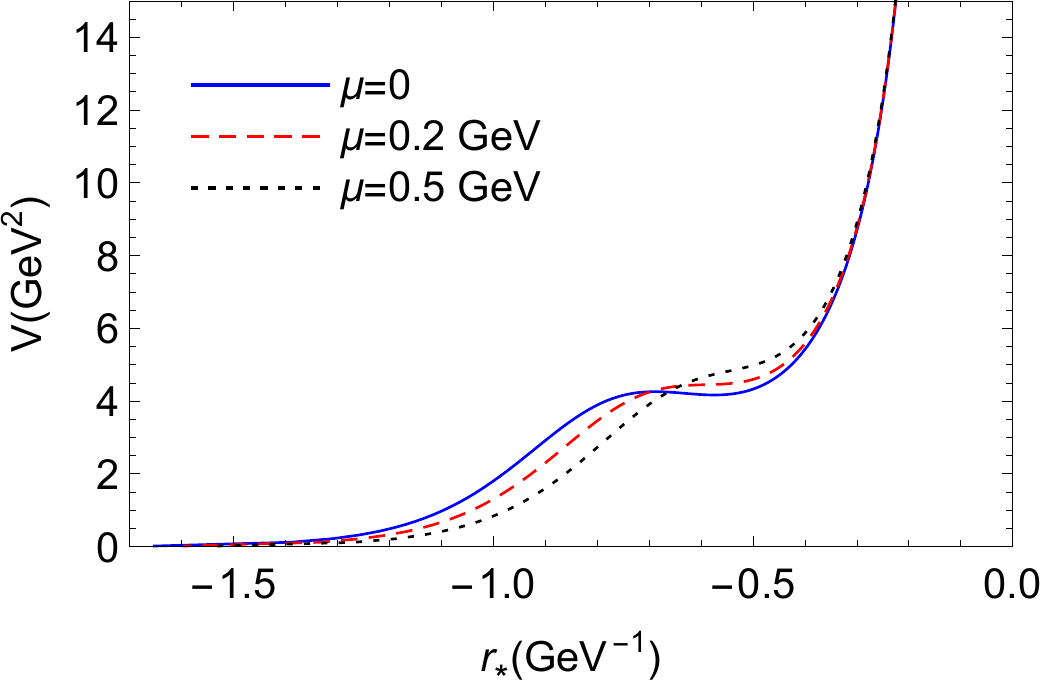}
\caption{
The potential as a function of the tortoise coordinate $r_*$ for $T=T_{\text{min}}=0.612\,\text{GeV}$, and different values of the chemical potential: $\mu=0$ (blue line), $\mu=0.2\,\text{GeV}$ (red dashed line), and $\mu=0.5\,\text{GeV}$ (black dashed line).
}
\label{Fig1:PotentialMu}
\end{figure}

Meanwhile, turning on the chemical potential, $\mu\neq0$, there are three branches arising in the plot of the temperature as a function of $z_h$, as seen on the left panel of Fig.~\ref{Fig1:TF}, depending on the value of $\mu$. We displayed our numerical results for $\mu=0.2\,\text{GeV}$ and the isotherm at $T=0.8\,\text{GeV}$ in the right panel of Fig.~\ref{Fig1:Potential}. The first branch of Fig.~\ref{Fig1:TF}, where the background is stable, is represented with blue line. Meanwhile, the second branch, where the background is unstable, is represented with red dashed line. While the third branch, where we got a stable solution, is represented with black dashed line. As can be seen, it is possible to find quasiparticle states in the third branch (black dashed line) because displays a potential well, while in the second branch they shall be in an unstable phase. In turn, it is also illustrative to show the effects of the chemical potential on the deformation of the potential well. In Fig.~\ref{Fig1:PotentialMu}, we displayed the potential fixing the temperature at the isotherm $T=T_{\text{min}}=0.612\,\text{GeV}$, and varying the chemical potential for $q=0$. As can be seen, the potential is also sensitive to the variation of the chemical potential.
\noindent

In conclusion, increasing the temperature and the chemical potential the melting process speeds out. It is also interesting pointing out that these results are qualitatively equivalents to results obtained within the bottom-up holographic QCD models in the literature \cite{Miranda:2009uw, Mamani:2013ssa, Braga:2016wkm, Braga:2017bml, Braga:2017oqw, Mamani:2018uxf, Braga:2019xwl, Braga:2019yeh, Cao:2021tcr}, see also references therein. The difference of our results in relation to those is that the background we are working with was obtained solving the Einstein-Maxwell-Dilaton equations.

\section{Spectral functions}
\label{Sec:SpectralFunctions}

To calculate the correlation functions we need to determine the on-shell action, then, we use the Son-Starinets prescription \cite{Son:2002sd} to read off the correlation functions. Let us start by writing the action \eqref{Eq:5Daction} in the form
\noindent
\begin{equation}
S_m=\frac{1}{16\pi G_5}\int d^5x\sqrt{-g}\frac{f(\phi)}{2}\partial_{m}A_{n}F^{mn}_V.
\end{equation}
\noindent
Plugging the components of the gauge field, the background metric \eqref{EqBHmetric}, and the Fourier transform \eqref{Eq:FourierTransf} the action may be rewritten in the form
\begin{equation}
S_{m}=\frac{1}{32\pi G_5}\int \frac{dqd\omega}{(2\pi)^2}\frac{f}{\zeta}\left(g\,\mathbf{A}(z,-k)\cdot\partial_z \mathbf{A}(z,k)-A_{t}(z,-k)\partial_zA_{t}(z,k)\right)\Bigg{|}_{z_0}^{z_h},
\end{equation}
\noindent
where $\mathbf{A}=(A_{x^1},A_{x^2},A_{x^3})$ is a spatial vector. In terms of the gauge invariant fields this action becomes
\noindent
\begin{equation}\label{Eq:ElectricFieldAction}
\begin{split}
S_m=-\frac{1}{32\pi G_5}\int\frac{dqd\omega}{(2\pi)^2}\frac{g\,f}{\zeta\, \omega^2}\bigg[&\frac{\omega^2}{\omega^2-q^2g}E_{\scriptscriptstyle{x^3}}(z,-k)E'_{\scriptscriptstyle{x^3}}(z,k)\\
&+E_{\scriptscriptstyle{x^1}}(z,-k)E'_{\scriptscriptstyle{x^1}}(z,k)+E_{\scriptscriptstyle{x^2}}(z,-k)E'_{\scriptscriptstyle{x^2}}(z,k)\bigg]\Bigg{|}_{z_0}^{z_h}.
\end{split}
\end{equation}
\noindent
To get the correlation functions it is useful to split up the gauge field as the product of two functions, one of them depending only on the holographic coordinate, $\mathcal{E}_j(z)$, and the other on the wave-number $E^{(-)}_j(k)$
\noindent
\begin{equation}
E_j(z,k)=\mathcal{E}_j(z)E^{(-)}_j(k),\qquad (j=x^1,x^2,x^3)
\end{equation}
\noindent
where the function $\mathcal{E}_j$ is normalized such that $\lim_{z_0\to 0}\mathcal{E}_j(z_0)=1$, we also  consider the ingoing solution at the horizon such that we are computing the retarded Green's function. Thus, the on-shell action can be rewritten in the form
\noindent
\begin{equation}
\begin{split}
S_m=&\frac{1}{32\pi G_5}\int \frac{dqd\omega}{(2\pi)^2}\frac{f\,g}{\zeta}\Bigg[\frac{\omega^2}{\omega^2-q^2g}\bigg(\frac{q^2}{\omega^2}A^{0}_t(-k)A^{0}_t(k)+\frac{q}{\omega}A^{0}_t(-k)A^{0}_{x^3}(k)+\\
&+\frac{q}{\omega}A^{0}_{x^3}(-k)A^{0}_{t}(k)+A^{0}_{x^3}(-k)A^{0}_{x^3}(k)\bigg)\mathcal{E}_{x^3}(z)\mathcal{E}'_{x^3}(z)+\\
&+\sum_{\alpha=x^1,x^2}A^{0}_{\alpha}(-k)A^{0}_{\alpha}(k)\mathcal{E}_{\alpha}(z)\mathcal{E}'_{\alpha}(z)\Bigg]\Bigg{|}_{z_0}^{z_h}
\end{split}
\end{equation}
\noindent
The last expression can be written in a compact form
\begin{equation}
S_m=\int \frac{dqd\omega}{(2\pi)^2} A^{0}_{\mu}(-k)\mathcal{F}^{\mu\nu}(z,k)A^{0}_{\nu}(k)\Bigg{|}_{z_0}^{z_h}
\end{equation}
Thus, we get the current-current correlators using the Son-Starinets prescription $C^{R}_{\mu\nu}(k)=-2\eta_{\mu\gamma}\eta_{\nu\beta}\lim_{z_0\to 0}\mathcal{F}^{\gamma\beta}(z_0,k)$,
\noindent
\begin{subequations}
\begin{align}
\frac{C^{R}_{tt}}{q^2}=\frac{C^{R}_{x^3x^3}}{\omega^2}=-\frac{C^{R}_{tz}}{q\,\omega}=-\frac{C^{R}_{zt}}{q\,\omega}=&-\frac{N_c^2}{16\pi^2 (\omega^2-q^2)}\lim_{z_0\to 0} \frac{1}{\zeta(z_0)}\mathcal{E}'_{x^3}(z)\bigg{|}_{z=z_0},\label{Eq:CurrentCurrentLong} \\
C^R_{\alpha\alpha}=&-\frac{N_c^2}{16\pi^2}\lim_{z_0\to 0} \frac{1}{\zeta(z_0)}\mathcal{E}'_{\alpha}(z)\bigg{|}_{z=z_0}.
\label{Eq:CurrentCurrentTrans}
\end{align}
\end{subequations}
\noindent
To get these results we have considered the fact that $g(z_0)\to 1$, $f(z_0)\to 1$, and $\mathcal{E}_{j}(z_0)\to 1$ in the limit of zero $z_0$. We also considered the relation $G_5=\pi/N_c^2$, where $N_c$ is the number of colors. It is instructive to write an explicit expression for $\mathcal{E}_{j}(z)$ which is obtained from the transformations:
\noindent
\begin{equation}\label{Eq:NormalSol}
\begin{split}
\mathcal{E}_{j}(z)=\,&\left(\frac{\zeta}{f}\right)^{1/2}\left(\frac{\omega^2-q^2g}{\omega^2-q^2}\right)^{\frac{1}{2}\delta_{jx^3}}\left(\psi^{(2)}_{j}+\frac{\mathcal{B}^{(-)}_{j}}{\mathcal{A}^{(-)}_{j}}\psi_{j}^{(1)}\right)
\end{split}
\end{equation}
\noindent
where we have considered $E_{j}^{(-)}(k)=(\omega^2-q^2)^{\frac{1}{2}\delta_{jx^3}}\mathcal{A}^{(-)}_{j}(k)$ in order to guarantee the condition $\mathcal{E}_j(0)=1$. Plugging \eqref{Eq:NormalSol} in \eqref{Eq:CurrentCurrentLong} and \eqref{Eq:CurrentCurrentTrans}, and using also \eqref{Eq:NormSol} ,\eqref{Eq:NonNormSol} we get
\noindent
\begin{subequations}
\begin{align}
C_{x^3x^3}^{R}(\omega,q)=\,&-\frac{N_c^2}{8\pi^2}\left(\frac{\omega^2}{\omega^2-q^2}\right)\lim_{z_0\to0}\left(\frac{c}{2}+d+2d\ln{z_0}+\frac{\mathcal{B}^{(-)}_{x^3}(\omega,q)}{\mathcal{A}^{(-)}_{x^3}(\omega,q)}+\cdots\right),\\
C_{\alpha\alpha}^{R}(\omega,q)=\,&-\frac{N_c^2}{8\pi^2}\lim_{z_0\to0}\left(\frac{c}{2}+d+2d\ln{z_0}+\frac{\mathcal{B}^{(-)}_{\alpha}(\omega,q)}{\mathcal{A}^{(-)}_{\alpha}(\omega,q)}+\cdots\right),
\end{align}
\end{subequations}
\noindent
the ellipses represent terms which are zero in the limit of zero $z_0$. We rule out the divergent terms, i.e., $\ln{z_0}$, adding appropriate counterterms in the action \eqref{Eq:5Daction}. Thus, one may calculate the spectral functions which are defined as the imaginary part of the retarded Green's functions
\noindent
\begin{subequations}
\begin{align}
\mathcal{R}_{x^3x^3}(\omega,q)\equiv\,&-2\text{Im}C_{x^3x^3}^{R}(\omega,q)=\frac{N_c^2}{4\pi^2}\left(\frac{\omega^2}{\omega^2-q^2}\right)\text{Im}\left[\frac{\mathcal{B}^{(-)}_{x^3}(\omega,q)}{\mathcal{A}^{(-)}_{x^3}(\omega,q)}\right],\\
\mathcal{R}_{\alpha\alpha}(\omega,q)\equiv\,&-2\text{Im}C_{\alpha\alpha}^{R}(\omega,q)=\frac{N_c^2}{4\pi^2}\text{Im}\left[\frac{\mathcal{B}^{(-)}_{\alpha}(\omega,q)}{\mathcal{A}^{(-)}_{\alpha}(\omega,q)}\right].
\end{align}
\end{subequations}
\noindent
As can be seen, the spectral functions depend on the ratio $\mathcal{B}^{(-)}_j/\mathcal{A}^{(-)}_j$ which are the coefficients related to the ingoing solution. One may rewrite this relation in terms of the normalizable and non-normalizable solutions using the matrix relation \eqref{Eq:MatrixCoeff}, for additional details see Refs.~\cite{Miranda:2009uw, Mamani:2013ssa},
\noindent
\begin{equation}\label{Eq:ratioBA}
\frac{\mathcal{B}^{(-)}_j}{\mathcal{A}^{(-)}_j}=-\frac{\partial_z\psi^{(-)}_j\,\psi^{(2)}_j-\psi^{(-)}_j\,\partial_z\psi^{(2)}_j}{\partial_z\psi^{(-)}_{j}\,\psi^{(1)}_j-\psi^{(-)}_{j}\,\partial_z\psi^{(1)}_j}.
\end{equation}
\noindent

In the following, our strategy shall be the following, we solve the differential Eqs. \eqref{Eq:SchrodingerTrans} and \eqref{Eq:SchrodingerLong} integrating from the boundary to the horizon using as ``initial conditions'' the asymptotic solutions \eqref{Eq:NormSol} and \eqref{Eq:NonNormSol}. Then, we plug these solutions in \eqref{Eq:ratioBA}, evaluate them at the horizon and finally we extract the imaginary part to get the retarded Green's function.

In the sequence we present our results for the spectral functions computed following the previous procedure. These results are closely related to the results displayed in Fig.~\ref{Fig1:TF}. For $\mu=0$ we obtained two branches for the temperature, large and small black holes. The former is a stable phase from the thermodynamic point of view, while the last is unstable. If we investigate the melting at zero chemical potential we realized that solutions are only possible for temperatures larger than the global minimum, $T\geq 0.612\,\text{GeV}$. As can be seen in the figure of the potential, see blue line in Fig.~\ref{Fig1:PotentialMu}, there is no potential well for this temperature. This means that no peaks are expected in the spectral functions meaning that the quasiparticles were dissociated.

\begin{figure}[ht!]
\centering
\includegraphics[width=7cm]{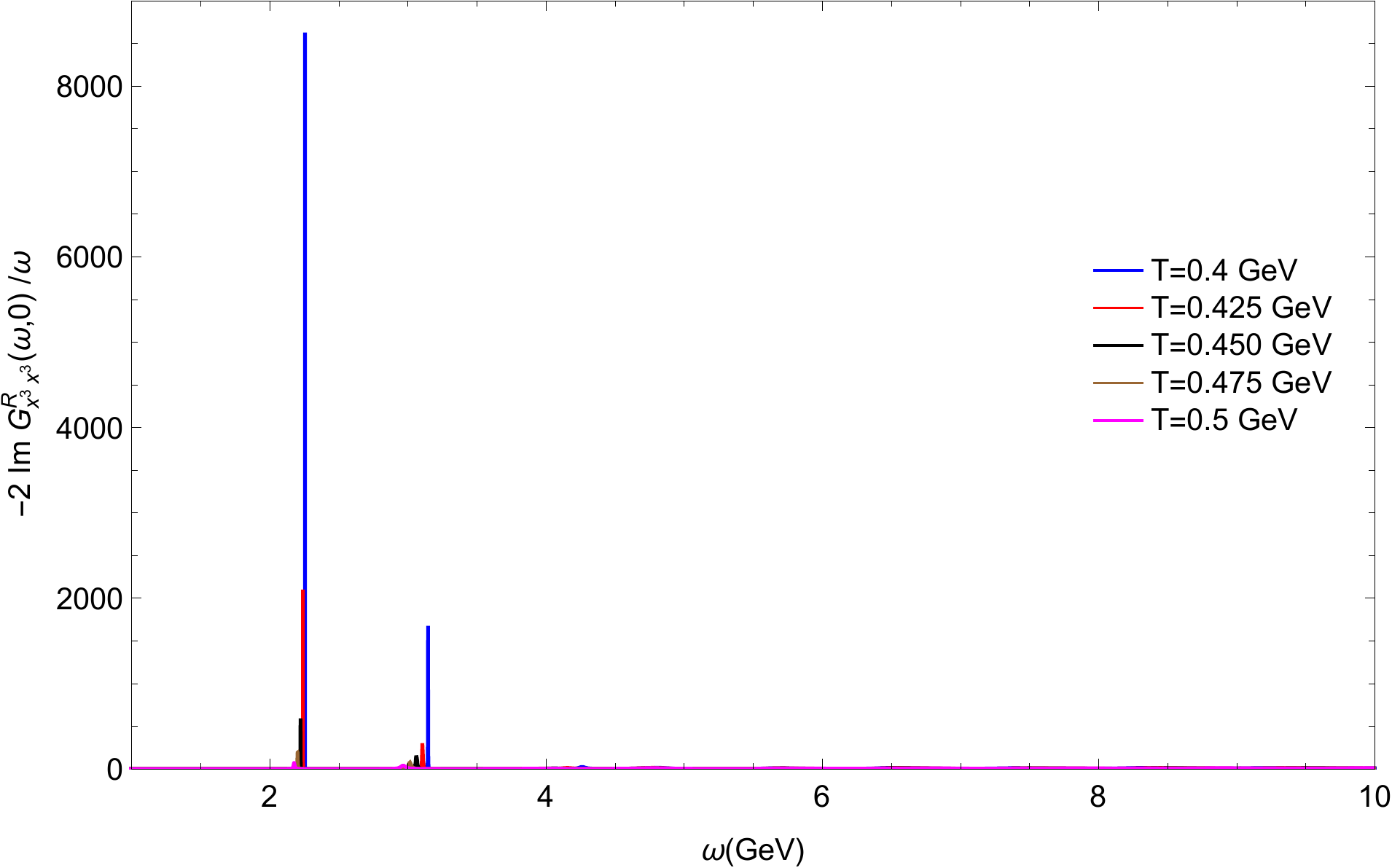}\hfill
\includegraphics[width=7cm]{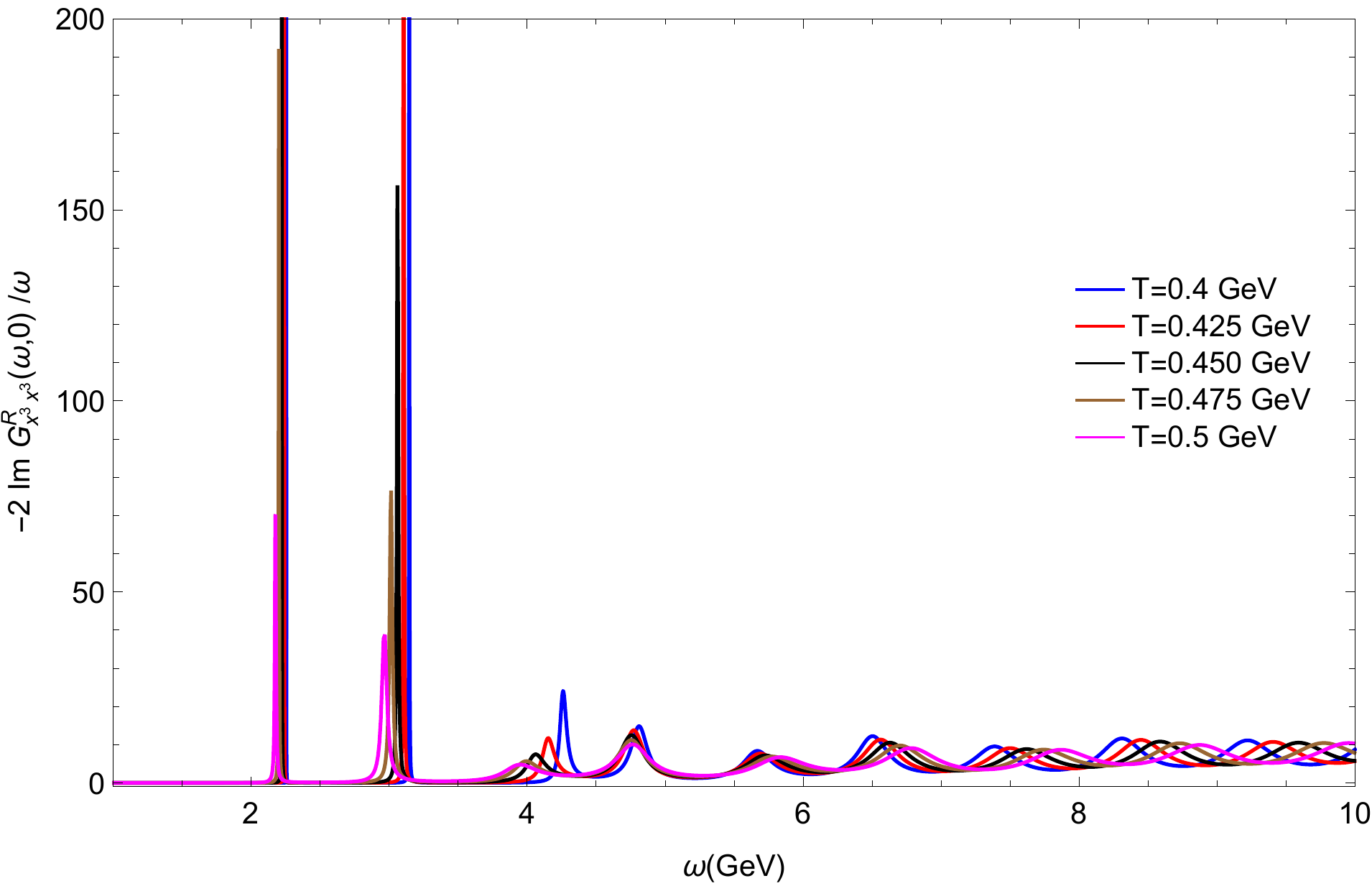}
\caption{
The spectral function for $\mu=0.2\,\text{GeV}$ and selected values of the temperature and $q=0$.
}
\label{Fig:Spectral1}
\end{figure}

On the other hand, from Fig.~\ref{Fig1:TF} we realized that the only way to reach low temperatures is turning on the chemical potential. To be more specific, for $0<\mu<\mu_c$ it is possible to get three phases for the same temperature, see left panel of Fig.~\ref{Fig1:TF} for $\mu=0.2\,\text{GeV}$. In the following we work on the third branch where we can reach low temperatures. For $\mu=0.2\,\text{GeV}$ the temperature in this branch belongs to the interval $0\leq T\leq0.936\,\text{GeV}$. Our numerical results for the spectral function for $q=0$ and selected values of the temperature are displayed in Fig.~\ref{Fig:Spectral1} considering different scales. The location of the peaks on the horizontal axis are interpreted as the mass of the quasiparticle states (which also corresponds to the real part of the frequency), while the width of the peaks are related to the inverse of the decay rate of these quasiparticle states (related to the imaginary part of the frequency). 
As can be seen in the left panel, the height of the peaks decreases with the increasing of the temperature, while the width of the peaks increases meaning that the decay time decreases, i.e., the quasiparticles melt faster when the temperature increases. These results are in agreement with previous results obtained in the literature, see for instance \cite{Miranda:2009uw, Mamani:2013ssa, Braga:2017bml, Braga:2017oqw}. It is worth mentioning that the background metric considered in those holographic models is always AdS, while in the model we are working with the metric is asymptotically AdS. Recently a background obtained solving the Einstein-Maxwell-Dilaton equations was investigated in Ref.~\cite{Zhao:2021ogc}. On the left panel of  Fig.~\ref{Fig:Spectral1} we chose a vertical scale such that the highest peak, corresponding to the first radial excitation is shown in his total height. This way one can see the relative sizes of the peaks. On the other hand, on the  right panel of the same figure we display the spectral function in an expanded scale, such that one can notice the presence of  a series of additional peaks arising in the spectral function, corresponding to the higher order excited states. This  means that the model we consider is capable, through numerical methods, to study high order excitations  of charmonium not  previously studied in the literature. From Fig.~\ref{Fig:Spectral1} we conclude that at temperatures larger than the confinement/deconfinement temperature, $0.170$ GeV, we still have the presence of charmonium states in the quark-gluon plasma. This result is in agreement with previous results in the literature indicating that heavy vector mesons melt at temperatures above $0.170$ GeV, see for instance \cite{Braga:2016wkm, Braga:2017bml}.

\begin{figure}[ht!]
\centering
\includegraphics[width=7cm]{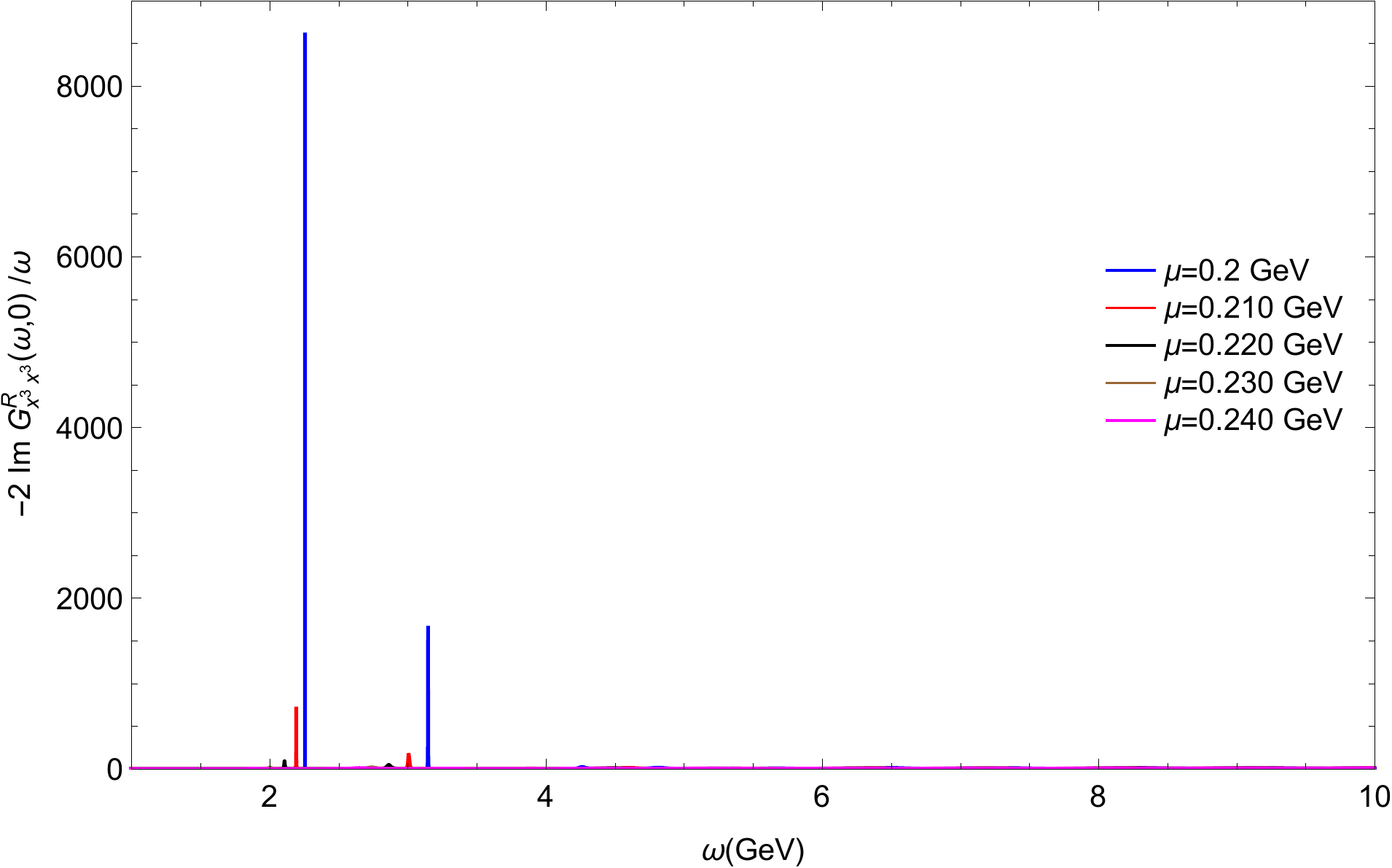}\hfill
\includegraphics[width=7cm]{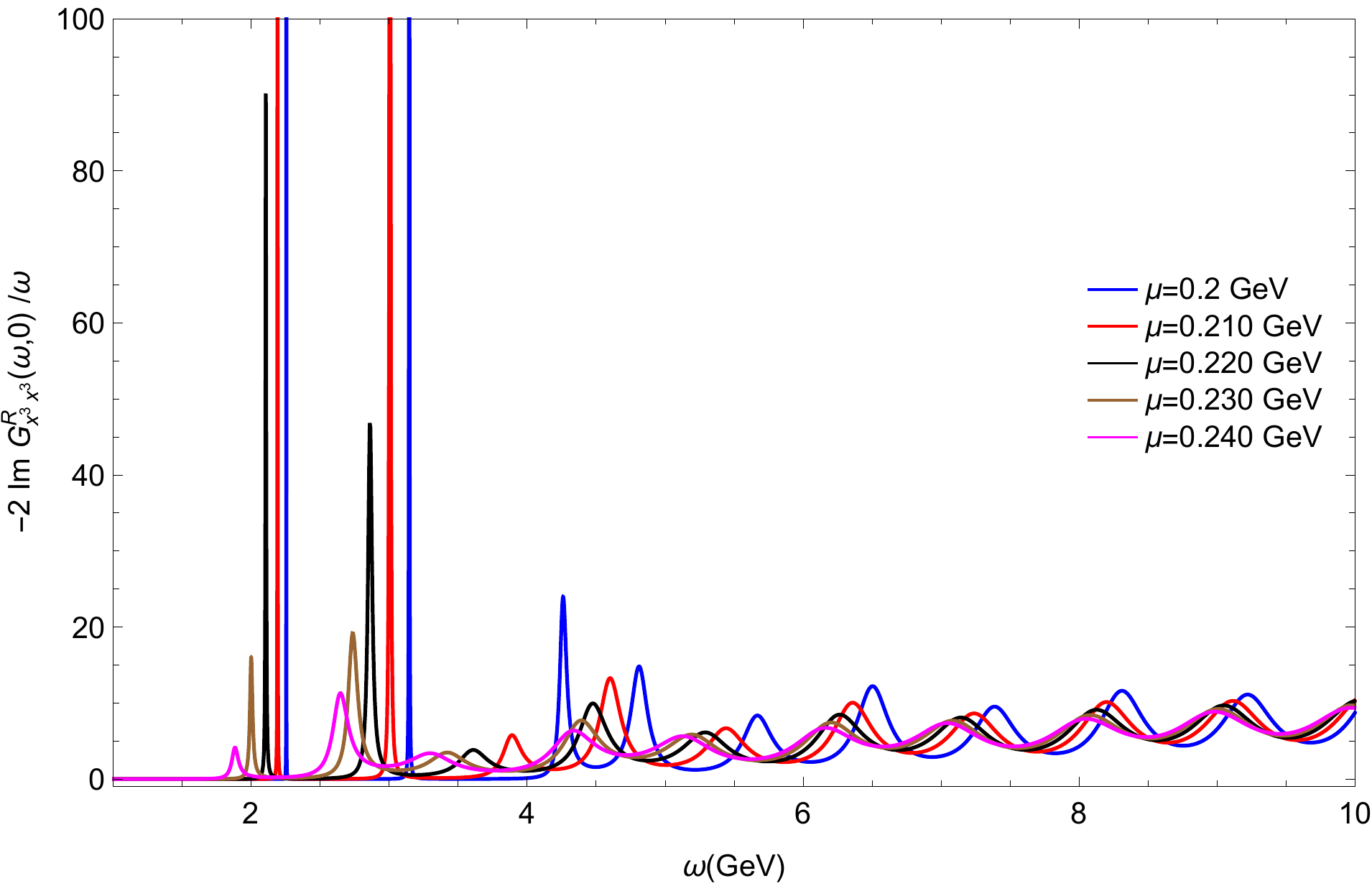}
\caption{
The spectral function for $T=0.4\,\text{GeV}$ and selected values of the chemical potential and $q=0$.
}
\label{Fig:Spectral2}
\end{figure}

Now we investigate the effects of the density on the spectral functions. For this analysis we fix the temperature at $T=0.4\,\text{GeV}$ and compute the spectral functions for selected values of the chemical potential. Our numerical results are displayed in Fig.~\ref{Fig:Spectral2} considering different scales. As can be seen on the left panel, the chemical potential speeds up the melting process because the height of the peaks decreases rapidly, as well as the width of the peaks increases, see right panel. We also realize that increasing  the chemical potential  produces a stronger dissociation effect on  the quasiparticles than increasing the  temperature by the same amount. Right panel also shows the additional peaks arising due to the sensitivity of the numerical procedure. These results are in agreement with previous results in the literature where holographic models for investigating melting of particles including finite density effects were investigated \cite{Braga:2017oqw, Braga:2019xwl}

\begin{figure}[ht!]
\centering
\includegraphics[width=8cm]{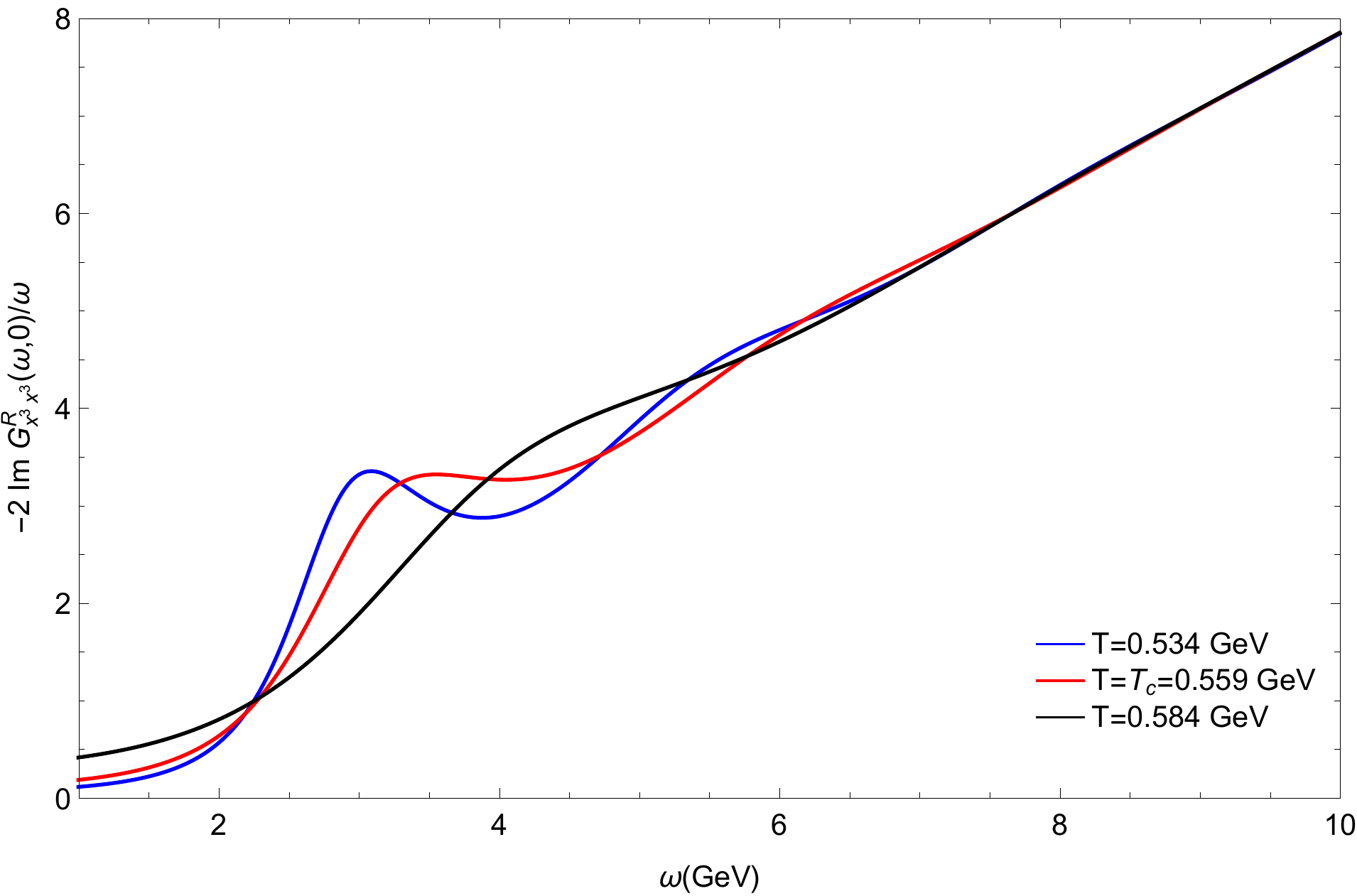}
\caption{
The spectral function for $\mu=\mu_c=0.708\,\text{GeV}$ and selected values of the temperature and $q=0$.
}
\label{Fig:Spectral3}
\end{figure}

Finally, it should be interesting to see the spectral function at the critical end point of the phase diagram, see right panel of Fig.~\ref{Fig1:TF}. For that reason we calculate the spectral functions for $\mu=\mu_c$ and selected values of the temperature: $T=0.534\,\text{GeV}$, $T=T_c=0.559\,\text{GeV}$, and $T=0.584\,\text{GeV}$. We display our numerical results in Fig.~\ref{Fig:Spectral3} where one can see that at the critical end point temperature (red line) the spectral function does not have peaks meaning that quasiparticles melted. In turn, for temperature lower than the critical one (blue line) the spectral function shows a few peaks meaning that a few quasiparticles states might be present in the plasma. For a temperature above the critical one (black line), we do not see peaks in the spectral function.

\section{Hydrodynamic limit}
\label{Sec:HydrodynamicLimit}

In the long-wave and low-energy regime the theory can be described by an effective hydrodynamic description, in this regime one may investigate important physical properties of the system like transport properties. The gauge/gravity duality provides us the theoretical framework to investigate this regime in the dual field theory by solving the perturbation equations, arising in the gravitational side, in the hydrodynamic limit. These perturbations are characterized by a set of complex frequencies, the quasinormal modes, which in the hydrodynamic limit are known as hydrodynamic quasinormal modes.

\subsection{Longitudinal sector}

As described in previous sections, longitudinal perturbation propagating along the direction $k^{\mu}=(\omega,0,0,q)$ is described by Eq.~\eqref{Eq:Longitudinal}. In this section we are going to solve this equation in the hydrodynamic limit, for doing so it is convenient to normalize the parameters by the temperature such that the new parameters and the holographic coordinate becomes dimensionless:
\begin{equation}
\begin{split}
\mathfrak{w}=\frac{\omega}{\pi T}, \quad \mathfrak{q}=\frac{q}{\pi T},\quad \mathfrak{u}=\frac{\mu}{\pi T}, \quad \mathfrak{c}=\frac{c}{(\pi T)^2}, \quad \mathfrak{b}=\frac{b}{(\pi T)^4},\quad u=(\pi T)\,z.
\end{split}
\end{equation}
\noindent
Then, the dimensionless version of Eq.~\eqref{Eq:Longitudinal} is given by
\noindent
\begin{equation}
{E}_{x^3}''(u)-\left(\frac{(1+2\mathfrak{c}\,u^2)g\left(\mathfrak{w}^2-\mathfrak{q}^2\,g\right)-\mathfrak{w}^2\,u\,g'(u)}{u\, g\,\left(\mathfrak{w}^2-\mathfrak{q}^2g\right)}\right){E}_{x^3}'(u)+\frac{\left(\mathfrak{w}^2-\mathfrak{q}^2g\right)}{g^2}{E}_{x^3}(u)=0.
\end{equation}
\noindent
Let us consider a transformation which takes into account the ingoing boundary condition at the horizon
\noindent
\begin{equation}\label{Eq:Ex3Hydro}
{E}_{\scriptscriptstyle{x^3}}(u)=g^{-\frac{i\mathfrak{w}}{4}}F_{\scriptscriptstyle{x^3}}(u),
\end{equation}
\noindent
where $F_{\scriptscriptstyle{x^3}}(u)$ is a regular function, then, the differential equation we must solve is
\noindent
\begin{equation}\label{Eq:Fx3}
\begin{split}
&F_{\scriptscriptstyle{x^3}}''(u)+\left(\frac{\mathfrak{w}^2g'(u)}{g(\mathfrak{w}^2-\mathfrak{q}^2g)}-\frac{1}{u}-2\mathfrak{c}\,u-\frac{i \mathfrak{w}g'(u)}{2g}\right)F_{\scriptscriptstyle{x^{3}}}'(u)+\bigg(\frac{\mathfrak{w}^2}{g^2}-\frac{\mathfrak{q}^2}{g}+\frac{i\mathfrak{w}g'(u)}{4u\,g}\\
&+\frac{i\mathfrak{c}\,\mathfrak{w}u\,g'(u)}{2g}+\frac{i \mathfrak{w} g'(u)^2}{4 g^2}-\frac{\mathfrak{w}^2 g'(u)^2}{16 g^2}-\frac{i\mathfrak{w}^3g'(u)^2}{4g^2(\mathfrak{w}^2-\mathfrak{q}^2g)}-\frac{i\mathfrak{w}g''(u)}{4g}\bigg)F_{\scriptscriptstyle{x^3}}(u)=0.
\end{split}
\end{equation}
\noindent

In the hydrodynamic limit the energy and wavenumber are smaller than the temperature such that $\mathfrak{w}\ll 1$ and $\mathfrak{q}\ll1$. Thus, we may build a multi-scale perturbative solution on these parameters. Nevertheless, here we will use an alternative expansion considering the new parameterization $\mathfrak{w}\to \lambda\, \mathfrak{w}$ and $\mathfrak{q}\to \lambda\, \mathfrak{q}$ such that $\lambda\ll 1$. Then, we use $\lambda$ as the parameter controlling the expansion,
\noindent
\begin{equation}\label{Eq:LongExpansion}
F_{\scriptscriptstyle{x^3}}(u)=F_{\scriptscriptstyle{x^3}}^{\scriptscriptstyle{(0)}}+\lambda\, F_{\scriptscriptstyle{x^3}}^{\scriptscriptstyle{(1)}}(u)+\lambda^2\,F_{\scriptscriptstyle{x^3}}^{\scriptscriptstyle{(2)}}(u)+\cdots.
\end{equation}
\noindent
Plugging \eqref{Eq:LongExpansion} in \eqref{Eq:Fx3} we get differential equations for $F_{\scriptscriptstyle{x^3}}^{\scriptscriptstyle{(0)}}$, $F_{\scriptscriptstyle{x^3}}^{\scriptscriptstyle{(1)}}$ and so on. We must solve these differential equations and fix the integration constants imposing regularity conditions at the horizon, at the end we get the solutions:
\noindent
\begin{subequations}
\begin{align}
F^{\scriptscriptstyle{(0)}}_{\scriptscriptstyle{x^3}}=&F_0,\\
F^{\scriptscriptstyle{(1)}}_{\scriptscriptstyle{x^3}}=&\frac{i F_0\mathfrak{w}}{4}\ln{\left[g(u)\right]}+\frac{i F_0\mathfrak{q}^2}{8\mathfrak{w}\,\mathfrak{c}}\frac{\left(e^{\mathfrak{c} u^2}-e^{\mathfrak{c} u_h^2}\right)}{\int_{0}^{u_h}\frac{e^{\mathfrak{c} x^2}x}{g(x)}dx}\ln{\left[g(u)\right]}-\frac{i F_0 \mathfrak{w}}{4}\ln{\left[g(u_h)\right]}\frac{\int_{0}^{u}\frac{e^{\mathfrak{c} x^2}x}{g(x)}dx}{\int_{0}^{u_h}\frac{e^{\mathfrak{c} y^2}y}{g(y)}dy},
\end{align}
\end{subequations}
\noindent
where $F_0$ is a constant. To calculate the dispersion relation is enough to solve up to $F^{\scriptscriptstyle{(1)}}_{\scriptscriptstyle{x^3}}$. Thus, plugging into \eqref{Eq:Ex3Hydro} and imposing the Dirichlet condition at the boundary ${E}_{\scriptscriptstyle{x^3}}(0)=0$, we get the dispersion relation
\noindent
\begin{equation}\label{Eq:DisperRelation1}
\mathfrak{w}=i\frac{\left(e^{\mathfrak{c}\,u_h^2}-1\right)}{8\mathfrak{c}}\left[\frac{\ln{g(u_h)}}{\int_{0}^{u_h}\frac{e^{\mathfrak{c} x^2}x}{g(x)}dx}\right]\mathfrak{q}^2
\end{equation}
\noindent
Note that the logarithm and integral blow up at the horizon,
\noindent
\begin{equation}
\lim_{u\to u_h}{\ln{g(u)}}=-\infty,\qquad\quad \lim_{u\to u_h}\int_{0}^{u}\frac{e^{\mathfrak{c} x^2}x}{g(x)}dx=\infty.
\end{equation}
\noindent
However, one may use the L'hospital rule to evaluate this ratio, such that the result is finite
\noindent
\begin{equation}
\frac{\ln{g(u_h)}}{\int_{0}^{u_h}\frac{e^{\mathfrak{c} x^2}x}{g(x)}dx}=\frac{e^{-\mathfrak{c}\, u_h^2}g'(u_h)}{u_h}.
\end{equation}
\noindent
Plugging this result in \eqref{Eq:DisperRelation1} and considering the definition of the temperature $g'(u_h)=-4$ one gets the dispersion relation
\noindent
\begin{equation}\label{Eq:DisperRelation2}
\omega=-i\frac{\left(1-e^{-\mathfrak{c}\,u_h^2}\right)}{2\mathfrak{c}\,u_h\,(\pi T)}q^2.
\end{equation}
\noindent
This result represents the most general solution for this kind of holographic models. As a check of consistency let us apply this formula for problems investigated previously in the literature. Considering $g(u)=1-u^4$ and $u_h=1$ we get Eq.~(4.16) of \cite{Kovtun:2005ev}. In turn, for the problem investigated in \cite{Mamani:2013ssa}, plugging the horizon function $g(u)=1-u^4$ and $u_h=1$ in \eqref{Eq:DisperRelation2} we get the same result obtained in Eq.~(4.6) of that paper. As the last check of consistency we get Eq.~(3.31) of \cite{Kim:2010zg}, where $g(u)=(1-u)(1+u-(2-4\pi T\, b)u^2)$. Applying this formula in our case we can write the dispersion relation as
\noindent
\begin{equation}
\omega=iD\,q^2,
\end{equation}
\noindent
where the coefficient $D$ is given by
\noindent
\begin{equation}
D=\frac{\#_1}{\#_2}.
\end{equation}
\noindent
The numerator and denominator are given by
\noindent
\begin{subequations}
\begin{align}
\#_1=&e^{3\mathfrak{b}\,u_h^4}u_h^2\bigg[3\sqrt{\mathfrak{b}}\left(e^{\mathfrak{c} \,u_h^2}-1\right)\left(6\mathfrak{b}\left[e^{\mathfrak{c}\,u_h^2}-1\right]+\mathfrak{c}\,\mathfrak{u}^2\right)-\sqrt{3}\mathfrak{c}^2\mathfrak{u}^2\bigg(e^{\mathfrak{c}\,u_h^2}\mathcal{F}\left[\frac{\mathfrak{c}}{2\sqrt{3\mathfrak{b}}}\right]-\\
&-2\mathcal{F}\left[\frac{\mathfrak{c}}{\sqrt{3\mathfrak{b}}}\right]\bigg)+e^{2\mathfrak{c}\,u_h^2+3\mathfrak{b}\,u_h^4}\bigg(2\mathcal{F}\left[\frac{\mathfrak{c}+3\mathfrak{b}u_h^2}{\sqrt{3\mathfrak{b}}}\right]-\mathcal{F}\left[\frac{\mathfrak{c}+6\mathfrak{b}u_h^2}{2\sqrt{3\mathfrak{b}}}\right]\bigg)\bigg],\nonumber\\
\#_2=&4\pi T\mathfrak{c}\left(e^{\mathfrak{c}\,u_h^2}-1\right)\bigg[3\sqrt{\mathfrak{b}}-\sqrt{3}\mathfrak{c}\mathcal{F}\left[\frac{\mathfrak{c}}{2\sqrt{3\mathfrak{b}}}\right]-e^{\mathfrak{c}u_h^2+3\mathfrak{b}\,u_h^4}\bigg(3\sqrt{\mathfrak{b}}-\sqrt{3}\mathfrak{c}\mathcal{F}\left[\frac{\mathfrak{c}+6\mathfrak{b}u_h^2}{2\sqrt{3\mathfrak{b}}}\right]\bigg)\bigg].
\end{align}
\end{subequations}
\noindent
Here, $\mathcal{F}[x]=e^{-x^2}\int_{0}^{x}e^{y^2}dy$, is the Dawson integral. 

Meanwhile, following Ref.~\cite{Kovtun:2005ev} one may rewrite the field component $E_{\scriptscriptstyle{x^3}}(k,u)$ given by \eqref{Eq:Ex3Hydro} close to the boundary in the form
\noindent
\begin{equation}\label{Eq:Ex3BoundaryN1}
E_{\scriptscriptstyle{x^3}}(u)=\mathfrak{A}_{\scriptscriptstyle{x^3}}(\mathfrak{w},\mathfrak{q})+\cdots\mathfrak{B}_{\scriptscriptstyle{x^3}}(\mathfrak{w},\mathfrak{q})u^2+\cdots
\end{equation}
\noindent
where the coefficients are given by
\noindent
\begin{subequations}\label{Eq:ABBoundary}
\begin{align}
\mathfrak{A}_{\scriptscriptstyle{x^3}}(\mathfrak{w},\mathfrak{q})=\,&F_0\left(1-i\mathfrak{q}^2\frac{\left(1-e^{-\mathfrak{c}\,u_h^2}\right)}{8\mathfrak{c}\,\mathfrak{w}}\frac{g'(u_h)}{u_h}\right),\\
\mathfrak{B}_{\scriptscriptstyle{x^3}}(\mathfrak{w},\mathfrak{q})=\,&i\,F_0\frac{\left(\mathfrak{q}^2-\mathfrak{w}^2\right)}{8\mathfrak{w}}\frac{e^{-\mathfrak{c}\,u_h^2}g'(u_h)}{u_h}.
\end{align}
\end{subequations}
\noindent
To write the last expressions we have used the L'hospital rule again. Thus, one may calculate the retarded Green function from \eqref{Eq:ElectricFieldAction}. Considering the decomposition of  $E_{\scriptscriptstyle{x^3}}(u)=E_{\scriptscriptstyle{x^3}}^{(0)}(k)\mathcal{E}_{\scriptscriptstyle{x^3}}(u)$, from \eqref{Eq:Ex3BoundaryN1} we get (setting $E_{\scriptscriptstyle{x^3}}^{(0)}(k)=\mathfrak{A}_{\scriptscriptstyle{x^3}}(k)$)
\noindent
\begin{equation}\label{Eq:Ex3Boundary}
\mathcal{E}_{\scriptscriptstyle{x^3}}(u)=1+\cdots\frac{\mathfrak{B}_{\scriptscriptstyle{x^3}}(\mathfrak{w},\mathfrak{q})}{\mathfrak{A}_{\scriptscriptstyle{x^3}}(\mathfrak{w},\mathfrak{q})}u^2+\cdots
\end{equation}
\noindent
Thus, the correlation functions may be calculated using the functional derivative of the action \eqref{Eq:ElectricFieldAction}. We have special interest in the component $C_{tt}^{R}(\mathfrak{w},\mathfrak{q})$ which is given by
\noindent
\begin{equation}
C^{R}_{tt}(\mathfrak{w},\mathfrak{q})=\frac{\delta^2 S}{\delta A_{t}^{(0)}(k) \delta A_{t}^{(0)}(-k)}=\frac{\mathfrak{q}^2\delta S}{\delta E^{(0)}_{\scriptscriptstyle{x^3}}(k) \delta E^{(0)}_{\scriptscriptstyle{x^3}}(-k)}.
\end{equation}
\noindent
Plugging \eqref{Eq:Ex3Boundary} in the on-shell action and taking the limit $u\to 0$ we rewrite the correlation function in the form depending on the coefficients of the asymptotic expansion close to the boundary,
\noindent
\begin{equation}\label{Eq:Ctt}
C^{R}_{tt}(\mathfrak{w},\mathfrak{q})=\frac{N_c^2}{8\pi^2}\frac{q^2\,(\pi T)^2}{\left(q^2-\omega^2\right)}\frac{\mathfrak{B}_{\scriptscriptstyle{x^3}}(\mathfrak{w},\mathfrak{q})}{\mathfrak{A}_{\scriptscriptstyle{x^3}}(\mathfrak{w},\mathfrak{q})}.
\end{equation}
\noindent
Finally, by plugging \eqref{Eq:ABBoundary} in the last equation one gets
\noindent
\begin{equation}\label{Eq:CttN2}
\begin{split}
C^{R}_{tt}(\omega,{q})=\frac{N_c^2}{16\pi^2}\frac{q^2}{z_h}\frac{e^{-{c}\,z_h^2}}{\left(i\omega-{q}^2\frac{\left(1-e^{-{c}\,z_h^2}\right)}{2{c}\,z_h}\right)}.
\end{split}
\end{equation}
\noindent
Note that the correlation function has a singular point when the denominator is zero, resulting in the dispersion relation obtained by imposing the Dirichlet boundary condition on the field component $E_{\scriptscriptstyle{x^3}}(u)$, see Eq.~\eqref{Eq:DisperRelation2}.

\subsection{Transversal perturbation}

The dimensionless version of the transverse sector is given by Eq.~\eqref{Eq:Transverse} which in dimensionless parameters is given by
\noindent
\begin{equation}
{E}_{\alpha}''(u)-\left(\frac{g'(u)}{g(u)}+\frac{1}{u}-2\mathfrak{c}\,u\right){E}_{\alpha}'(u)+\frac{\left(\mathfrak{w}^2-\mathfrak{q}^2g\right)}{g^2}{E}_{\alpha}(u)=0,\quad (\alpha=x^1,x^2).
\end{equation}
\noindent
Once again, we consider the transformation which takes into account the ingoing boundary condition at the horizon
\noindent
\begin{equation}\label{Eq:Ex1x3Hydro}
{E}_{\scriptscriptstyle{\alpha}}(u)=g^{-\frac{i\mathfrak{w}}{4}}F_{\scriptscriptstyle{\alpha}}(u),
\end{equation}
\noindent
generating the differential equation we must solve
\noindent
\begin{equation}\label{Eq:Fx1x2}
\begin{split}
&F_{\alpha}''(u)+\left(\frac{g'(u)}{g(u)}-\frac{1}{u}-2\mathfrak{c}\,u-\frac{i \mathfrak{w}g'(u)}{2g}\right)F_{\alpha}'(u)+\bigg(\frac{\mathfrak{w}^2}{g^2}-\frac{\mathfrak{q}^2}{g}+\frac{i\mathfrak{w}g'(u)}{4u\,g}\\
&+\frac{i\mathfrak{c}\,\mathfrak{w}u\,g'(u)}{2g}-\frac{\mathfrak{w}^2 g'(u)^2}{16\, g^2}-\frac{i\mathfrak{w}g''(u)}{4g}\bigg)F_{\alpha}(u)=0.
\end{split}
\end{equation}
\noindent
As before, we build the perturbation solution which is controlled by the parameter $\lambda\ll 1$. Thus, we expand $F_{\alpha}$ in the form
\noindent
\begin{equation}\label{Eq:TransExpansion}
F_{\alpha}(u)=F_{\alpha}^{\scriptscriptstyle{(0)}}+\lambda\, F_{\alpha}^{\scriptscriptstyle{(1)}}(u)+\lambda^2\,F_{\alpha}^{\scriptscriptstyle{(2)}}(u)+\cdots.
\end{equation}
\noindent

Plugging \eqref{Eq:TransExpansion} in \eqref{Eq:Fx1x2} and solving order-by-order we get the solutions:
\noindent
\begin{subequations}
\begin{align}
F^{\scriptscriptstyle{(0)}}_{\alpha}=&F_0,\\
F^{\scriptscriptstyle{(1)}}_{\alpha}=&\frac{i F_0\mathfrak{w}}{4}\ln{\left[g(u)\right]}-\frac{i F_0 \mathfrak{w}}{4}\ln{\left[g(u_h)\right]}\frac{\int_{0}^{u}\frac{e^{\mathfrak{c} x^2}x}{g(x)}dx}{\int_{0}^{u_h}\frac{e^{\mathfrak{c} y^2}y}{g(y)}dy},
\end{align}
\end{subequations}
\noindent
To calculate the dispersion relation is enough to solve up to $F^{\scriptscriptstyle{(1)}}_{\alpha}$. Thus, plugging into \eqref{Eq:Ex1x3Hydro} and imposing the Dirichlet condition at the boundary, ${E}_{\alpha}(0)=0$, we do not get solutions in the hydrodynamic limit, i.e., $\mathfrak{w}\ll 1$ and $\mathfrak{q}\ll 1$. This means that the correlation function does not have poles. It is not difficult to show this statement by expanding the solution, Eq.~\eqref{Eq:Ex1x3Hydro}, close to the boundary,
\noindent
\begin{equation}\label{Eq:Ex1x2BoundaryN1}
E_{\alpha}(u)=\mathfrak{A}_{\alpha}(\mathfrak{w},\mathfrak{q})+\cdots\mathfrak{B}_{\alpha}(\mathfrak{w},\mathfrak{q})u^2+\cdots
\end{equation}
\noindent
where the coefficients are given by
\noindent
\begin{subequations}\label{Eq:ABBoundaryTrans}
\begin{align}
\mathfrak{A}_{\alpha}(\mathfrak{w},\mathfrak{q})=\,&F_0,\\
\mathfrak{B}_{\alpha}(\mathfrak{w},\mathfrak{q})=\,&-i\,F_0\frac{\mathfrak{w}}{8}\frac{e^{-\mathfrak{c}\,u_h^2}g'(u_h)}{u_h}.
\end{align}
\end{subequations}
\noindent
To write the last expression we have used the L'hospital rule. Thus, one may calculate the retarded Green's function from \eqref{Eq:ElectricFieldAction}, and the fact that \eqref{Eq:Ex1x2BoundaryN1} may be decomposed as $E_{\alpha}(u)=E_{\alpha}^{(0)}(k)\mathcal{E}_{\alpha}(u)$, where 
\noindent
\begin{equation}\label{Eq:Ex1x2Boundary}
\mathcal{E}_{\alpha}(u)=1+\cdots\frac{\mathfrak{B}_{\alpha}(\mathfrak{w},\mathfrak{q})}{\mathfrak{A}_{\alpha}(\mathfrak{w},\mathfrak{q})}u^2+\cdots
\end{equation}
\noindent
Thus, the correlation functions may be calculated using the functional derivative of the action \eqref{Eq:ElectricFieldAction}. We are interested in the component $C_{\alpha\alpha}^{R}(\mathfrak{w},\mathfrak{q})$ which is given by
\noindent
\begin{equation}
C^{R}_{\alpha\alpha}(\mathfrak{w},\mathfrak{q})=\frac{\delta^2 S}{\delta A_{\alpha}^{(0)}(k) \delta A_{\alpha}^{(0)}(-k)}=\frac{\mathfrak{w}^2\delta S}{\delta E^{(0)}_{\alpha}(k) \delta E^{(0)}_{\alpha}(-k)}.
\end{equation}
\noindent
Plugging \eqref{Eq:Ex1x2Boundary} into the on-shell action and taking the limit $u\to 0$ we rewrite the correlation function in the form depending on the coefficients of the asymptotic expansion close to the boundary,
\noindent
\begin{equation}\label{Eq:Cx1x1}
C^{R}_{\alpha\alpha}(\mathfrak{w},\mathfrak{q})=-\frac{N_c^2\, T^2}{8}\frac{\mathfrak{B}_{\alpha}(\omega,{q})}{\mathfrak{A}_{\alpha}(\omega,q)}.
\end{equation}
\noindent
Finally, by plugging \eqref{Eq:ABBoundaryTrans} in the last equation one gets
\noindent
\begin{equation}
\begin{split}
C^{R}_{\alpha\alpha}(\omega,{q})=\,&-\frac{N_c^2}{16\pi^2}\frac{i\,\omega\,e^{-{c}\,z_h^2}}{z_h}.
\end{split}
\end{equation}
\noindent
Note that the correlation function has no poles, which is consistent with the solution in the hydrodynamic limit.

\section{Quark number susceptibility}
\label{Sec:QuarkNumber}

In QCD, the response of the system to a change in the chemical potential is measured through the quark number susceptibility $\chi$, and it was investigated in holographic QCD models in Refs.~\cite{Kim:2006uta, Kim:2010zg, He:2009zzp, Jo:2008jrj, Kim:2010ag, Dudal:2018rki} (see also references therein). Once we have solved the differential equations in the hydrodynamic limit and obtained the retarded Green's functions we may now calculate the quark number susceptibility following the procedure implemented in Refs.~\cite{Kim:2006uta, Kim:2010zg}, where they used the prescription
\noindent
\begin{equation}
\chi(T,\mu)=-\lim_{\substack{q\to 0}}\text{Re}\{C_{tt}(0,q)\}
\end{equation}
\noindent
Thus, from \eqref{Eq:CttN2} we get
\noindent
\begin{equation}\label{Eq:QuarkSuceptibility}
\begin{split}
\frac{\chi(T,\mu)}{N_c^2}=\frac{1}{8\pi^2}\frac{c}{\left(e^{c\,z_h^2}-1\right)}.
\end{split}
\end{equation}
\noindent

Note that even though this result is similar to the one obtained in Refs.~\cite{Kim:2010zg, Dudal:2018rki}, the model we are working with has an additional parameter $b$, and an intricate form for the horizon function $g(z)$. In turn, one may compare this result against the one obtained using the baryon density read off from Eqs.~\eqref{Eq:GaugeBoundary} and \eqref{Eq:AtCloseBoundary}, which is given by \cite{He:2013qq}
\noindent
\begin{equation}\label{Eq:BaryonDensity}
\rho=\frac{c\,\mu}{e^{c\,z^2_h}-1}
\end{equation}
\noindent
Then, we calculate the quark number susceptibility
\noindent
\begin{equation}\label{Eq:QuarkSuceptibilityN2}
\chi=\frac{\partial \rho}{\partial \mu}=\frac{c}{e^{c\,z_h^2}-1}-\frac{2\,c^2\mu\,z_he^{c\,z_h^2}}{\left(e^{c\,z_h^2}-1\right)^{2}}\frac{1}{\partial_{z_h} \mu}.
\end{equation}
\noindent

\begin{figure}[ht!]
\centering
\includegraphics[width=7cm]{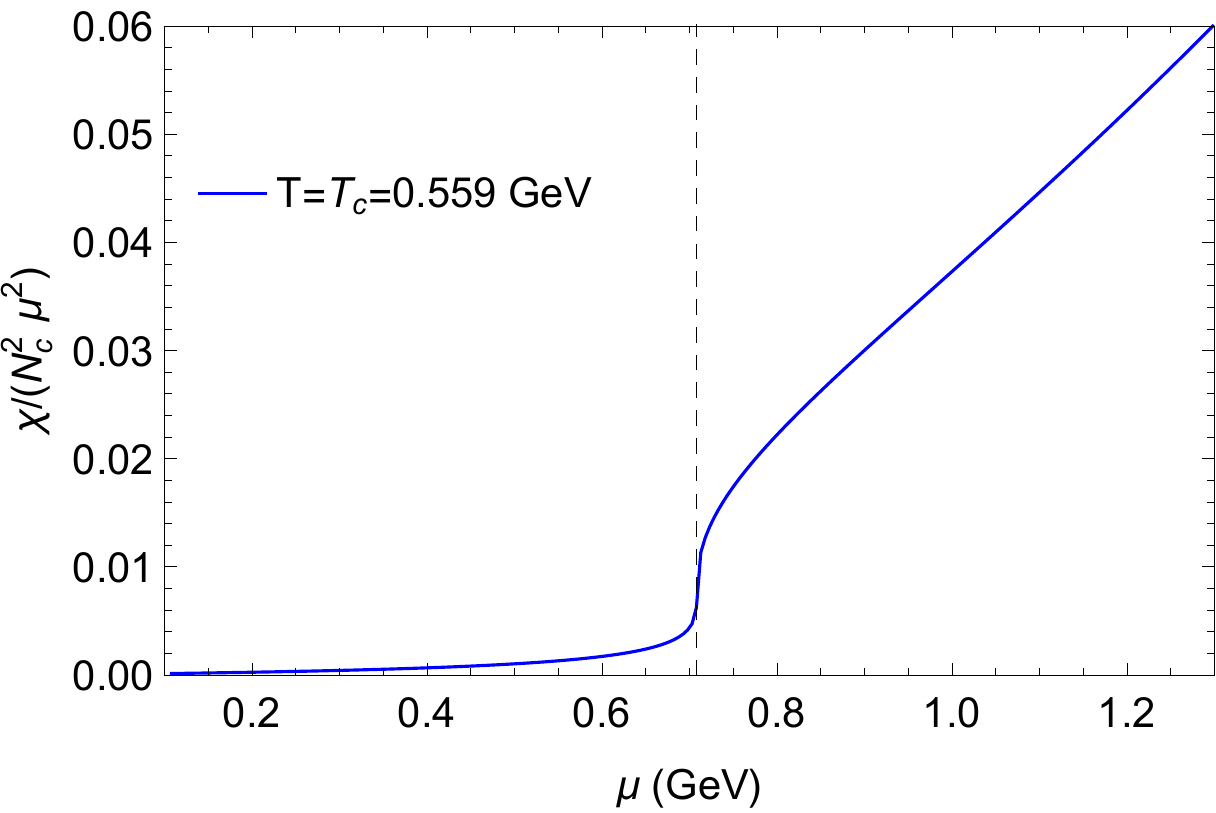}\hfill
\includegraphics[width=7cm]{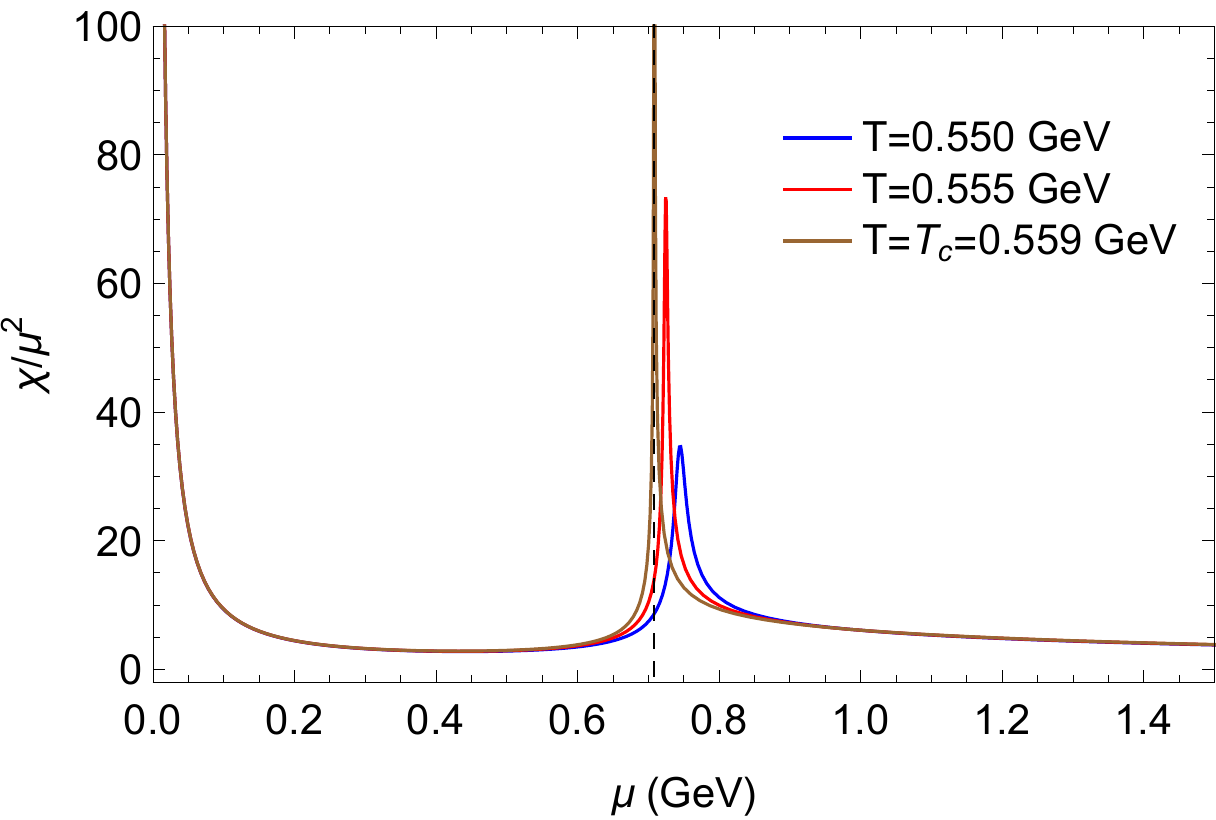}
\caption{
Left: The quark number susceptibility as a function of the chemical potential for $T=T_c=0.599$ GeV given by Eq.~\eqref{Eq:QuarkSuceptibility}. Right: The quark number susceptibility as a function of the chemical potential for $T=0.550\,\text{GeV}$ (blue), $T=0.555\,\text{GeV}$ (red), and $T=0.559\,\text{GeV}$ (brown) given by Eq.~\eqref{Eq:QuarkSuceptibilityN2}. Both panels also show the critical value for the chemical potential represented by vertical dashed line.
}
\label{Fig:TMuq1}
\end{figure}

Let us calculate the behavior of Eqs.~\eqref{Eq:QuarkSuceptibility} and \eqref{Eq:QuarkSuceptibilityN2} in the conformal limit, i.e., high temperatures regime, where we get the relation $z_h=1/(\pi\, T)$ (see left panel of Fig.~\ref{Fig1:TF}). Plugging this result in \eqref{Eq:QuarkSuceptibility} and \eqref{Eq:BaryonDensity}, then, considering the approximation $e^{c/(\pi\, T)^2}\approx 1+c/(\pi\, T)^2$, we get
\noindent
\begin{equation}
\chi=\frac{N_c^2}{8}\,T^2 \qquad\text{and}\qquad \chi=\pi^2 T^2,
\end{equation}
respectively. This result is in agreement with the result found in the literature, see for instance Refs.~\cite{Kim:2010zg, He:2009zzp, Jo:2008jrj}, where the quark number susceptibility goes like $\chi\sim T^2$. Let us plot Eqs.~\eqref{Eq:QuarkSuceptibility} and \eqref{Eq:QuarkSuceptibilityN2} as a function of the chemical potential for selected values of the temperature. Our numerical results are displayed in Fig.~\ref{Fig:TMuq1}, where the left panel shows Eq.~\eqref{Eq:QuarkSuceptibility}, while right panel shows the results for Eq.~\eqref{Eq:QuarkSuceptibilityN2}. It is worth mentioning that previous investigation works showed that the quark susceptibility must blows up at the critical end point, for results obtained solving Dyson-Schwinger equation see Ref.~\cite{He:2008yr} and references therein. Our results are showing two different behaviors for the quark number susceptibility, one obtained from Eq.~\eqref{Eq:QuarkSuceptibility} and another from Eq.~\eqref{Eq:QuarkSuceptibilityN2}. The result displayed in the left panel of Fig.~\ref{Fig:TMuq1} does not diverge at the critical end point. This result may be explained because the matter action we are considering to get Eq.~\eqref{Eq:QuarkSuceptibility} represents the action of probe fields. In turn, right panel of Fig.~\ref{Fig:TMuq1} shows that $\chi$ blows up at the critical end point. Note that Eq.~\eqref{Eq:QuarkSuceptibilityN2} was obtained using the holographic dictionary on $A_t$, which was obtained by solving Maxwell's equations \eqref{Eq:Maxwell2}.

\noindent

\section{Quasinormal modes}
\label{Sec:QuasinormalModes}

In this section we calculate the quasinormal frequencies by solving the differential equations numerically. We split up the problem in two parts:  the longitudinal and the transverse sectors. In order too calculate the quasinormal frequencies we are going to use the pseudo-spectral method, for a discussion see Ref.~\cite{boyd01}, see also Refs.~\cite{Jansen:2017oag, Rougemont:2018ivt, Finazzo:2016psx, Mamani:2018uxf, Lucas007} and references therein, where this method was applied to calculate quasinormal frequencies in different scenarios.

\subsection{Longitudinal sector}

Our starting point is the Schrodinger-like equation \eqref{Eq:SchrodingerLong}, with potential \eqref{Eq:SchrodingerPotLong}. From here on we follow the procedure implemented in Ref.~\cite{Lucas007}. In order to write the differential equation suitable to apply the pseudo-spectral method we first implement the transformation $\psi_\alpha=e^{-i\,\omega\,r_*}\varphi$. Then, we replace the tortoise coordinate and $B_L$ to get
\noindent
\begin{equation}\label{Eq:LongitudinalEq1}
\begin{split}
&(4q^2z^2\omega^4-8q^4z^2\omega^2g+3\omega^4g+4c^2z^4\omega^4+4q^6z^2g^2-6q^2\omega^2g^2-8c^2q^2z^4\omega^2g^2\\
&+3q^4g^3+4c^2q^4z^4g^3-2z\omega^4g'-4cz^3\omega^4g'+2q^2z\omega^2gg'+4cq^2z^3\omega^2gg'\\
&+2q^2z^2\omega^2g'^2+q^4z^2gg'^2+2q^2z^2\omega^2gg''-2q^4z^2g^2g'')\varphi(z)\\
&-(8iz^2\omega (\omega^2-q^2g)^2+4z^2\,(\omega^2-q^2g)^2g')\varphi'(z)\\
&-g(4z^2\omega^4-8q^2z^2\omega^2g+4q^4z^2g^2)\varphi''(z)=0.
\end{split}
\end{equation}
\noindent
In order to compare our results against the results obtained in Ref.~\cite{Kovtun:2005ev}, for $\mu=0$, and $c=0=b$, we normalize the coordinate and parameters by the temperature
\noindent
\begin{equation}\label{Eq:NormalizationQNMs}
u=z\,\pi\, T;\qquad \widetilde{\omega}=\frac{\omega}{2 \pi T};\qquad \widetilde{q}=\frac{q}{2 \pi T};\qquad \widetilde{c}=\frac{c}{(\pi T)^2};\quad \widetilde{b}=\frac{b}{(\pi T)^4}; \qquad \widetilde{\mu}=\frac{\mu}{\pi T}.
\end{equation}
\noindent
One may calculate the asymptotic solutions of the last differential equation close to the horizon considering the ansatz $\varphi\sim(1-u)^\alpha$. Plugging in Eq.~\eqref{Eq:LongitudinalEq1} we get

\noindent
\begin{equation}
\alpha_1=0,\qquad\qquad\qquad \alpha_2=4i\,\widetilde{\omega}.
\end{equation}
\noindent
The first solution is interpreted as the ingoing solution, falling into the black hole event horizon, while the second solution is interpreted as the outgoing solution. As we are interested in retarded Green`s functions we choose the ingoing solution in the following analysis. In turn, to calculate the asymptotic solution close to the boundary we consider the ansatz $\varphi\sim u^\beta$. Plugging in \eqref{Eq:LongitudinalEq1} we get the solutions
\noindent
\begin{equation}
\beta_1=-\frac{1}{2},\qquad\qquad\qquad \beta_2=\frac{3}{2},
\end{equation}
\noindent
where the first solution is interpreted as the non-normalizable solution, while the second the normalizable one. As we are looking for normalizable solutions of the eigenvalue problem we consider the normalizable solution in the following analysis. Then, to get the final differential equation to solve we consider the additional transformation which takes into account the information about the asymptotic solutions we got, $\varphi(u)=u^{3/2}\phi(u)$
\noindent
\begin{equation}
\begin{split}
&\bigg(8(2\widetilde{q}^{\,2}u-3i\widetilde{\omega})\widetilde{\omega}^{\,4}+4\widetilde{c}^{\,2}\widetilde{q}^{\,4}u^3g^3-4(2+\widetilde{c}u^2)\widetilde{\omega}^{\,4}g'+2\widetilde{q}^{\,2}u\widetilde{\omega}^{\,2}g'^2+2g^2(8\widetilde{q}^{\,6}u-12i\widetilde{q}^{\,4}\widetilde{\omega}\\
&-4\widetilde{c}^{\,2}\widetilde{q}^{\,2}u^3\widetilde{\omega}^{\,2}-3\widetilde{q}^{\,4}g'-\widetilde{q}^{\,4}ug'')+g(2\widetilde{\omega}^{\,2}(-16\widetilde{q}^{\,4}u+24i\widetilde{q}^{\,2}\widetilde{\omega}+2\widetilde{c}^{\,2}u^3\widetilde{\omega}^{\,2}+\widetilde{q}^{\,2}ug'')\\
&+2\widetilde{q}^{\,2}(7+2\widetilde{c} u^2)\widetilde{\omega}^{\,2}g'+\widetilde{q}^{\,4}ug'^{\,2})\bigg)\phi(u)-4i(\widetilde{\omega}^2-\widetilde{q}^2\,g)^2(4u\widetilde{\omega}-3ig-iu\,g')\phi'(u)\\
&-4ug(\widetilde{\omega}^{\,2}-\widetilde{k}^{\,2}g)^2\phi''(u)=0.
\end{split}
\end{equation}
\noindent
As can be seen, this is a fifth-order eigenvalue problem, see the largest power of $\omega$, to solve it we are going to use the pseudo-spectral method. Nevertheless, in the holographic model we are working with three parameters: $\mu$, $c$ and $b$. To see how the QN frequencies change with the chemical potential $\widetilde{\mu}$ we set $\widetilde{c}=0=\widetilde{b}$. It is worth mentioning that in this case the horizon function reduces to the Reissner-Nordstr\"om AdS solution
\noindent
\begin{equation}
g=1-\left(1+\frac{\widetilde{\mu}^{\,2}}{3}\right)u^4+\frac{\widetilde{\mu}^{\,2}}{3}u^6.
\end{equation}
\noindent
One of the zeros of the equation $g(u)=(u^2-u_1)(u^2-u_2)(u^2-u_3)=0$ represent the location of the event horizon, which we fix at $u_h=1$ due to the normalization we are working with \eqref{Eq:NormalizationQNMs}. Then, the solutions are:
\noindent
\begin{equation}
u_1^2=1,\qquad u_2^2=\frac{3-\sqrt{9+12\widetilde{\mu}^2}}{2\widetilde{\mu}^2},\qquad u_3^2=\frac{3+\sqrt{9+12\widetilde{\mu}^2}}{2\widetilde{\mu}^2}.
\end{equation}
\noindent
As can be seen, $u_1^2$ does not depend on the chemical potential, while $u_2^2$ is always negative for $\widetilde{\mu}>0$. Meanwhile, $u_3^2>1$ for $\widetilde{\mu}>0$, while it becomes $u_3^2=1$ for $\widetilde{\mu}=\sqrt{6}$, and $u_3^2<1$ for $\widetilde{\mu}>\sqrt{6}$. For that reason we restrict our numerical analysis to the region where $\widetilde{\mu}\leq \sqrt{6}$. This means that the horizon lies at $u_h=1$. 

\begin{figure}[ht!]
\centering
\includegraphics[width=7cm]{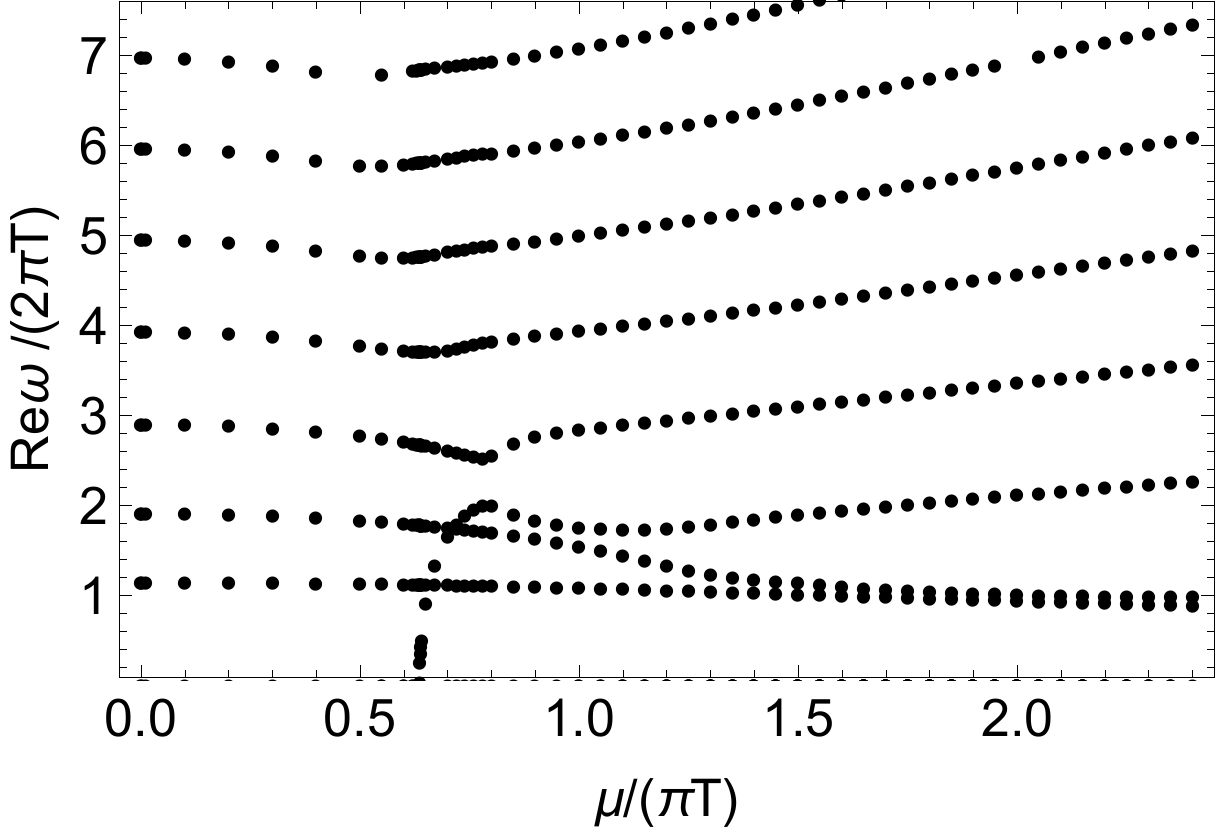}\hfill
\includegraphics[width=7cm]{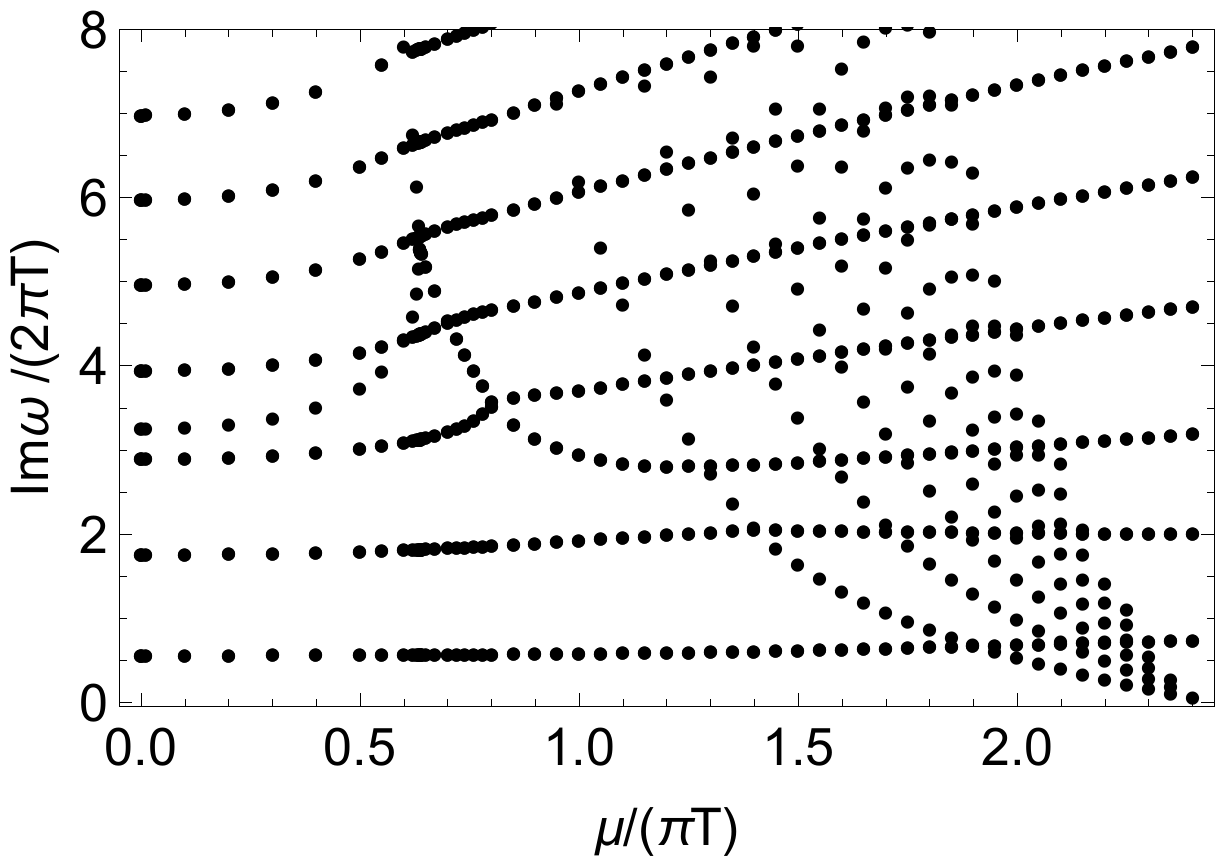}
\caption{
Left: The real part of the frequency as a function chemical potential. Right: The imaginary part of the frequency as a function of the chemical potential. These results were obtained for $q/(2\pi T)=1$ and $c=0=b$.
}
\label{Fig:MuwrwiLongitudinal}
\end{figure}

Our numerical results for $\widetilde{q}=1$ are displayed in Table \ref{Tab:QNMsLongitudinal}. We point out that the results for $\widetilde{\mu}=0$ are in agreement with those results obtained in Ref.~\cite{Kovtun:2005ev}. As can be seen, the real part of the frequency decreases with the increasing of the chemical potential in the region of small values for the chemical potential, then, it increases with the increasing of the chemical potential. 
However, the ground and first two states deserve an additional comment, as can be seen in Fig.~\ref{Fig:MuwrwiLongitudinal}, there is a new mode whose real part starts to grow up at approximately $\widetilde{\mu}\approx 0.637$, it grows up rapidly and crosses the ground state, however it is not clear if this mode crosses or stays below the first state, see left panel of Fig.~\ref{Fig:MuwrwiLongitudinal}. In turn, the imaginary part always increases with the increasing of the chemical potential, see right panel of Fig.~\ref{Fig:MuwrwiLongitudinal}.

\begin{table}[ht]
\centering
\begin{tabular}{l |c|c|c}
\hline
\hline
$n$ & $\widetilde{\mu}=0$ & $\widetilde{\mu}=0.01$ & $\widetilde{\mu}=0.1$ \\
\hline 
$0$ & $\pm 1.1478314-0.5592036i$  & $\pm1.1478256-0.5592051i$ & $\pm1.1472503-0.5593576i$  \\
$1$ & $\pm 1.9100059-1.7580648i$  & $\pm1.9099775-1.7580789i$ & $\pm1.9071552-1.7594740i$ \\
$2$ & $\pm 2.9032931-2.8916809i$  & $\pm2.9032453-2.8917233i$ & $\pm2.8984967-2.8959528i$ \\
$3$ & $\pm 3.9285553-3.9433859i$  & $\pm3.9284951-3.9434557i$ & $\pm3.9225172-3.9504090i$  \\
$4$ & $\pm 4.9468182-4.9651851i$  & $\pm4.9467456 + 4.9652802i$ & $\pm4.9395555-4.9747681i$ \\
\hline 
\hline
  $n$ & $\widetilde{\mu}=0.2$ & $\widetilde{\mu}=0.3$ & $\widetilde{\mu}=0.5$ \\
\hline 
 $0$ & $\pm 1.1455051-0.5598240i$  & $\pm1.1425904-0.5606158i$ & $\pm1.1332155-0.5632753i$  \\
 $1$ & $\pm 1.8985348-1.7637362i$  & $\pm1.8839332-1.7709573i$ & $\pm1.8350724-1.7950779i$ \\
 $2$ & $\pm 2.8839215-2.9090645i$  & $\pm2.8589691-2.9319886i$ & $\pm2.7723821-3.0169965i$ \\
 $3$ & $\pm 3.9042584-3.9721219i$  & $\pm3.8733809-4.0107026i$ & $\pm3.7734185-4.1623459i$  \\
 $4$ & $\pm 4.9177089-5.0045069i$  & $\pm4.8812902-5.0577870i$ & $\pm4.7755298-5.2719366i$ \\
\hline\hline
\end{tabular}
\caption{
The quasinormal frequencies of the longitudinal sector for selected values of the chemical potential for $\widetilde{q}=1$, $\widetilde{c}=0$, and $\widetilde{b}=0$. The results for $\widetilde{\mu}=0$ are equivalent to those of Ref.~\cite{Kovtun:2005ev}. 
}
\label{Tab:QNMsLongitudinal}
\end{table}

The hydrodynamic frequency has imaginary part different from zero, $\widetilde{\omega}=-3.2506370 i$, which is also in agreement with the result of Ref.~\cite{Kovtun:2005ev} for $\widetilde{\mu}=0$. The dependence of the hydrodynamic frequency with chemical potential deserves an additional analysis. Our numerical results for this frequency considering selected values of the chemical potential are displayed in Table \ref{Tab:HydroQNM}. As can be seen, the frequency increases with the increasing of the chemical potential. However, at $\widetilde{\mu}\approx 0.637$ it seems that the hydrodynamic mode merges with a new mode coming from above, see right panel of Fig.~\ref{Fig:MuwrwiLongitudinal}. It is also worth mentioning that a family of purely imaginary modes shows up in the spectrum. They have a characteristic behavior, decreasing with the increasing of the chemical potential and are purely imaginary, see right panel of the same figure. This kind of behavior was previously observed in the literature, see for instance Ref.~\cite{Mamani:2018qzl} (see also \cite{Cook:2016fge} to see how this kind of modes behave as a function of the rotation parameter).

\begin{table}[]
\centering
\begin{tabular}{l|c}
\hline\hline
$\widetilde{\mu}=0$ & $-3.2506370i$ \\
\hline
$\widetilde{\mu}=0.01$ & $-3.2507627i$\\
\hline
$\widetilde{\mu}=0.1$ & $-3.2633614i$\\
\hline
$\widetilde{\mu}=0.2$ & $-3.3035845i$\\
\hline
$\widetilde{\mu}=0.3$ & $-3.3786204i$\\
\hline
$ \widetilde{\mu}=0.5$ & $-3.7348719i$\\
 \hline
 $\widetilde{\mu}=0.6$ & $-4.2947702i$\\
  \hline
 $\widetilde{\mu}=0.62$ & $-4.5834402i$\\
\hline\hline
\end{tabular}
\caption{The hydrodynamic frequency for different values of the chemical potential. These results were obtained considering $q/(2\pi T)=1$ and ${c}=0={b}$.}
\label{Tab:HydroQNM}
\end{table}

\noindent

\begin{figure}[ht!]
\begin{center}
\includegraphics[width=7cm]{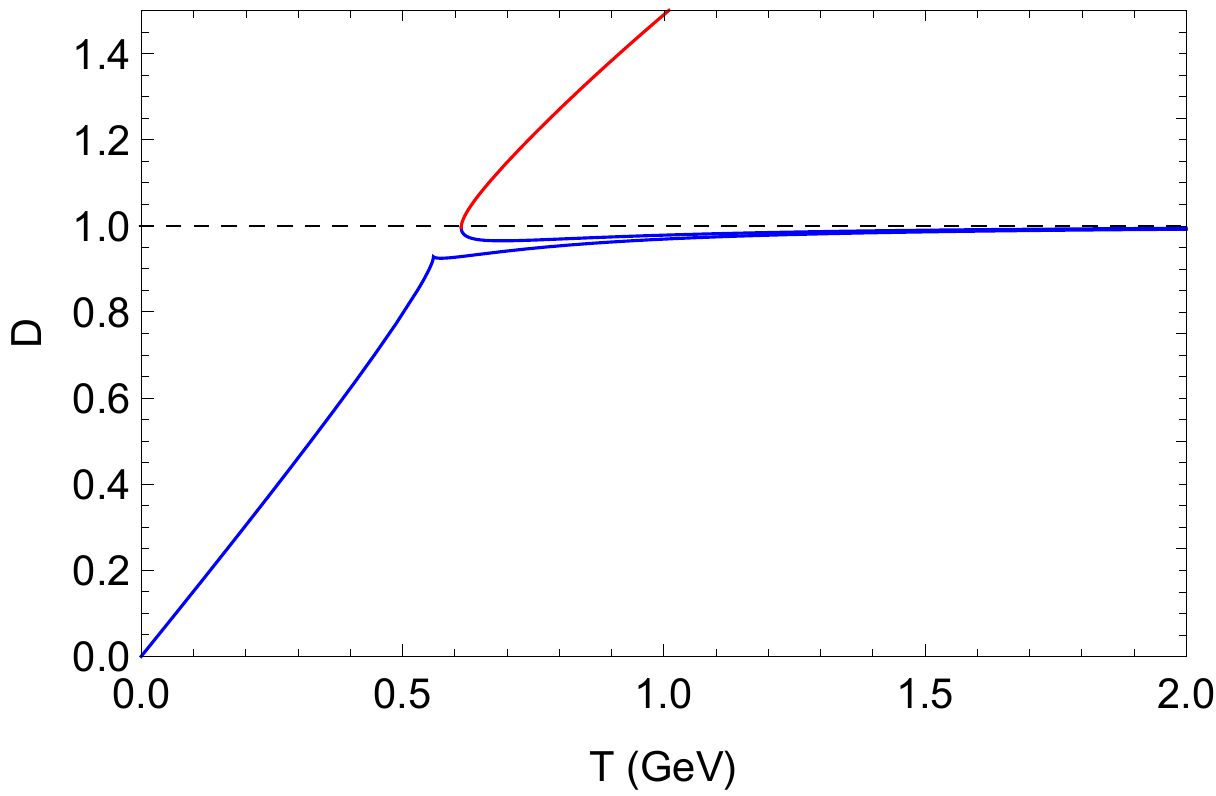}\hfill
\includegraphics[width=7cm]{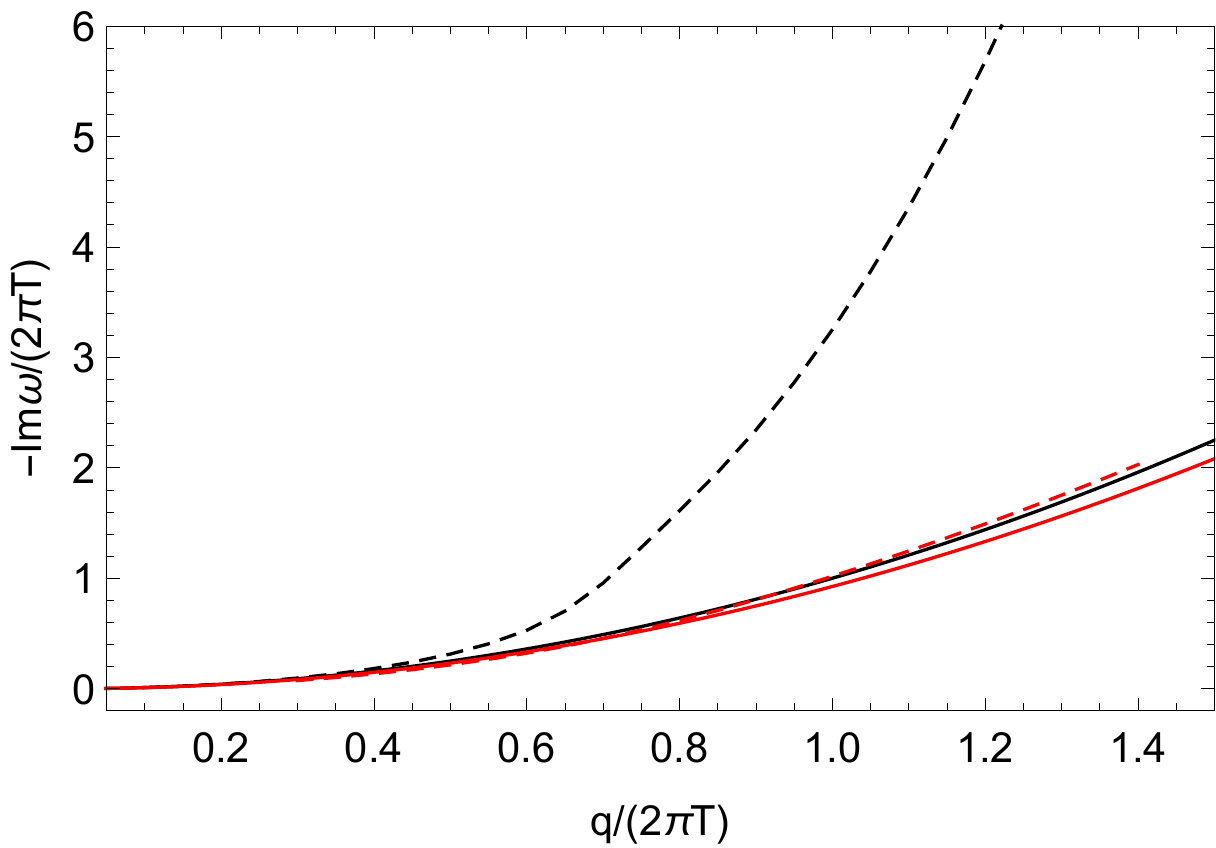}
\end{center}
\caption{
Left: The diffusion coefficient as a function of the temperature, the colors correspond to the notation used in Fig.~\ref{Fig1:TF}, the conformal case is represented by horizontal dashed line. Right: The numerical (dashed lines) and analytic (solid lines) solutions for the hydrodynamic frequency for ${\mu}=0$ and ${c}=0={b}$ (black) and for ${\mu}={\mu}_c$, $c=1.46\,\text{GeV}^2$, $b=0.273\,\text{GeV}^4$ and $T=T_c$ (red).
}
\label{Fig:DispersionHydro}
\end{figure}

Let us now compare the analytic solution obtained in the hydrodynamic limit, i.e., Eq.~\eqref{Eq:DisperRelation2}, against the numerical solution. For $\widetilde{\mu}=0$ we realized that the coefficient $D=\pi T(1-e^{-cz_h^2})/(cz_h)$, lies close to the unity when $T\geq T_{\text{min}}$ for the big black hole branch, see blue line in the left panel of Fig.~\ref{Fig:DispersionHydro}. In turn, the coefficient $D$ increases when $T\geq T_{\text{min}}$ for the small black hole branch, see red line in the left panel of Fig.~\ref{Fig:DispersionHydro}. In turn, for $\mu\geq \mu_c$ the diffusion coefficient becomes zero in the limit of zero temperature, then, it increases converging to the unity in the limit of high temperatures where conformal symmetry must be restored, see blue line in the same figure. Meanwhile, the results for the dispersion relation are displayed in the right panel of Fig.~\ref{Fig:DispersionHydro}, where continuous lines represent analytic results for the conformal (black line) and non-conformal case (red line), while numerical solutions for the conformal are represented by dashed black line and non-conformal are represented by dashed red. In the non-conformal case we fixed the parameters $\mu=\mu_c$, $c=1.46\,\text{GeV}^2$, $b=0.273\,\text{GeV}^4$ and $T=T_c$. As can be seen, in the non-conformal case the frequency decreases. It is also worth mentioning that the precision of the numerical method gets poor in the non-conformal case.

\noindent

\begin{figure}[ht!]
\centering
\includegraphics[width=7cm]{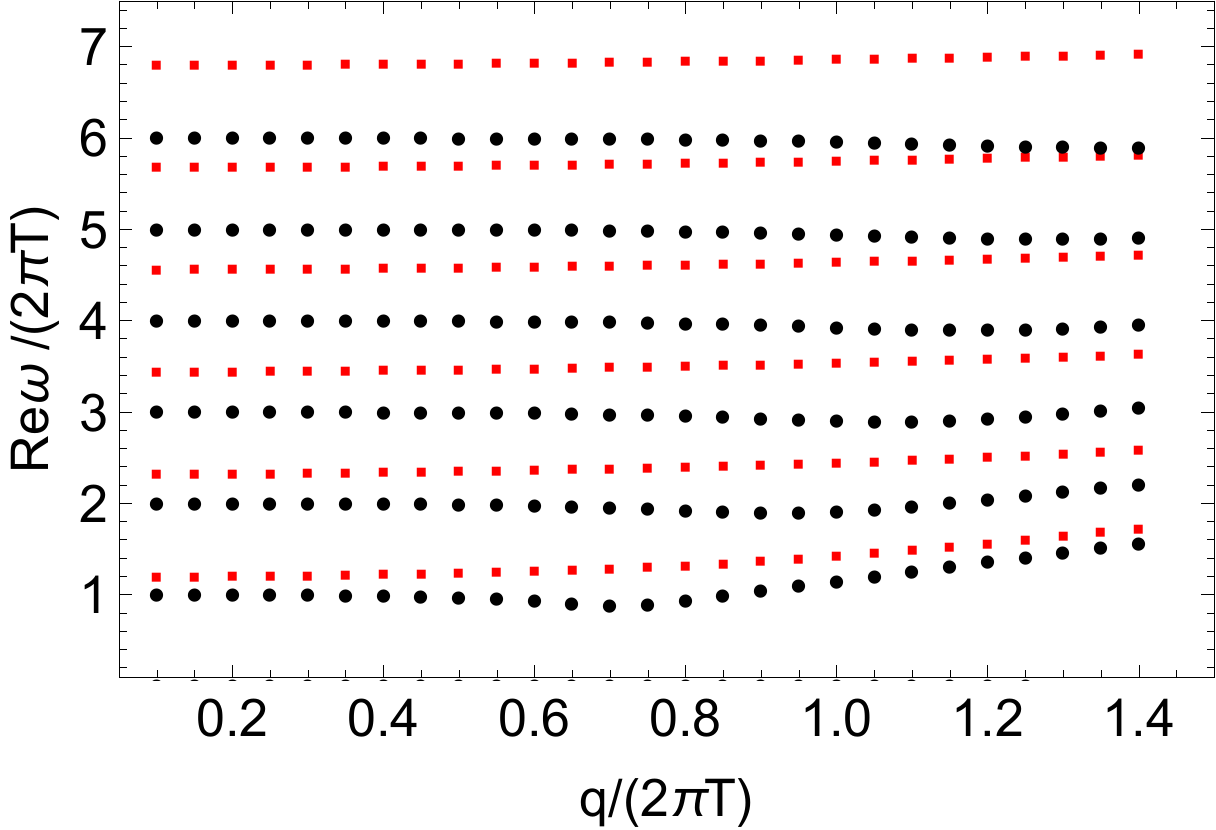}\hfill
\includegraphics[width=7cm]{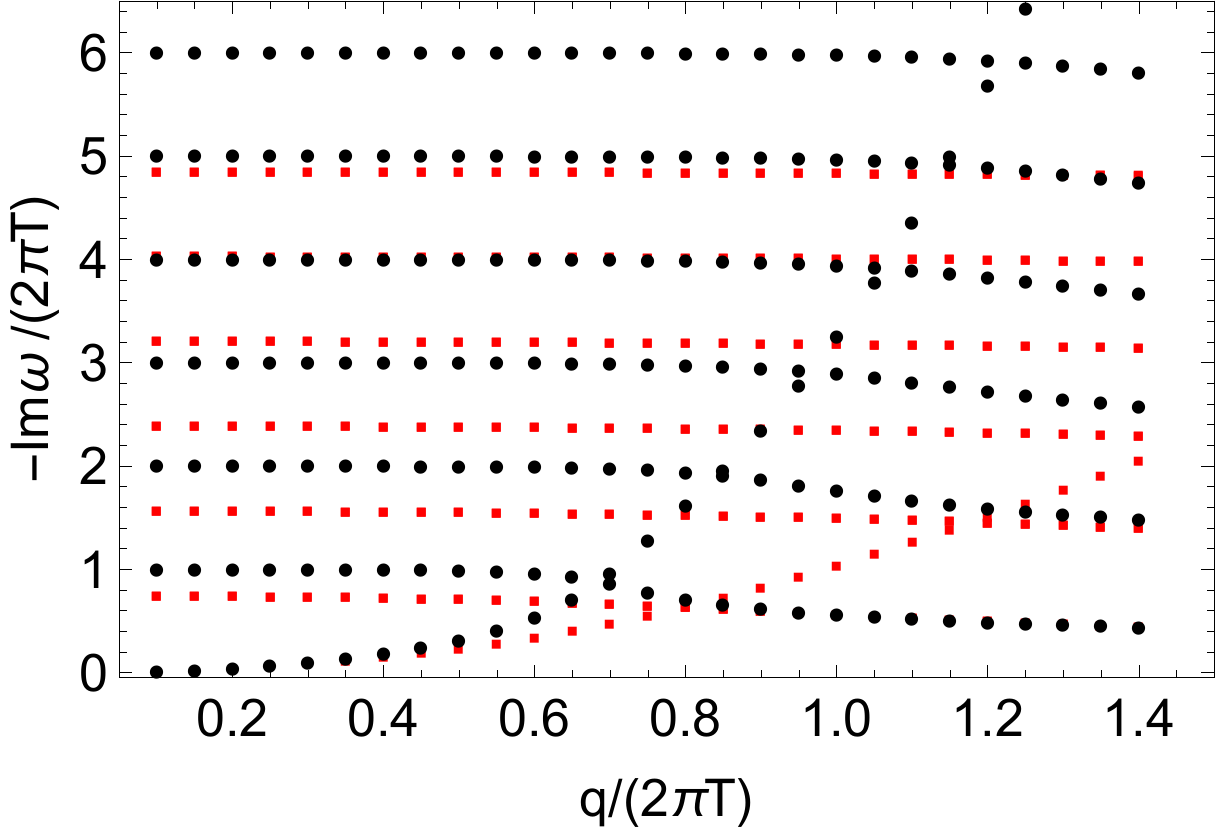}
\caption{
Left: The real part of the frequency as a function of the wavenumber. Right: The imaginary part of the frequency as a function of the wavenumber. Black dots represent the results for ${\mu}=0$ and ${c}=0={b}$ (conformal case), while red squares were obtained for ${\mu}={\mu}_c$, $c=1.46\,\text{GeV}^2$, $b=0.273\,\text{GeV}^4$ and $T=T_c$ (non-conformal case).
}
\label{Fig:qwrwiLongitudinal}
\end{figure}

We also calculated the quasinormal frequencies as a function of the momentum. Our numerical results are displayed in Fig.~\ref{Fig:qwrwiLongitudinal}, where black dots represent the results obtained in the conformal case, while red squares represent the results obtained in the non-conformal case. As can be seen, the real part of the frequency increases when we turn on the parameters, we also observe that higher states are more sensitive to the parameters than lower states. In turn, the imaginary part decreases when we turn on the parameters, again, higher states are more sensitive to the parameters than lower states.

\noindent

\subsection{Transverse sector}

Our starting point is the Schrodinger-like equation \eqref{Eq:SchrodingerTrans}, with the potential given by \eqref{Eq:SchrodingerPotTrans}. First, we implement the transformation $\psi_\alpha=e^{-i\,\omega\,r_*}\varphi$. Then, we replace the tortoise coordinate and $B_T$ to get
\noindent
\begin{equation}
\begin{split}
\left(4q^2z^2+3g+4c^2z^4g-2zg'-4c\,z^3g'\right)\varphi(z)-z^2\left(8\,i\omega+4g'\right)\varphi'(z)-4z^2g\varphi''(z)=0.
\end{split}
\end{equation}
\noindent

In the following we normalize the variable and parameters as in \eqref{Eq:NormalizationQNMs}. To solve numerically the problem we also need to calculate the asymptotic solutions close to the horizon, $u=1$. Considering the ansatz $\varphi=(1-u)^{\alpha}$, then, plugging it in the last equation we get the solutions:
\noindent
\begin{equation}
\alpha_1=0,\qquad\qquad \alpha_2=4\,i\,\widetilde{\omega}.
\end{equation}
\noindent
The first solution is interpreted as waves falling into the black hole through the event horizon, while the second solution represents waves coming out from the black hole interior. Classically nothing comes from the black hole interior, for that reason we work with the first solution. Then, the solution at the horizon reduces to a constant that we can set to the unit. 

Repeating the procedure close to the boundary, $u=0$, considering the ansatz $\varphi=u^{\beta}$, plugging into the differential equation and solving the leading equation we get the solutions:
\noindent
\begin{equation}\label{Eq:BoundarySolTransverse}
\beta_1=-\frac{1}{2},\qquad\qquad \beta_2=\frac{3}{2}.
\end{equation}
\noindent
The first solution is interpreted as the non-normalizable (or source), while the second is interpreted as the normalizable (or vacuum expectation value - VEV). The QN frequencies are solutions of the differential equation by imposing Dirichlet condition at the boundary, such that the poles of the retarded Green's functions correspond to the QN frequencies, this means that we must neglect the non-normalizable solution in \eqref{Eq:BoundarySolTransverse}. Thus, the last transformation is given by $\varphi=u^{3/2}\phi(u)$, yielding  us to the final differential equation 
\noindent
\begin{equation}\label{Eq:TransversalQNMs}
\begin{split}
\left(4\,\widetilde{q}^{\,2}u-6\,i\,\widetilde{\omega}+\widetilde{c}^{\,2}u^3g-(2+\widetilde{c}\,u^2)g'\right)\phi(u)-\left(4\,i\,\widetilde{\omega}u+3g+ug'\right)\phi'(u)-ug\phi''(u)=0.
\end{split}
\end{equation}
\noindent
As a check of consistency, let us calculate the QN frequencies setting $\widetilde{\mu}=0$, $\widetilde{c}=0$, and $\widetilde{b}=0$. The problem reduces to the case investigated in Ref.~\cite{Kovtun:2005ev}. Our numerical results for the frequencies are displayed in Table~\ref{Tab:QNMsTransverse}, for $\widetilde{\mu}=0$. As can be seen, the first four QN frequencies are in good agreement with the results displayed in the first Table of Appendix B in Ref.~\cite{Kovtun:2005ev}. In turn, the fifth frequency is in agreement up to the fourth decimal. It is worth pointing out that the results of Ref.~\cite{Kovtun:2005ev} were obtained using Frobenius method.

\begin{table}[ht!]
\centering
\begin{tabular}{l |c|c|c}
\hline 
\hline
 $n$ & $\widetilde{\mu}=0$ & $\widetilde{\mu}=0.01$ & $\widetilde{\mu}=0.1$ \\
\hline 
 $0$ & $\pm 1.5471870-0.8497232i$  & $\pm1.5471814-0.8497309i$ & $\pm1.5466297-0.8504960i$  \\
 $1$ & $\pm 2.3989034-1.8743432i$  & $\pm2.3988866-1.8743668i$ & $\pm2.3972277-1.8767126i$ \\
 $2$ & $\pm 3.3232289-2.8949008i$  & $\pm3.3231995-2.8949440i$ & $\pm3.3202938-2.8992378i$ \\
 $3$ & $\pm 4.2764313-3.9095832i$  & $\pm4.2763887-3.9096479i$ & $\pm4.2721793-3.9160847i$  \\
 $4$ & $\pm 5.2440583-4.9203464i$  & $\pm5.2440021-4.9204338i$ & $\pm5.2384548-4.9291372i$ \\
 \hline 
\hline
  $n$ & $\widetilde{\mu}=0.2$ & $\widetilde{\mu}=0.5$ & $\widetilde{\mu}=1$ \\
\hline 
 $0$ & $\pm 1.5449644-0.8528302i$  & $\pm1.5336234-0.8698423i$ & $\pm1.5010485-0.9401700i$  \\
 $1$ & $\pm 2.3922258-1.8839098i$  & $\pm2.3587417-1.9382540i$ & $\pm2.3025258-2.1792915i$ \\
 $2$ & $\pm 3.3115486-2.9124735i$  & $\pm3.2546584-3.0154195i$ & $\pm3.2350527-3.4380124i$ \\
 $3$ & $\pm 4.2595354-3.9360044i$  & $\pm4.1803200-4.0948307i$ & $\pm4.2135273-4.6768033i$  \\
 $4$ & $\pm 5.2218242-4.9561634i$  & $\pm5.1223650-5.1761809i$ & $\pm5.2107958-5.9057699i$ \\
  \hline 
\hline
  $n$ & $\widetilde{\mu}=1.2$ & $\widetilde{\mu}=1.5$ & $\widetilde{\mu}=2$ \\
\hline 
 $0$ & $\pm1.4917208-0.9860494i$  & $\pm1.4946232-1.0638620i$ & $\pm1.5262256-1.1876869i$  \\
 $1$ & $\pm2.3258501-2.3022563i$  & $\pm2.3822490-2.4727024i$ & $\pm2.5070989-2.7321848i$ \\
 $2$ & $\pm3.2897756-3.6134830i$  & $\pm3.3970290-3.8642567i$ & $\pm3.6112874-4.2453557i$ \\
 $3$ & $\pm4.2964798-4.9072267i$  & $\pm4.4515234-5.2357684i$ & $\pm4.7516006-5.7369090i$  \\
 $4$ & $\pm5.3214206-6.1898699i$  & $\pm5.5227181-6.5960159i$ & $\pm5.9070037-7.2171368i$ \\
\hline\hline
\end{tabular}
\caption{
The quasinormal frequencies for selected values of the chemical potential for $q/(2\pi T)=1$, ${c}=0$, and ${b}=0$. The results for ${\mu}=0$ are equivalent to those of Ref.~\cite{Kovtun:2005ev}.
}
\label{Tab:QNMsTransverse}
\end{table}

\begin{figure}[ht!]
\centering
\includegraphics[width=7cm]{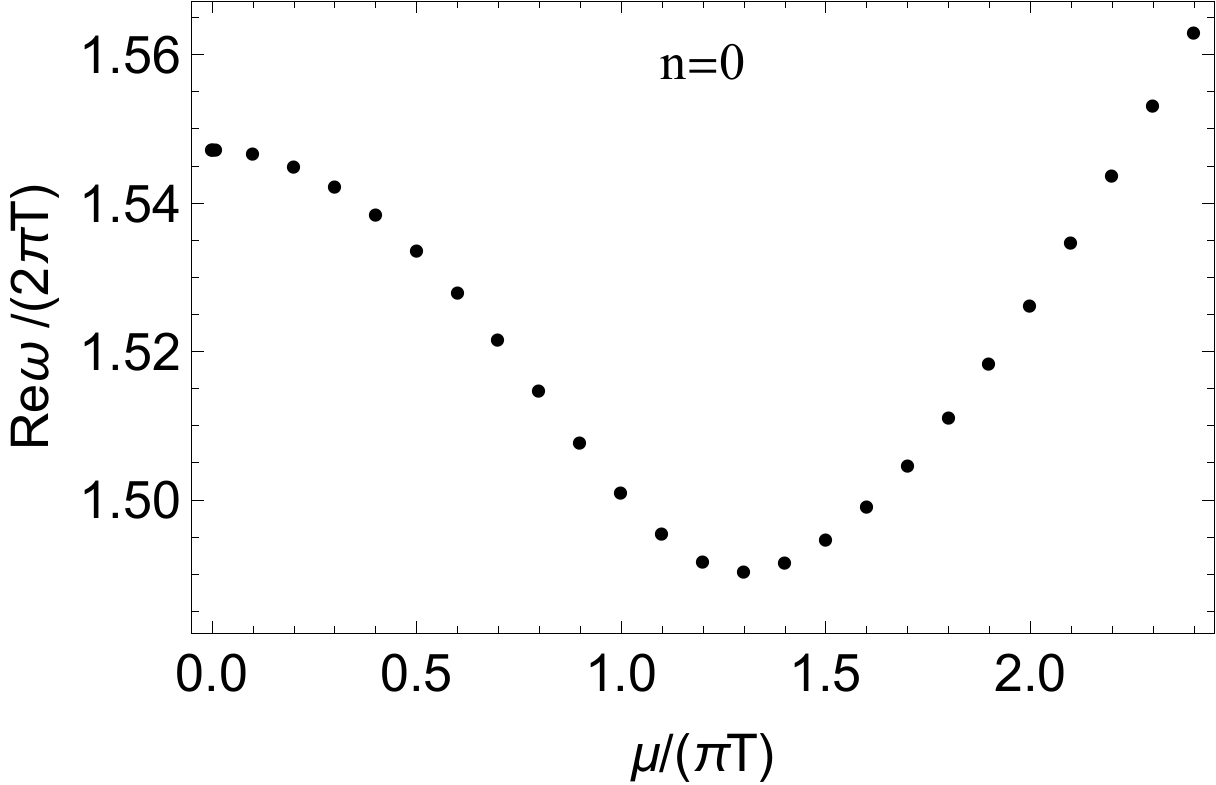}\hfill
\includegraphics[width=7cm]{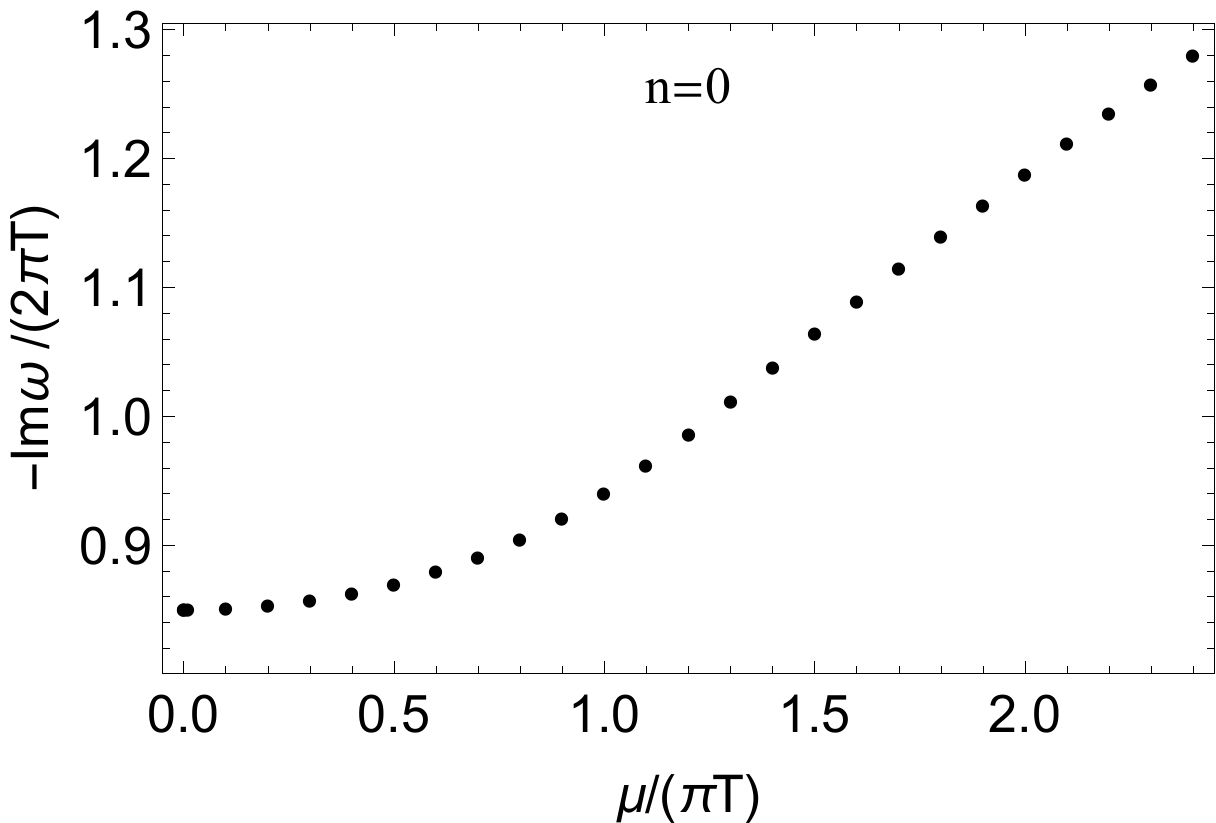}
\newline
\includegraphics[width=7cm]{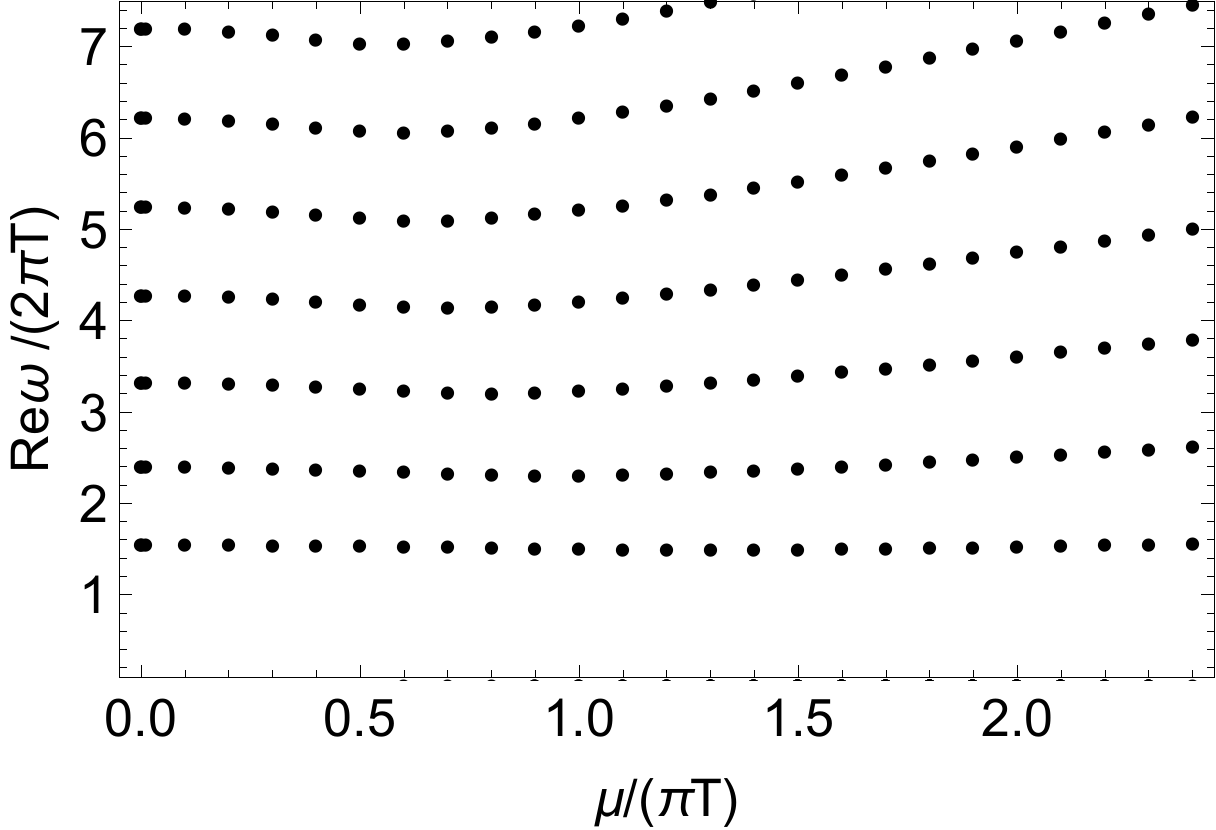}\hfill
\includegraphics[width=7cm]{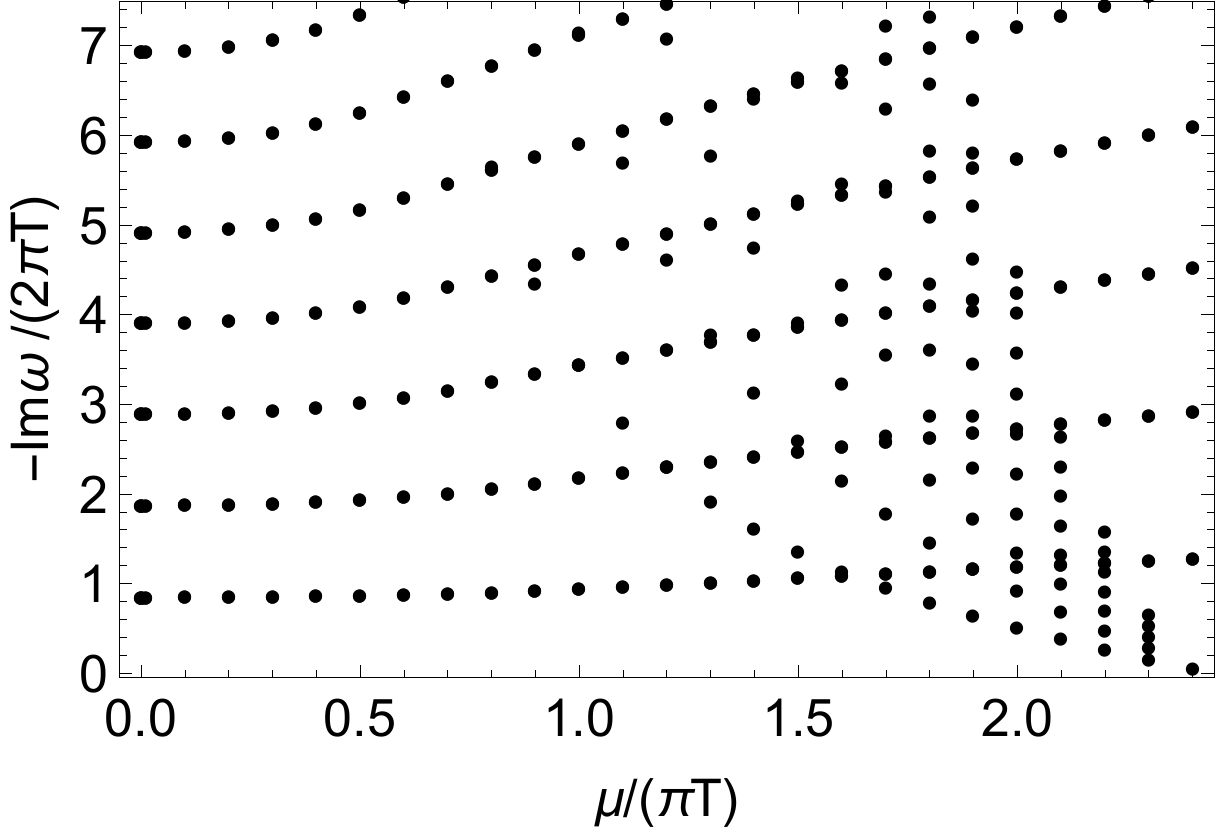}
\caption{
Left: The real part of the frequency as a function of the chemical potential. Right: The imaginary part of the frequency as a function of the chemical potential. These results were obtained setting $q/(2\pi T)=1$, and ${c}=0={b}$.
}
\label{Fig:MuwrwiTransversal}
\end{figure}

A plot of the real and imaginary parts of the frequency as a function of the chemical potential is displayed in Fig.~\ref{Fig:MuwrwiTransversal}. As can be seen, the real part decreases in the region of small values of the chemical potential, reaches a minimum, then, it increases, see top-left panel. This behavior is shared by the real part of the other quasinormal frequencies. Meanwhile, the imaginary part increases monotonically with the increasing of the chemical potential, see top-right panel for $n=0$ state, and bottom-right panel for the first six quasinormal frequencies. These results were obtained considering $\widetilde{q}=1$ and $\widetilde{c}=0=\widetilde{b}$. As 
mentioned above, we restrict our results to the region $\widetilde{\mu}\leq \sqrt{6}$ where the pseudo-spectral method provides reliable results. Observing carefully the imaginary part of the frequency in Fig.~\ref{Fig:MuwrwiTransversal} we can see additional frequencies which are purely imaginary. These frequencies decrease with the increasing of the chemical potential.

\noindent

\begin{figure}[ht!]
\centering
\includegraphics[width=7cm]{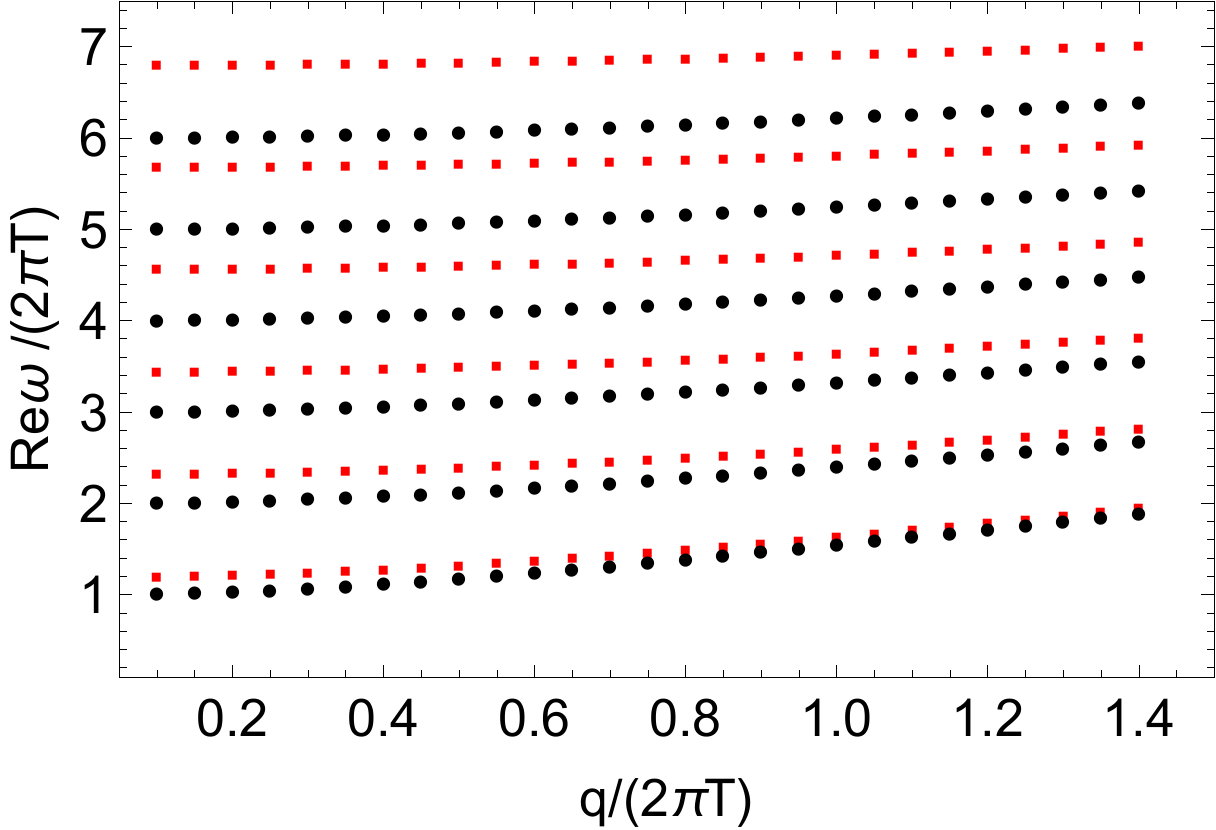}\hfill
\includegraphics[width=7cm]{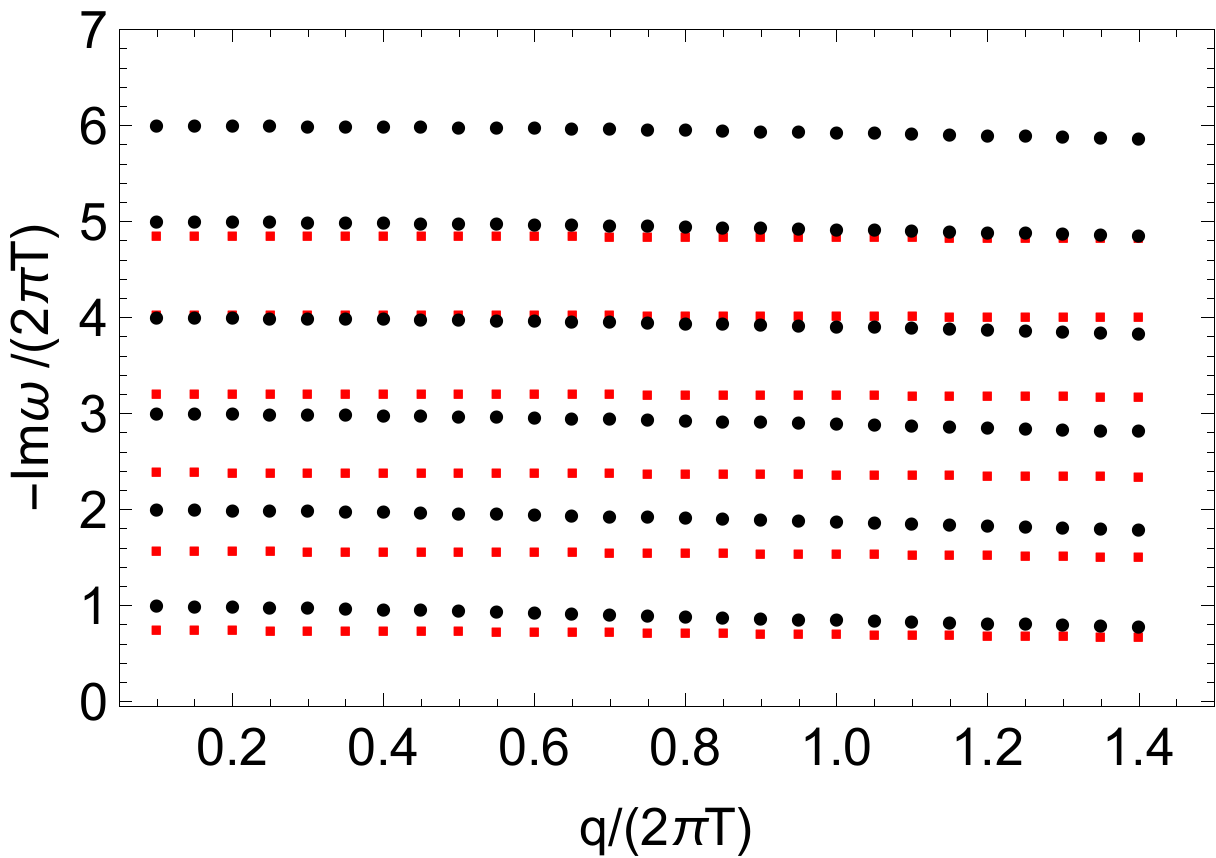}
\caption{
Left: The real part of the frequency as a function of the wavenumber. Right: The imaginary part of the frequency as a function of the wavenumber. Black dots represent the results for ${\mu}=0$ and ${c}=0={b}$ (conformal case), while red squares were obtained for ${\mu}={\mu}_c$, $c=1.46\,\text{GeV}^2$, $b=0.273\,\text{GeV}^4$ and $T=T_c$ (non-conformal case).
}
\label{Fig:qwrwiTransversal}
\end{figure}

Now, one can ask about the behavior of the quasinormal frequencies when one turn on the parameters $c=1.46\,\text{GeV}^2$ and $b=0.273\,\text{GeV}^4$. To see how the quasinormal frequencies change in the non-conformal case we also consider the value of the chemical potential and temperature as being $\mu=\mu_c=0.708\,\text{GeV}$ and $T=T_c=0.559\,\text{GeV}$, respectively.
Our numerical results for the real part of the frequency are displayed in left panel of Fig.~\ref{Fig:qwrwiTransversal}, while the imaginary part of the frequency are displayed in right panel of Fig.~\ref{Fig:qwrwiTransversal}. In this figure black dots represent the results for conformal symmetry case, while red squares represent the results for non-conformal case. As can be seen, the real part of the frequency increases while the imaginary part decreases when we turn on the parameters of the model. Note that lower states are less sensitive to the parameters than higher states. It is also worth mentioning that the precision of the numerical results gets poor when we increase the value of the parameters $\mu$, $c$ and $b$.

\noindent

\section{Conclusion}
\label{Sec:Conclusion}

In this paper we investigated the melting of charmonium in a holographic model describing heavy quark anti-quark  systems \cite{He:2013qq}. The holographic model provides a fit of charmonium masses at zero temperature. For investigating the finite temperature behaviour of these states,  a black hole was embedded in the gravitational background. In this way, we wrote the perturbation equations in the Schr\"odinger-like form. This analysis allows us to investigate how the potential well is deformed by the temperature and chemical potential, which is interpreted as the melting/dissociation process. To complement the analysis we calculated the spectral functions, where peaks are interpreted as the quasiparticle states. We show that the height and width of the peaks are affected by the temperature and chemical potential. These results represent the dissociation of the charmonium states in the finite density plasma. By comparing spectral functions for different temperatures and values of the chemical potential we observed that the chemical potential speeds up the melting process.

In the second part of this paper, we solved the perturbation equations in the hydrodynamic limit. By imposing Dirichlet condition at the boundary we calculated the dispersion relations. By comparing the dispersion relation obtained in the longitudinal sector against the dispersion relation obtained in fluid dynamics we were able to read off the diffusion coefficient. We also showed that the dispersion relation of the longitudinal sector is related to the pole of the $C_{tt}^{R}(\omega,q)$ retarded function in the dual field theory. Then, we calculated the quark number susceptibility following two approaches. The first approach considers the retarded function $C_{tt}^{R}(\omega,q)$ in the limit of zero wave number and frequency. The second approach considers the baryon density and its derivative. The quark number susceptibility does not blows up at the critical end point in the first approach, while in the second it does. The explanation for this apparent discrepancy is that  in the first approach one considers the matter action that was introduced as probe fields, while in the second approach, the quark number susceptibility is obtained from the background equations. This result suggests us that considering probe fields for describing the mesons maybe is not enough to extract full information of the dual field theory.

In the last part of this paper we solved the perturbation equations numerically using the pseudo-spectral method. Imposing ingoing condition at the horizon and Dirichlet at the boundary we obtained a discrete set of complex frequencies. In the longitudinal sector we observed a very interesting results when we fixed the momentum and varying  the chemical potential considering also the parameters $c=0=b$. The hydrodynamic mode increases with the chemical potential up to some value of the chemical potential where it merges with a mode whose frequency decreases with the increasing of the chemical potential, see right panel of Fig.~\ref{Fig:MuwrwiLongitudinal}. This new mode has a real part arising at exactly the same value of the chemical potential where the hydrodynamic mode merges with this mode. This kind of behavior was previously observed in the literature and is interpreted as a double pole in the retarded Green function \cite{Miranda:2008vb}. We found additional frequencies which are purely imaginary and they decrease with the increasing of the chemical potential. In turn, we also calculated the QN frequencies as a function of the wave number in the conformal and non-conformal limit. The real part of the frequency increases when we compare results obtained in the conformal case, i.e., $c=0$, $b=0$ and $\mu=0$, against results obtained in the non-conformal case, i.e., $c\neq 0$, $b\neq0$ and $\mu \neq0$, while the imaginary part decreases when we compared the conformal against the non-conformal case. These conclusions can be extended for the transverse sector, where the quasinormal frequencies do not bring any new information.

Finally, it would be interesting to investigate how the results obtained in this paper are affected by magnetic field or angular momentum \cite{Chen:2020ath, Braga:2022yfe}. We also are interested in the transport coefficients, which are obtained in the hydrodynamic limit. For example, we believe that the diffusion coefficient will be affected by the rotation parameter in the same form as obtained in Ref.~\cite{Mamani:2018qzl}. These problems and further extensions will be addressed in the future.

\section*{Acknowledgments}
The authors would like to acknowledge Song He, Alfonso Ballon Bayona and Alex Miranda for discussions along the development of this work. L.~A.~H.~M. is partially founded by the Universidade Estadual da Regi\~ao Tocantina do Maranh\~ao (UEMASUL, Brazil). D. F. Hou  is supported  in part by the National Natural Science Foundation of  China (NSFC) 
under Grant Nos. 11735007, 11890711, 35111890710. N.R.F.B. is partially supported by  CNPq - Conselho Nacional de Desenvolvimento Cientifico e Tecnologico grant  307641/2015-5 and by  Coordena\c{c}\~ao de Aperfei\c{c}oamento de Pessoal de N\'ivel Superior - Brasil (CAPES) - Finance Code 001.  

\appendix

\section{ Charmonium Spectrum }
\label{Sec:SpectrumT0}

Introducing the Fourier transform on the gauge field it transforms as $A^{\nu}(x^{\mu},z)\to A^{\nu}(k^{\mu},z)$. The resulting equation may be written in the Schr\"odinger-like form using the transformation $A_{\nu}=\xi_{\nu}e^{-B}\psi$, where $\xi_{\nu}$ is a polarization vector and $2B=\ln{\left(f/\zeta\right)}$, the resulting equation is
\noindent
\begin{equation}
-\partial_z^2\psi+V\,\psi=m^2\,\psi,
\end{equation}
\noindent
where we have replaced $\square\to m^2$, the mass of the particles, and $V$ is the potential given by
\noindent
\begin{equation}
V=\left(\partial_z\,B\right)^2+\partial_z^2B.
\end{equation}
\noindent
As the background was already fixed, we may solve the eigenvalue problem using a shooting method, for example. It is worth pointing out that the ratio $f/\zeta$ does not depend on the parameter $b$, for that reason the spectrum is insensitive to this parameter. Thus, in this case the problem has an analytic solution given by
\noindent
\begin{equation}
m_n^2=4\,c\,(n+1),\qquad\qquad n=0,1,2,\cdots
\end{equation}
\noindent
In the sequence, we fix the free parameter by fitting our formula with the first two resonances of the experimental data $3686.109\pm 0.012$ MeV and $4039\pm 1$ MeV, thus, we get $c=1.46\,\text{GeV}^2$. We decided to fix the parameter in this wave to avoid the lightest states. The numerical results of the spectrum compared against the results of Ref.~\cite{He:2013qq} and experimental data are displayed in Table.~\ref{Tab:VectorSF}.

Here, we presented the asymptotic solutions of the differential equations. Let us start with the model at zero temperature. Plugging the warp factor and kinetic function \eqref{Eq:HologrphicModel} in \eqref{Eq:PotentialT0}, the potential of the Sch\"rodinger-like equation becomes
\noindent
\begin{equation}
V=\frac{3}{4z^2}+c^2z^2.
\end{equation}
\noindent
As can be seen, the spectrum does not depend on the parameter $b$. Plugging the potential in the Schr\"odinger-like equation \eqref{Eq:SchrodingerT0} and considering the ansatz, $\psi=z^{\alpha}$, close to the boundary, we get the solution
\noindent
\begin{equation}
\psi=c_1 z^{-1/2}+c_2 z^{3/2}.
\end{equation}
\noindent
As we are looking for normalizable solutions of the eigenvalue problem we set $c_1=0$.

In turn, in the IR regime, the asymptotic solution may be obtained considering the leading term of the potential, thus, solving the Schrodinger-like equation we get 
\noindent
\begin{equation}
\psi=c_3\, e^{-c z^2/2}.
\end{equation}
\noindent
As the background does not have any singular behavior in the intermediate region, we conclude that the solutions of the Schr\"odinger-like equation are regular and normalizable.

\begin{table}[ht]
\centering
\begin{tabular}{l |c|c|c}
\hline 
\hline
 $n$ & Model & Model  & Quarkonium experimental \cite{Tanabashi:2018oca} \\
  & $c=1.46\,\text{GeV}^2$& $c=1.16\,\text{GeV}^2$ \cite{He:2013qq}& (MeV) \\
\hline 
 $0$ & 2420 & 2154 & $3096.916\pm 0.011$  \\
 $1$ & 3422 & 3046 & $3686.109\pm 0.012$  \\
 $2$ & 4191 & 3731 & $4039\pm 1$ \\
 $3$ & 4839 & & $4421\pm 4$  \\
\hline\hline
\end{tabular}
\caption{
The mass of the heavy vector mesons (in MeV) obtained in the holographic model, compared against the holographic model \cite{He:2013qq} and experimental results from PDG \cite{Tanabashi:2018oca}.
}
\label{Tab:VectorSF}
\end{table}

On the other hand, the problem change at finite temperature. At the horizon the potential is zero due to $g(z_h)=0$. Then, the Schr\"odinger-like equations have the asymptotic solution
\noindent
\begin{equation}
\psi_{k}=C_{k}\,e^{-i\,\omega\,r_*}+D_{k}\,e^{+i\,\omega\,r_*},\qquad (k=x^1, x^2, x^3)
\end{equation}

\begin{figure}[ht!]
\centering
\includegraphics[width=7cm]{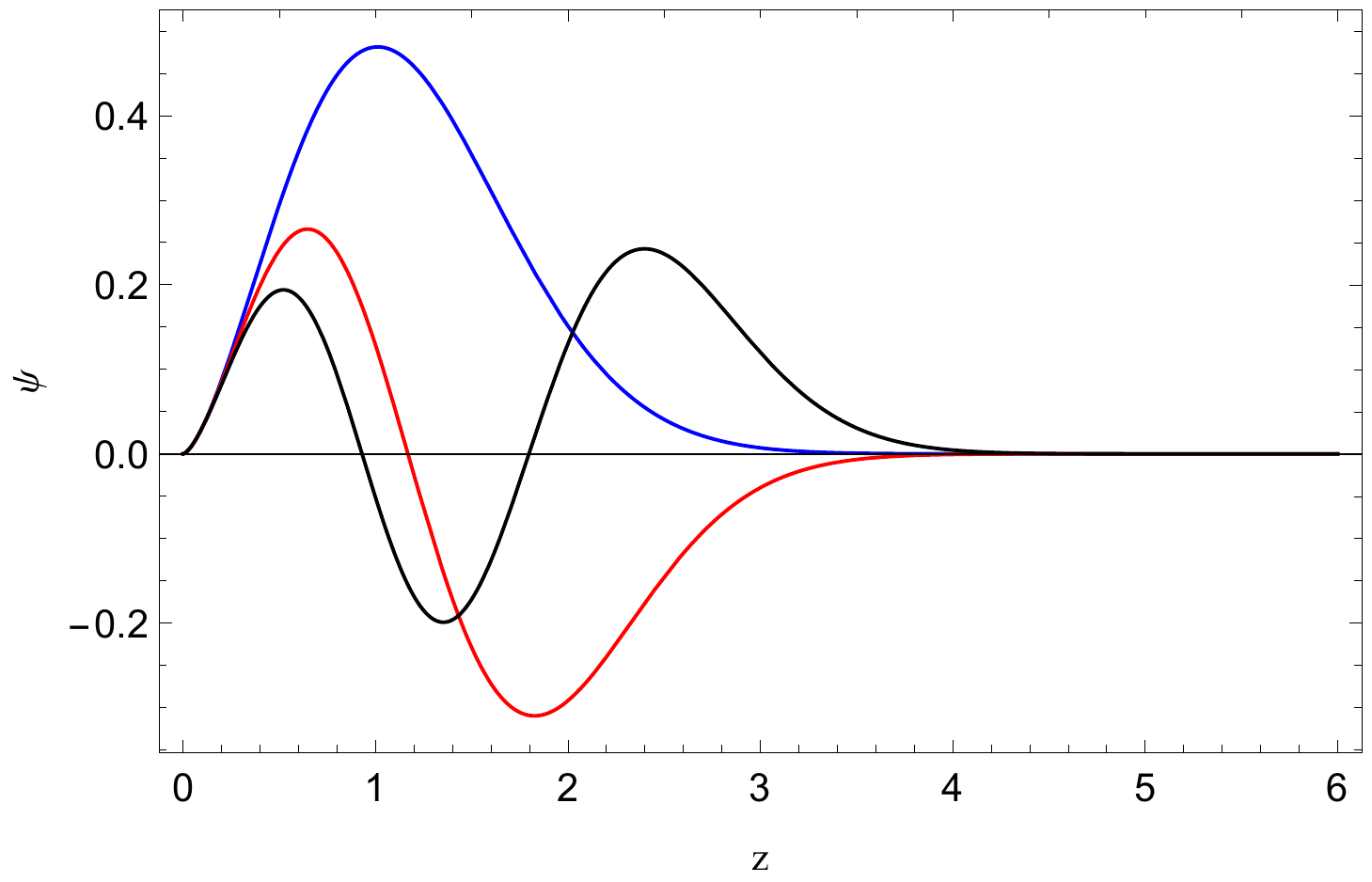}\hfill 
\includegraphics[width=7cm]{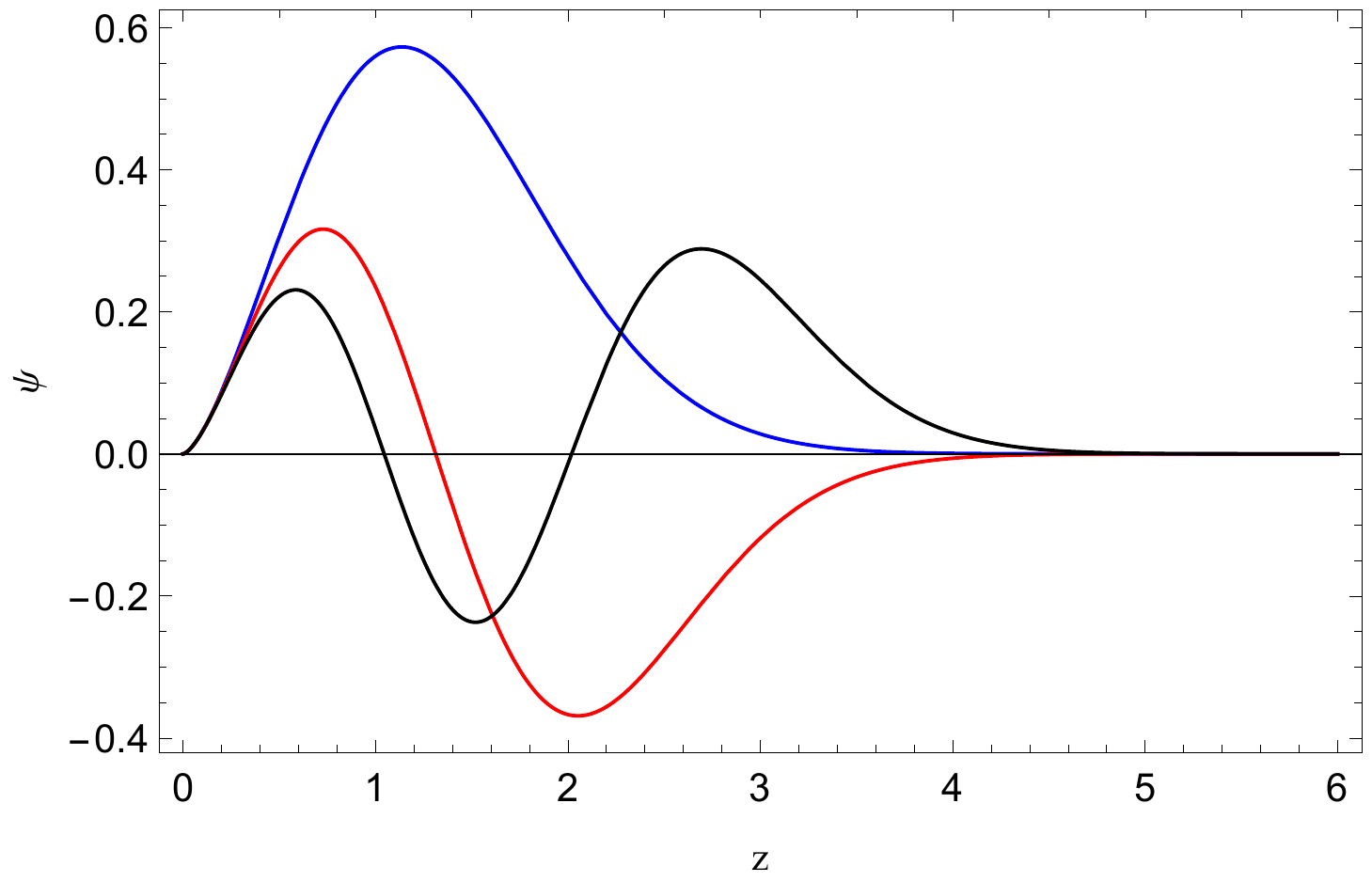}
\caption{
Left: The wave functions for $c=1.46\,\text{GeV}^2$. Right: The wave functions for $c=1.16\,\text{GeV}^2$.
}
\label{Fig1:WaveFunct}
\end{figure}

Considering the parameter $c=1.46\,\text{GeV}^2$ we calculate the wave functions. Our numerical results are displayed in the left panel of Fig.~\ref{Fig1:WaveFunct}, while the right panel shows the wave functions for $c=1.16\,\text{GeV}^2$ calculated in Ref.~\cite{He:2013qq}.

\noindent

\bibliographystyle{utphys}
\bibliography{References}

\end{document}